\documentclass[]{aa}

\input{psfig.sty}
\usepackage{lscape}
\begin{document}
%
% 10: Galaxy (section number)
% 05.01.1 astrometry
% 08.04.1 stars: distances
% 08.06.3 stars: fundamental parameters
% 08.08.1 stars: HRD
% 10.15.2 open clusters and associations: individual: Hyades
\thesaurus{10(05.01.1; 08.04.1; 08.06.3; 08.08.1; 10.15.2 Hyades)}
\title{A Hipparcos study of the Hyades open cluster%
%\thanks{Based on data from ESA's Hipparcos astrometry satellite}
}
\subtitle{Improved colour-absolute magnitude and Hertzsprung--Russell diagrams}

\author{J.H.J.\ de Bruijne, R.\ Hoogerwerf \and P.T.\ de Zeeuw}
\authorrunning{J.H.J.\ de Bruijne et al.}

\offprints{P.T.\ de Zeeuw}
\mail{tim@strw.leidenuniv.nl}

\institute{Sterrewacht Leiden, Postbus 9513, 2300 RA Leiden, the Netherlands}

\date{Accepted for publication in A\&A}

\hyphenation{Bruij-ne Hoog-er-werf}

\maketitle

\begin{abstract}

Hipparcos parallaxes fix distances to individual stars in the Hyades
cluster with an accuracy of $\sim$6~percent. We use the Hipparcos
proper motions, which have a larger relative precision than the
trigonometric parallaxes, to derive $\sim$3 times more precise
distance estimates, by assuming that all members share the same space
motion. An investigation of the available kinematic data confirms that
the Hyades velocity field does not contain significant structure in
the form of rotation and/or shear, but is fully consistent with a
common space motion plus a (one-dimensional) internal velocity
dispersion of $\sim$0.30~km~s$^{-1}$. The improved parallaxes as a set
are statistically consistent with the Hipparcos parallaxes. The
maximum expected systematic error in the proper motion-based
parallaxes for stars in the outer regions of the cluster (i.e., beyond
$\sim$2 tidal radii $\sim$20~pc) is $\la$0.30~mas. The new parallaxes
confirm that the Hipparcos measurements are correlated on small
angular scales, consistent with the limits specified in the Hipparcos
Catalogue, though with significantly smaller `amplitudes' than claimed
by Narayanan \& Gould. We use the Tycho--2 long time-baseline
astrometric catalogue to derive a set of independent proper
motion-based parallaxes for the Hipparcos members.

The new parallaxes provide a uniquely sharp view of the
three-dimensional structure of the Hyades. The colour-absolute
magnitude diagram of the cluster based on the new parallaxes shows a
well-defined main sequence with two `gaps'/`turn-offs'. These features
provide the first direct observational support of B\"ohm--Vitense's
prediction that (the onset of) surface convection in stars
significantly affects their $(B-V)$ colours. We present and discuss
the theoretical Hertzsprung--Russell diagram ($\log L$ versus $\log
T_{\rm eff}$) for an objectively defined set of 88 high-fidelity
members of the cluster as well as the $\delta$~Scuti star $\theta^2$
Tau, the giants $\delta^1$, $\theta^1$, $\epsilon$, and $\gamma$ Tau,
and the white dwarfs V471 Tau and HD 27483 (all of which are also
members). The precision with which the new parallaxes place individual
Hyades in the Hertzsprung--Russell diagram is limited by (systematic)
uncertainties related to the transformations from observed colours and
absolute magnitudes to effective temperatures and luminosities. The
new parallaxes provide stringent constraints on the calibration of
such transformations when combined with detailed theoretical stellar
evolutionary modelling, tailored to the chemical composition and age
of the Hyades, over the large stellar mass range of the cluster probed
by Hipparcos.

\keywords{astrometry --
          stars: distances --
          stars: fundamental parameters --
          stars: Hertzsprung--Russell diagram --
          open clusters and associations: individual: Hyades}

\end{abstract}

%--- Figure 1 -----------------------------------------------------------------
\def\placefigureOne{
\begin{figure*}[t!]
\centerline{\psfig{file=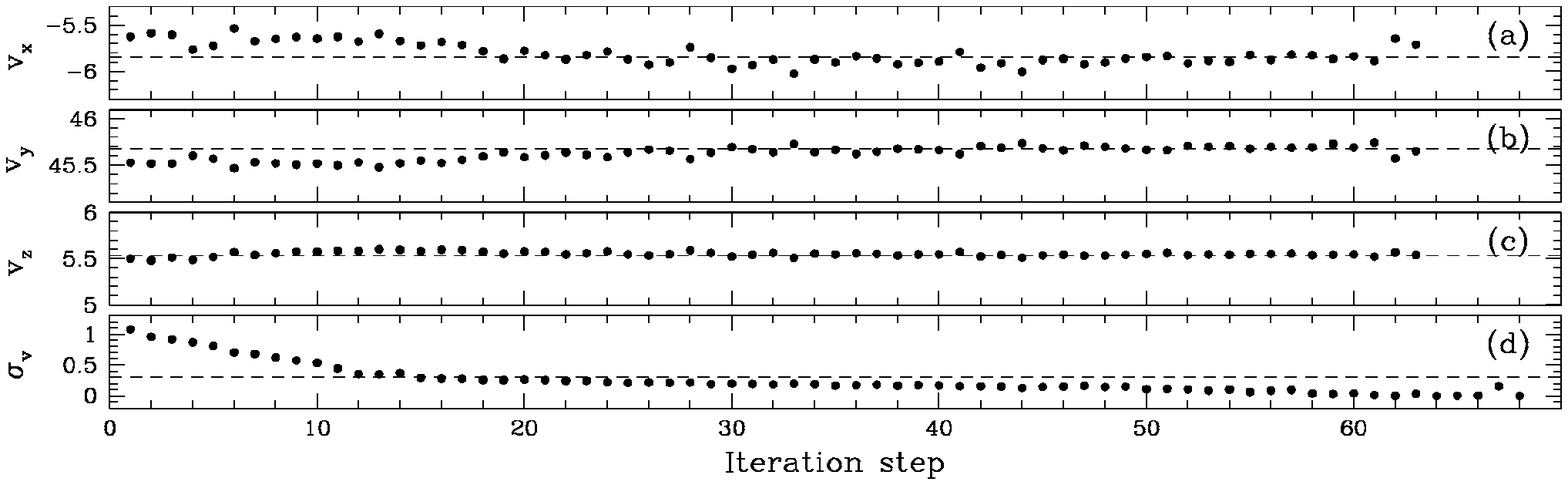,width=0.90\textwidth,silent=}}
\caption[]{{\it Panels a, b, c:\/} the evolution of the cluster space
motion components $v_x$, $v_y$, and $v_z$ (in km~s$^{-1}$) in the
equatorial Cartesian ICRS frame during the rejection of stars in the
{\it revised\/} iterative procedure using $\sigma_v =
0.30$~km~s$^{-1}$, $\sigma_\Delta = 0.5$, and a $g$ rejection limit of
9 (\S\S \ref{subsec:revision} and \ref{subsec:space_motion}). The dots
show the values of the space motion components at a given iteration
step. The dashed lines denote the median values of the space motion
components: $(v_x, v_y, v_z) = (-5.84, 45.68, 5.54)$~km~s$^{-1}$. {\it
Panel d:\/} the evolution of the velocity dispersion $\sigma_v$ during
the rejection of stars in the {\it unrevised\/} procedure using a $g$
rejection limit of 9 (\S\S \ref{subsec:outline} and
\ref{subsec:vel_disp}). The dots show the median values of the
velocity dispersion at a given iteration step. The discontinuity of
the slope of the relation exhibited by the dots around step $\sim$12
coincides with the physical (one-dimensional) velocity dispersion of
the Hyades ($\sim$$0.30$~km~s$^{-1}$; dashed
line).\label{fig:velocity}}
\end{figure*}
}
%--- Figure 1 -----------------------------------------------------------------

%--- Figure 2 -----------------------------------------------------------------
\def\placefigureTwo{
\begin{figure}[t!]
\psfig{file=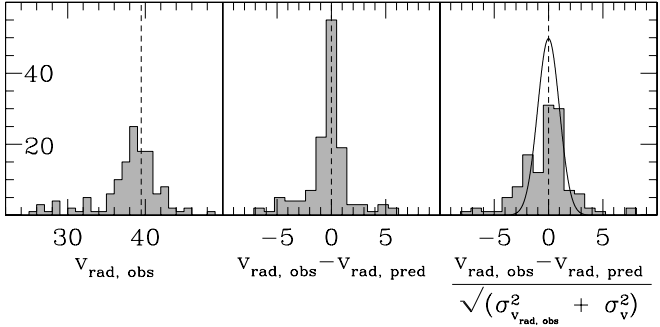,width=8.7truecm,silent=}
\caption[]{Observed ($v_{\rm rad, obs}$) and predicted ($v_{\rm rad,
pred}$; \S \ref{subsec:revision}) radial velocities for 131 secure
members (\S \ref{sec:space_motion}). {\it Left:\/} distribution of
observed radial velocities; the dashed line denotes the radial
component of the Hyades cluster motion derived in this study
($39.48$~km~s$^{-1}$) using $\sigma_v = 0.30$~km~s$^{-1}$ and
$\sigma_\Delta = 0.5$ ([0.30, 0.5] in Table~\ref{tab:space_mot_comp}).
The spread and skewness of the distribution are caused by the
perspective effect. {\it Middle:\/} distribution of observed minus
predicted radial velocities; 71/60 stars have negative/positive
$v_{\rm rad, obs} - v_{\rm rad, pred}$. {\it Right:\/} normalized
distribution of observed minus predicted radial velocities, taking
into account a velocity dispersion of $\sigma_v =
0.30$~km~s$^{-1}$. The black curve is a properly scaled zero-mean
unit-variance Gaussian; the mismatch between the observations and
prediction can be due to non-members and/or undetected close binaries,
an underestimated velocity dispersion, underestimated radial velocity
errors, or a combination of these effects (\S
\ref{subsec:space_motion}). Five stars fall outside the plotted
range.\label{fig:v_rad} }
\end{figure}
}
%--- Figure 2 -----------------------------------------------------------------

%--- Figure 3 -----------------------------------------------------------------
\def\placefigureThree{
\begin{figure}[t!]
\psfig{file=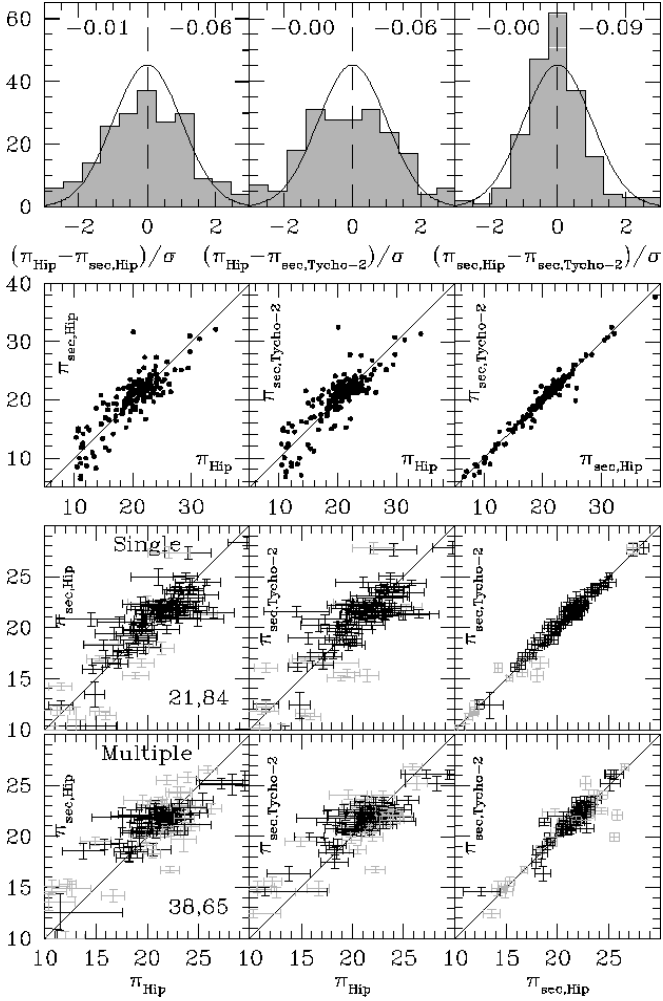,width=8.7truecm,silent=}
\caption[]{{\it First and second row:\/} the 208 entries which have
Hipparcos trigonometric ($\pi_{\rm Hip}$), Hipparcos secular
($\pi_{\rm sec, Hip}$), and Tycho--2 secular ($\pi_{\rm sec,
Tycho-2}$) parallaxes (in mas). The top row shows properly normalized
difference distributions for $\pi_{\rm Hip} - \pi_{\rm sec, Hip}$
(column 1), $\pi_{\rm Hip} - \pi_{\rm sec, Tycho-2}$ (column 2), and
$\pi_{\rm sec, Hip} - \pi_{\rm sec, Tycho-2}$ (column 3). The black
curves are zero-mean unit-variance Gaussian distributions. The numbers
in the top of each top panel denote the mean (left) and median (right)
of the plotted difference. The second row compares the different
parallaxes. {\it Third and fourth row:\/} trigonometric and secular
parallaxes, and their random errors (\S \ref{subsec:accuracy}), for
the same stars. The third row compares the different parallaxes for
single stars, while the bottom row shows multiple stars (i.e., either
column (s) in P98's table 2 is `SB' or `RV', column (t) is `H', `I',
or `M', or column (u) is one of `C$\,$G$\,$O$\,$V$\,$X$\,$S'). Black
symbols have $g_{\rm Hip} \leq 9$; gray symbols have $g_{\rm Hip} >
9$. The numbers in the left panels indicate the relevant numbers of
stars (gray, black). Stars having large goodness-of-fit parameters
$g_{\rm Hip}$ are $\sim$2 times more likely multiple than low-$g_{\rm
Hip}$ stars (cf.\ \S \ref{sec:space_motion}).\label{fig:prlxs} }
\end{figure}
}
%--- Figure 3 -----------------------------------------------------------------

%--- Figure 4 -----------------------------------------------------------------
\def\placefigureFour{
\begin{figure*}[t!]
\centerline{\psfig{file=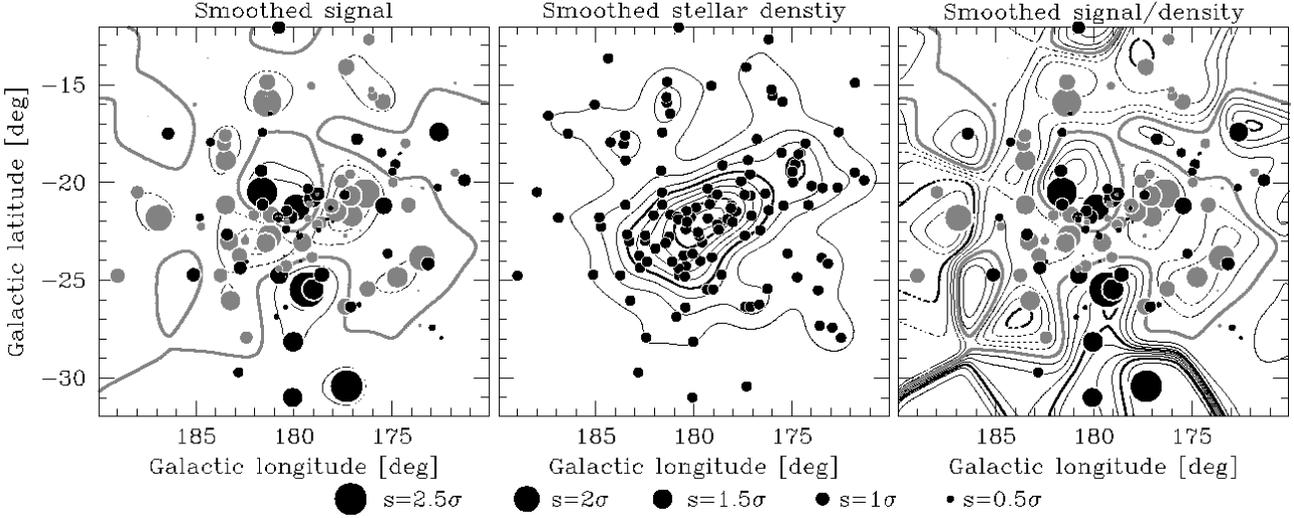,width=0.95\textwidth,silent=}}
\caption[]{{\it Left:\/} the trigonometric-minus-secular parallax
difference field $s(\ell, b)$ (eq.~\ref{eq:signal}), smoothed using
the Gaussian kernel $G(\ell, b)$ (eq.~\ref{eq:kernel}) with smoothing
length $\sigma_{\rm s} = 1^\circ\!$. Solid contours correspond to $s
\geq 0$; dotted contours correspond to $s < 0$; gray contours denote
$s = 0$. Heavy/light contours are spaced by $1.0$/$0.25$. The dots
indicate the positions of 127 Hyades with non-suspect secular
parallaxes ($g_{\rm Hip} \leq 9$). Black symbols have $s \geq 0$ (60
stars), while gray symbols have $s < 0$ (67 stars). The symbol sizes
correspond linearly to the strength of the signal $s$ (larger symbols
denote larger $|s|$; see legend). {\it Middle:\/} as left panel, but
for the projected stellar number density $\rho(\ell, b) \equiv
\delta_{\rm D}(\ell, b)$. The lowest contour level equals
$0.25$~star~${\rm deg}^{-2}$. {\it Right:\/} as left panel, but for
the signal $s(\ell, b)$ divided by the density $\rho(\ell,
b)$.\label{fig:NG99b_fig09} }
\end{figure*}
}
%--- Figure 4 -----------------------------------------------------------------

%--- Figure 5 -----------------------------------------------------------------
\def\placefigureFive{
\begin{figure*}[t!]
\centerline{\psfig{file=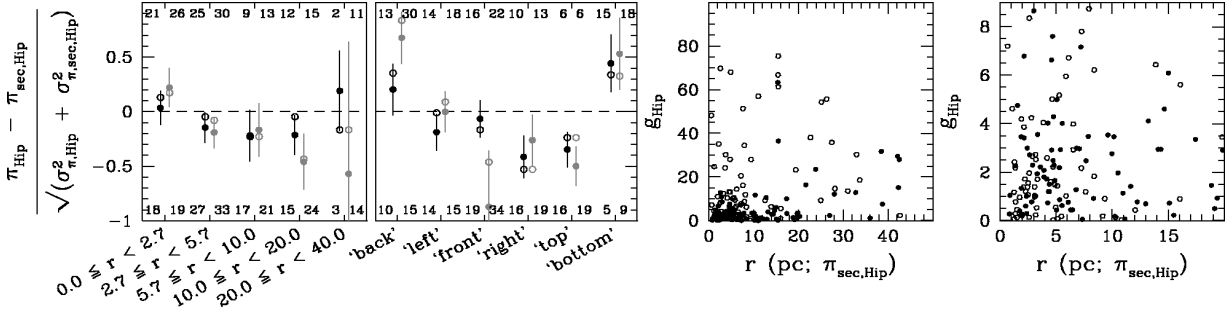,width=0.90\textwidth,silent=}}
\caption[]{{\it Left two panels:\/} normalized difference between
Hipparcos trigonometric and secular parallaxes for the 218 P98 members
as function of location in the cluster. The left panel shows a spatial
division according to $r$ (the three-dimensional distance to the
cluster center using Hipparcos secular parallaxes) in five spherical
annuli (from center to core radius [2.7~pc] to half-mass radius
[5.7~pc] to tidal radius [10~pc] to two tidal radii [20~pc] to 40~pc).
The cluster center is $(b_u, b_v, b_w) = (-43.37, 0.40, -17.46)$~pc in
Galactic Cartesian coordinates (cf. table~3 in P98). The right panel
shows a spatial division according to the six equal-volume pyramids
(\S \ref{subsec:3d_pos}; orientation in Galactic coordinates).
Black/gray symbols show results for stars with $g_{\rm Hip} \leq
9/\infty$. Filled symbols and vertical lines denote mean and standard
deviation ($\pm$1$\sigma$ uncertainty), while open symbols denote
median values; the numbers in the upper/lower halves of the panels
denote the corresponding numbers of stars with $\pi_{\rm Hip} -
\pi_{\rm sec, Hip} >$/$\leq 0$. {\it Right two panels:\/} the
goodness-of-fit parameter $g_{\rm Hip}$ as function of $r$ (left; 202
stars) and a magnification of this panel for 144 high-fidelity (i.e.,
low-$g_{\rm Hip}$) members in the central parts of the cluster
(right). Open symbols denote multiple stars (i.e., either column (s)
in P98's table 2 is `SB' or `RV', column (t) is `H', `I', or `M', or
column (u) is one of `C$\,$G$\,$O$\,$V$\,$X$\,$S'; 103/64 entries in
the left/right panel).\label{fig:prlxs_and_gs} }
\end{figure*}
}
%--- Figure 5 -----------------------------------------------------------------

%--- Figure 6 -----------------------------------------------------------------
\def\placefigureSix{
\begin{figure*}[t!]
\centerline{\psfig{file=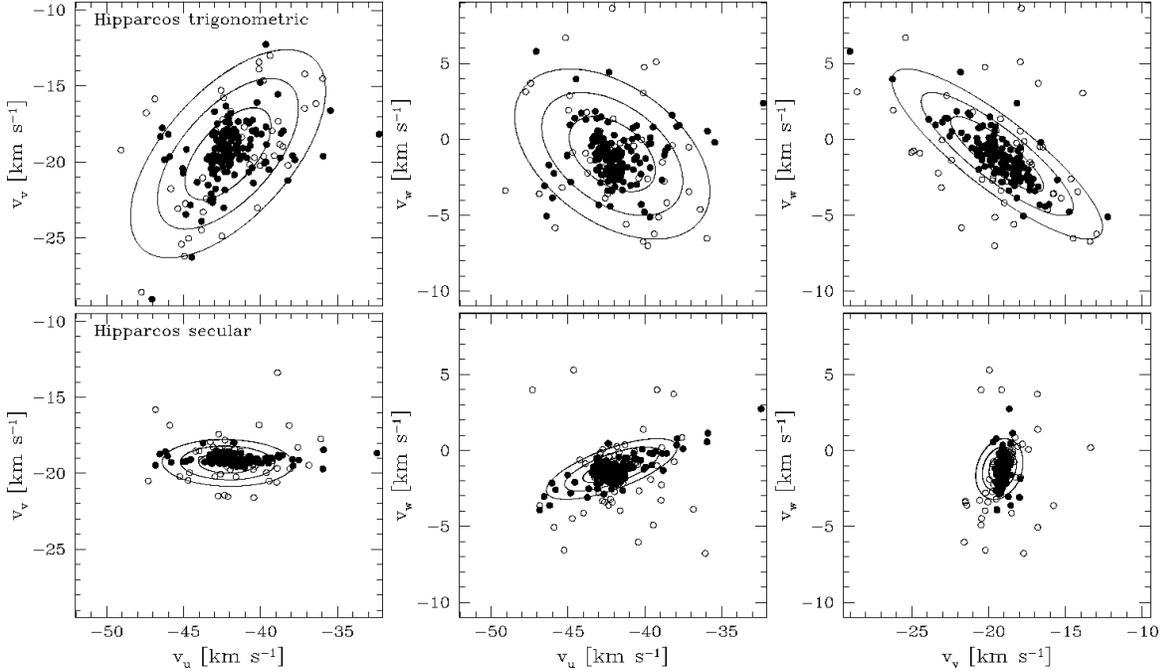,width=0.85\textwidth,silent=}}
\caption[]{{\it Top row:\/} three-dimensional velocity distribution in
Galactic Cartesian coordinates $(v_u, v_v, v_w)$, based on Hipparcos
trigonometric parallaxes, for the 197 P98 members with known radial
velocities (cf.\ figure~16 in P98; symbol coding as in bottom
row). The contours are centered on the (arithmetic) mean motion of all
stars $(-42.07, -19.45, -0.96)$~km~s$^{-1}$ (cf.\ table~3 in P98);
they show the 1, 2, and 3$\sigma$ (i.e., 68.3, 95.4, and
99.73~per~cent) confidence limits of the mean covariance matrix
associated with the mean space motion (\S
\ref{subsubsec:vel_field_Hip_P98}). All `outliers' have space motions
(based on Hipparcos proper motions and trigonometric parallaxes) which
are consistent with the mean motion of the cluster. {\it Bottom
row:\/} as top row, but using Hipparcos secular parallaxes (mean
motion $(-42.15, -19.31, -1.21)$~km~s$^{-1}$). Open symbols denote
suspect secular parallaxes ($g_{\rm Hip} > 9$;
50~stars).\label{fig:vel_field} }
\end{figure*}
}
%--- Figure 6 -----------------------------------------------------------------

%--- Figure 7 -----------------------------------------------------------------
\def\placefigureSeven{
\begin{figure*}[t!]
\centerline{\psfig{file=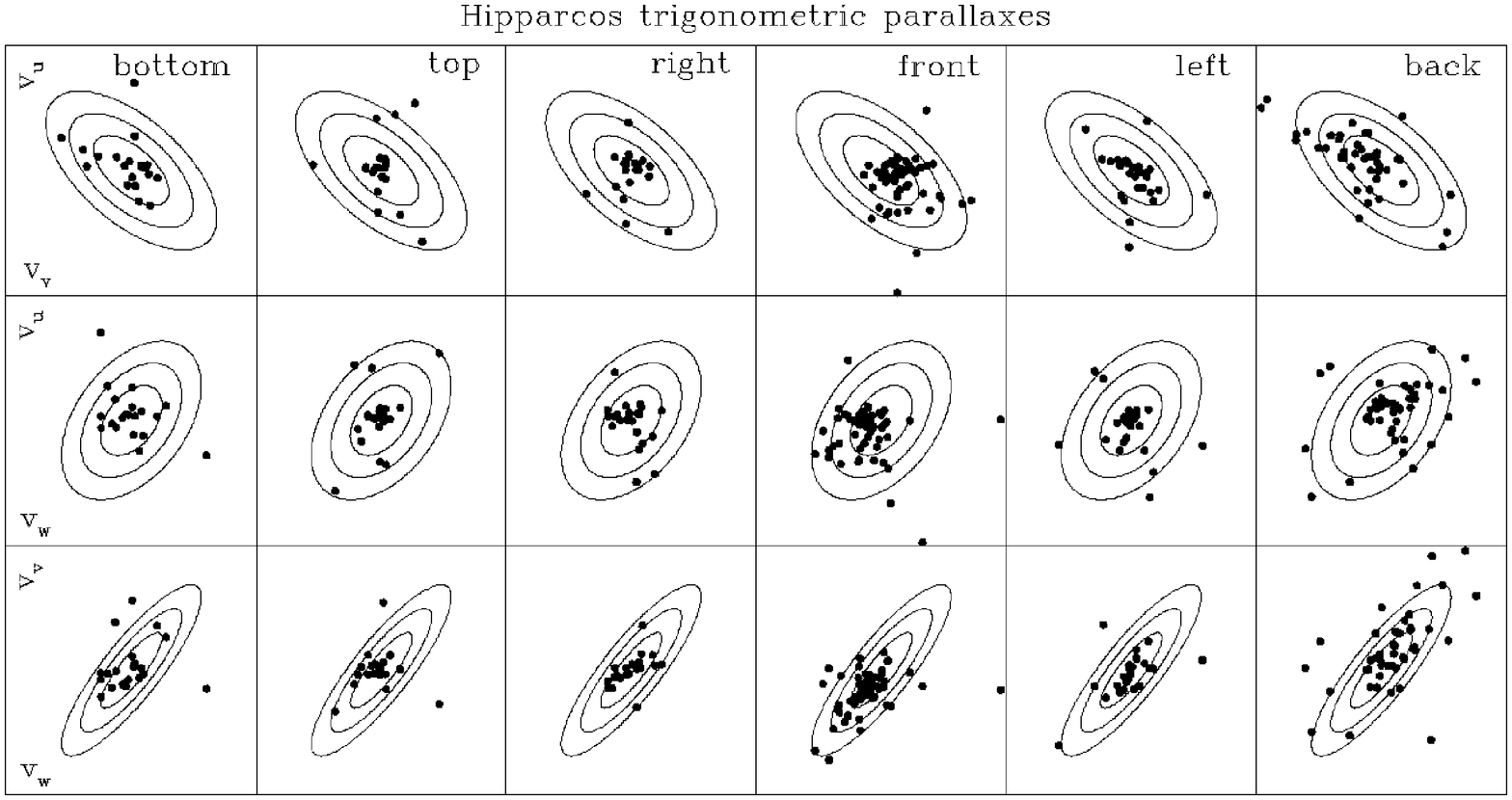,width=0.65\textwidth,angle=270.0,silent=}}
\centerline{\psfig{file=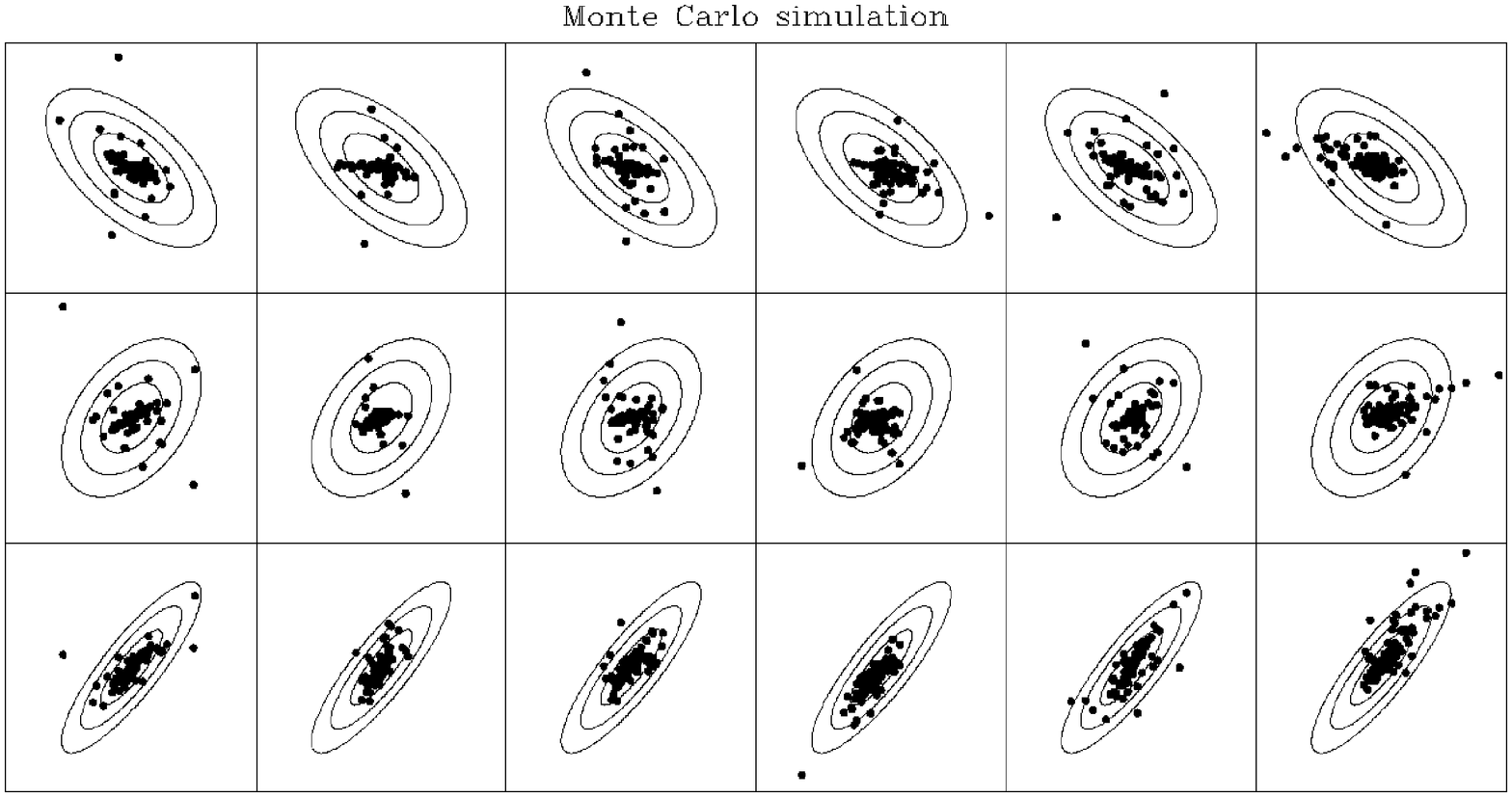,width=0.65\textwidth,angle=270.0,silent=}}
\caption[]{{\it Top three rows:\/} Hipparcos trigonometric
parallax-based velocity field decomposition, with respect to the mean
velocity $(v_u, v_v, v_w) = (-42.07, -19.45, -0.96)$~km~s$^{-1}$ in
$20$~km~s$^{-1} \times 20$~km~s$^{-1} \times 20$~km~s$^{-1}$ boxes,
for the 197 P98 members with known radial velocities according to the
location of the stars within the cluster (from the left column right:
`bottom' (21 stars), `top' (21 stars), `right' (22 stars), `front' (59
stars), `left' (28 stars), and `back' (46 stars; \S
\ref{subsec:3d_pos}) and from the top down: $v_u$ versus $v_v$, $v_u$
versus $v_w$, and $v_v$ versus $v_w$; velocity components in a
Galactic Cartesian coordinate frame). The ellipses denote 1, 2, and
3$\sigma$ confidence regions of the mean motion and associated
covariance matrix (\S \ref{subsubsec:vel_field_Hip_P98}). The {\it
bottom series of panels} are similar to the top series, but show 500
Monte Carlo stars which share a common space motion exclusively (\S
\ref{subsubsec:vel_field_Hip}).\label{fig:vel_sys} }
\end{figure*}
}
%--- Figure 7 -----------------------------------------------------------------

%--- Figure 8 -----------------------------------------------------------------
\def\placefigureEight{
\begin{figure}[t!]
\psfig{file=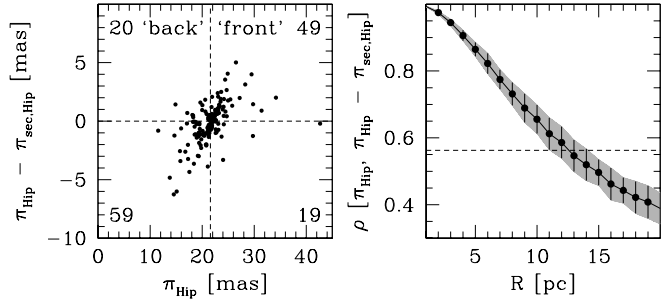,width=8.7truecm,silent=}
\caption[]{{\it Left:\/} an estimate of the Hipparcos trigonometric
parallax error $\sigma_{\pi, {\rm Hip}}$ ($\pi_{\rm Hip} - \pi_{\rm
sec, Hip} \approx \pi_{\rm Hip} - \pi_{\rm true} \equiv \Delta_{\pi,
{\rm Hip}}$) versus the Hipparcos trigonometric parallax $\pi_{\rm
Hip}$ for the 147 P98 Hyades members with non-suspect secular
parallaxes ($g_{\rm Hip} \leq 9$). Every sample of stars with (nearly)
equal true parallaxes shows a correlation between $\Delta_{\pi, {\rm
Hip}}$ (or: $\sigma_{\pi, {\rm Hip}}$) and $\pi_{\rm Hip}$; we find a
correlation coefficient $\rho = +0.56$. The dashed vertical line
denotes the mean distance of the cluster ($D = 46.34$~pc; $\pi =
21.58$~mas; table~3 in P98). The numbers in the corners of the four
quadrants denote the numbers of stars in the corresponding
regions. {\it Right:\/} the mean correlation coefficient for 100 Monte
Carlo realizations of a Hyades-like cluster as function of the cluster
radius $R$. Each cluster has a homogeneous number density. The dots
and vertical lines denote the mean value of $\rho$ and the
corresponding standard deviation; the gray band denotes the $\pm 1
\sigma$ region. The dashed horizontal line indicates the observed
value $\rho = +0.56$.\label{fig:pearson_r_etc} }
\end{figure}
}
%--- Figure 8 -----------------------------------------------------------------

%--- Figure 9 -----------------------------------------------------------------
\def\placefigureNine{
\begin{figure*}[t!]
\begin{minipage}{6.5812766truecm}
\centerline{\psfig{file=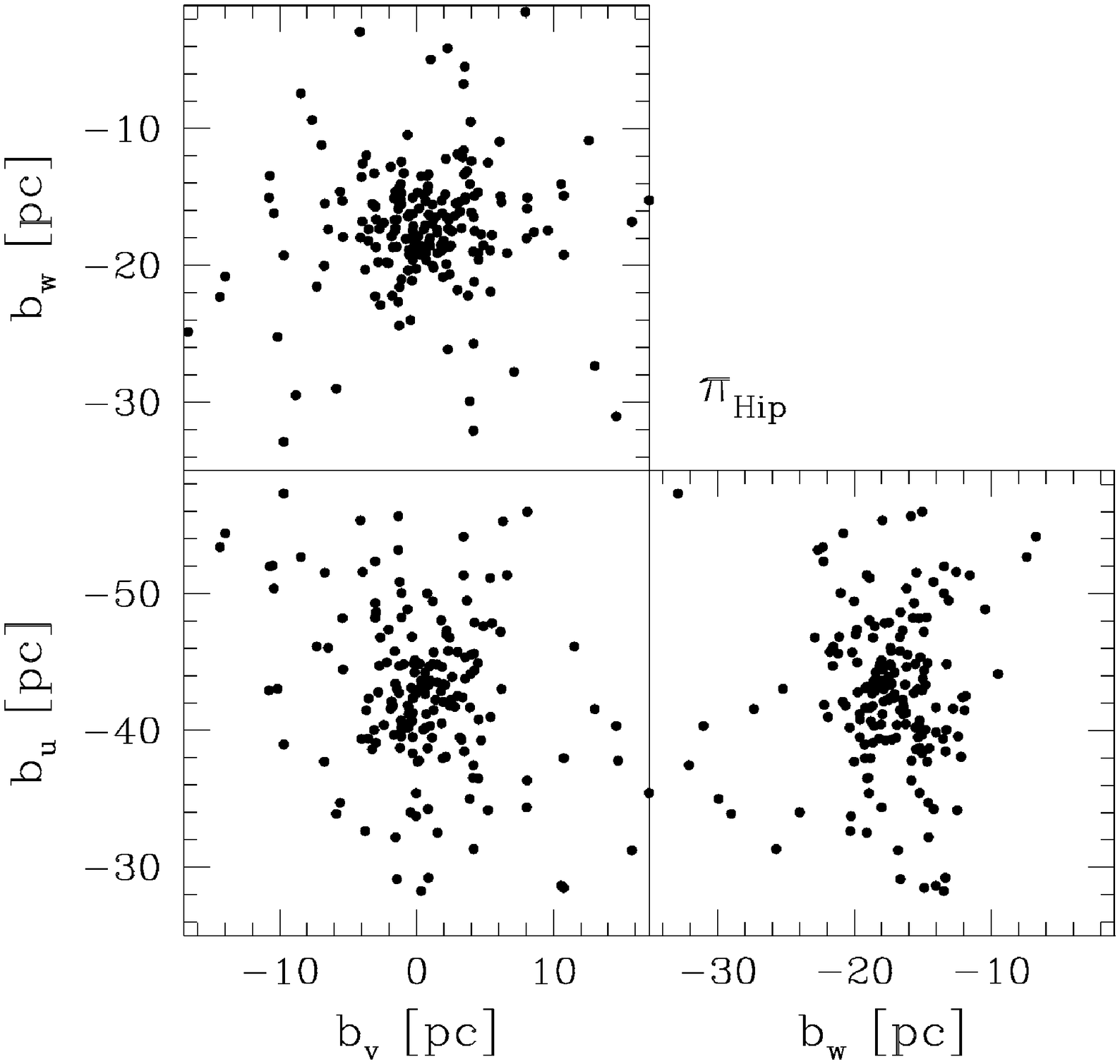,width=6.5812766truecm,silent=}}
\end{minipage}
\begin{minipage}{5.5893617truecm}
\centerline{\psfig{file=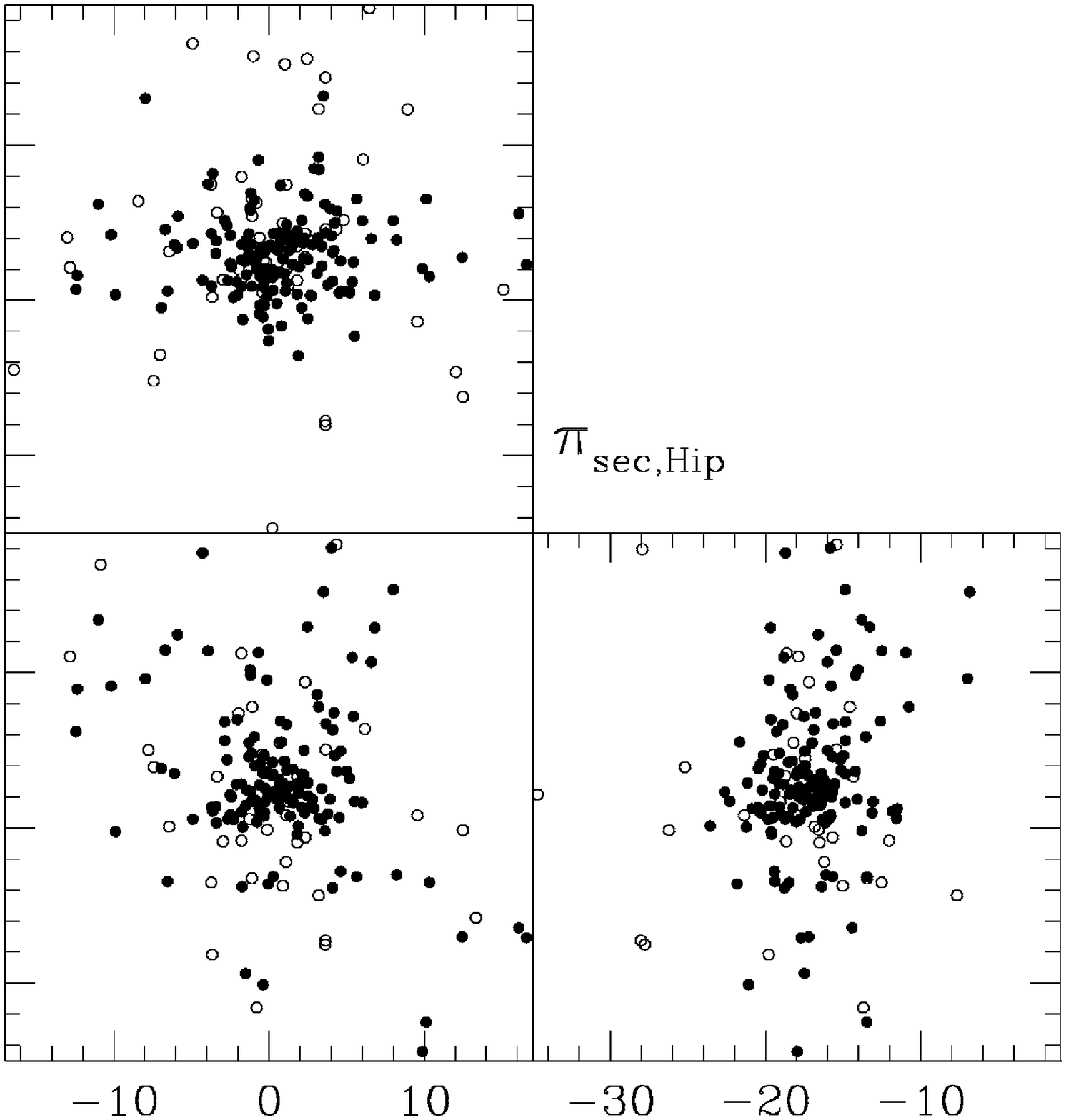,width=5.5893617truecm,silent=}}
\end{minipage}
\begin{minipage}{5.5893617truecm}
\centerline{\psfig{file=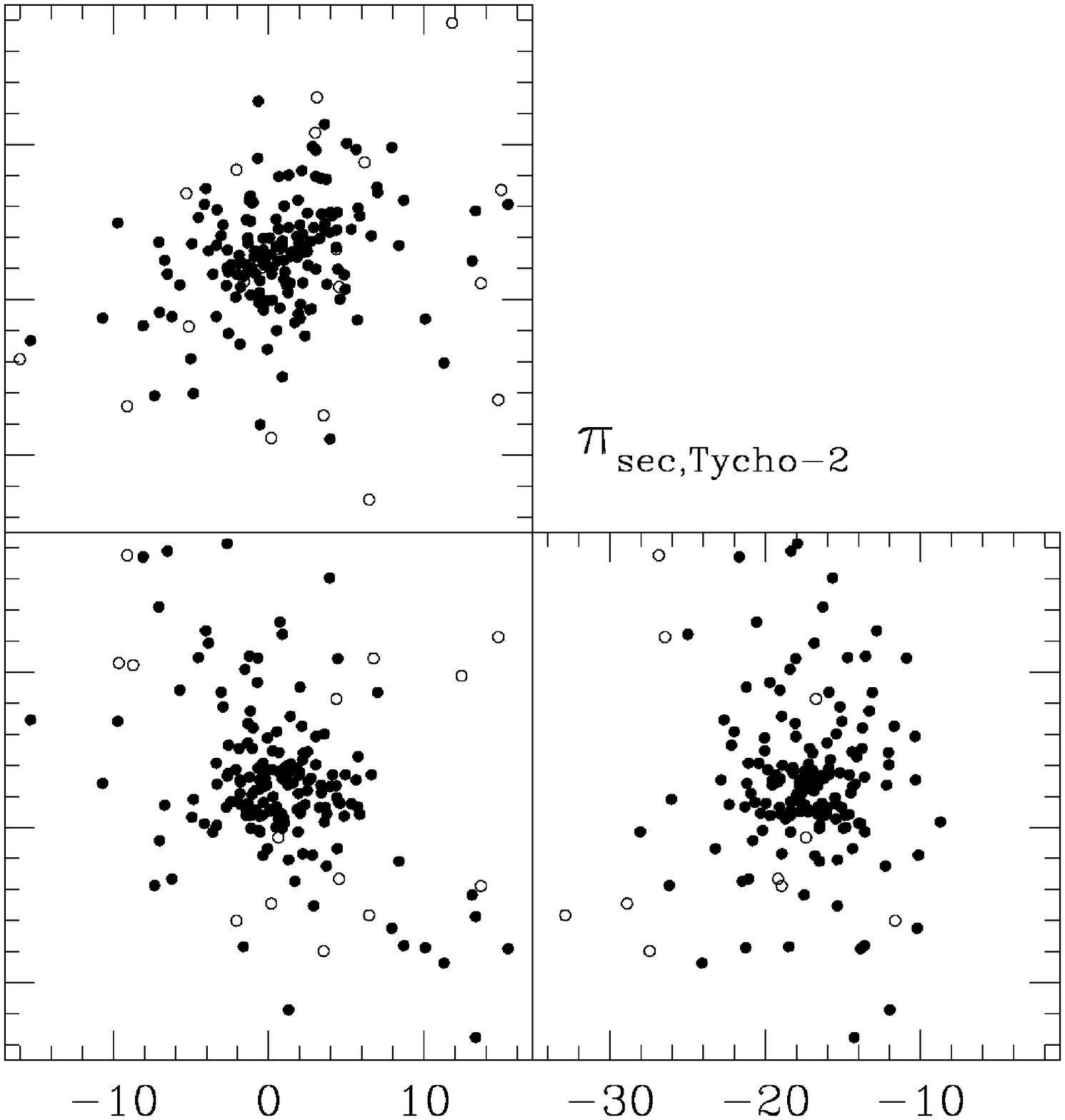,width=5.5893617truecm,silent=}}
\end{minipage}
\caption[]{{\it Left:\/} three-dimensional distribution, based on
Hipparcos trigonometric parallaxes, of the 218 P98 members in Galactic
Cartesian coordinates $(b_u, b_v, b_w)$ (in pc; cf.\ figures 8a--9 in
P98). Some stars fall outside the plotted range. {\it Middle:\/} as
left, but using Hipparcos secular parallaxes (154 stars with $g_{\rm
Hip} \leq 9$ [filled symbols] and 64 stars with $g_{\rm Hip} > 9$
[open symbols]). {\it Right:\/} as middle, but using Tycho--2 secular
parallaxes (176/32 filled/open
symbols).\label{fig:spatial_distribution} }
\end{figure*}
}
%--- Figure 9 -----------------------------------------------------------------

%--- Figure 10 ----------------------------------------------------------------
\def\placefigureTen{
\begin{figure*}[t!]
\psfig{file=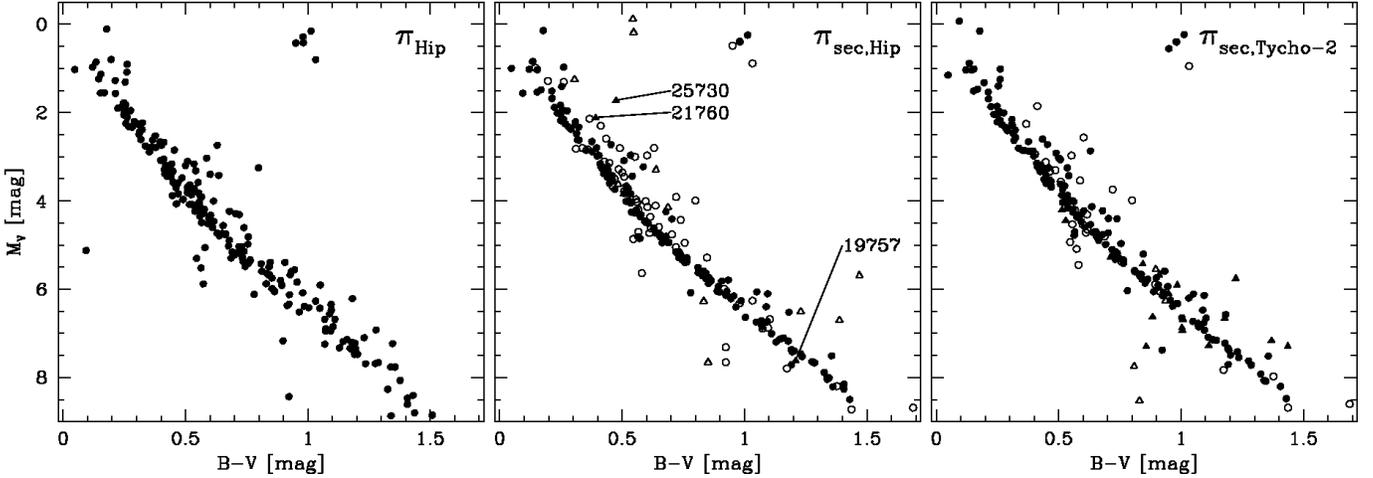,width=\textwidth,silent=}
\caption[]{Colour-absolute magnitude diagrams of the Hyades based on
Hipparcos trigonometric parallaxes ({\it left}; 218 P98 members) and
Hipparcos secular ({\it middle}; 218 P98 members plus 15 new
candidates) and Tycho--2 secular parallaxes ({\it right}; 208 P98
members plus 23 photometric BDA members; $V$ magnitude (field H5) and
$(B-V)$ colour (field H37) from the Hipparcos Catalogue). Most P98
stars below the main sequence in the left panel are `possible members'
(i.e., column (x) is `?' in their table~2; cf.\ figure~21 in
P98). Filled symbols in the right panels have $g_{\rm Hip/Tycho-2}
\leq 9$ while open symbols have $g_{\rm Hip/Tycho-2} > 9$. The 15
triangular symbols in the middle panel are the new Hipparcos
candidates (\S \ref{subsec:Hip_additional_members}); the three stars
with $g_{\rm Hip} \leq 9$ are labeled with their Hipparcos number. The
giant region contains three(!) filled and two open symbols. The
triangular symbols in the right panel represent 23 photometric BDA
members (\S \ref{subsec:Tycho2}; filled triangles for $g_{\rm Tycho-2}
\leq 9$ and open triangles for $g_{\rm Tycho-2} > 9$). The giant
region contains four(!) filled and one open symbol. Some faint stars
($V \mathrel{{\hbox to 0pt{\lower 3pt\hbox{$\sim$}\hss}} \raise
2.0pt\hbox{$>$}} 8.5$~mag, $M_V \mathrel{{\hbox to 0pt{\lower
3pt\hbox{$\sim$}\hss}} \raise 2.0pt\hbox{$>$}} 5.2$~mag) in the right
panel have significant $(B-V)$ errors, up to several tenths of a
magnitude (\S \ref{sec:cmd}).\label{fig:HRD_1} }
\end{figure*}
}
%--- Figure 10 ----------------------------------------------------------------

%--- Figure 11 ----------------------------------------------------------------
\def\placefigureEleven{
\begin{figure*}[t!]
\centerline{\psfig{file=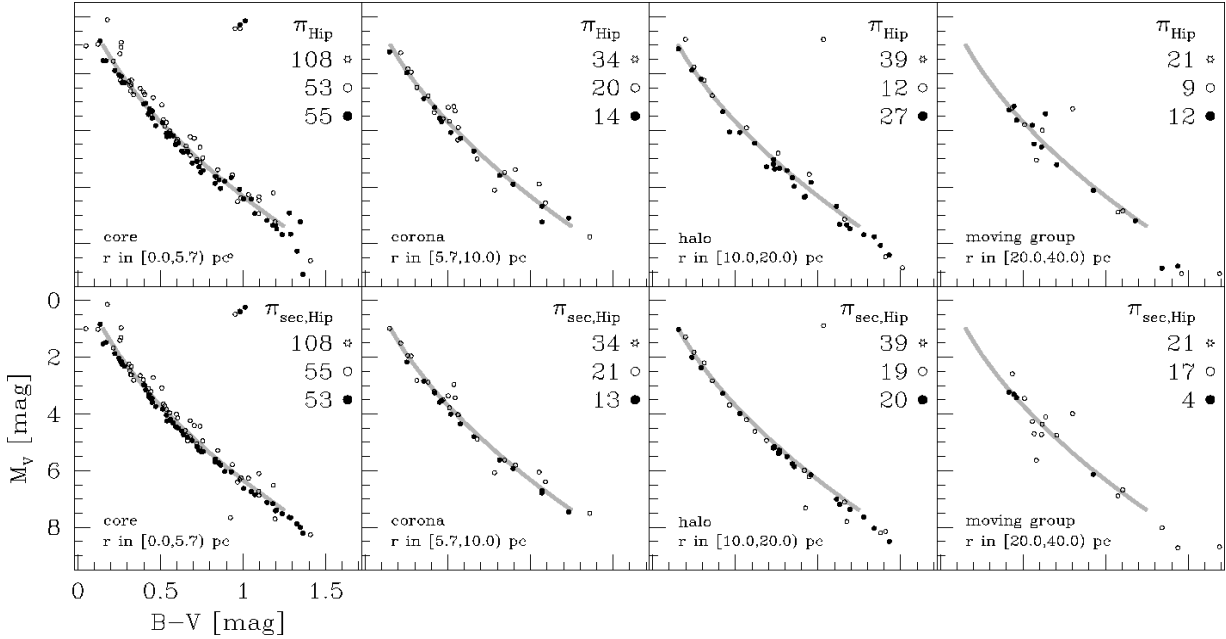,width=0.90\textwidth,silent=}}
\caption[]{Colour-absolute magnitude diagrams of the Hyades (218 P98
members) based on Hipparcos trigonometric ({\it top row}) and
Hipparcos secular ({\it bottom row}) parallaxes for four different
spatial regions within the cluster: the core ($r < 5.7$~pc; {\it first
column}), the corona ($5.7 \leq r < 10$~pc; {\it second column}), the
halo ($10 \leq r < 20$~pc; {\it third column}), and the moving group
population ($20 \leq r < 40$~pc; {\it fourth column}). In each panel,
the division of stars according to their three-dimensional distance
$r$ to the cluster center $(b_u, b_v, b_w) = (-43.37, 0.40,
-17.46)$~pc (in Galactic Cartesian coordinates) is based on the
Hipparcos secular parallax. The gray lines outline Schwan's (1991; his
table 3) Hyades main sequence; this calibration is not necessarily the
optimal one but is only shown for reference. The numbers preceding the
open asterisks denote the total number of stars in each panel. Open
symbols denote stars which {\it might\/} have `peculiar' HR diagram
positions: (1) kinematically deviant stars, i.e., $g_{\rm Hip} > 9$;
(2) (close) multiple stars, i.e., either column (s) in P98's table 2
is `SB or RV' or column (u) is one of `C$\,$G$\,$O$\,$V$\,$X$\,$S';
(3) photometrically variable stars, i.e., Hipparcos field H52 is one
of `D$\,$M$\,$P$\,$R$\,$U'; (4) stars with inaccurate Hipparcos
$(B-V)$ photometry, i.e., $\sigma_{(B-V)} > 0.05$~mag; or (5) suspect
objects (HIP 20901, 21670, 20614; \S 9.2 in P98). Filled symbols thus
denote kinematically high-fidelity photometrically non-variable single
members with reliable photometry.\label{fig:HRD_2} }
\end{figure*}
}
%--- Figure 11 ----------------------------------------------------------------

%--- Figure 12 ----------------------------------------------------------------
\def\placefigureTwelve{
\begin{figure}[t!]
\centerline{\psfig{file=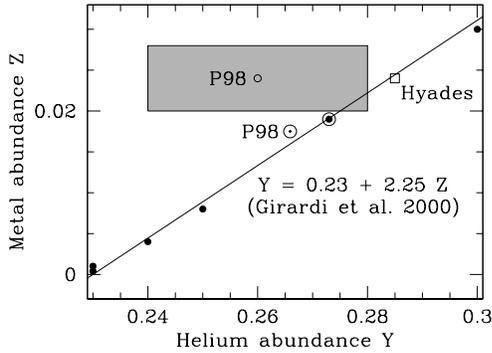,width=6.5truecm,silent=}}
\caption[]{The six $(Y, Z)$ values (solid dots) for which Girardi et
al.\ (2000a) present isochrones; the solid line shows the underlying
relation $Y = Y_{\rm p} + (\Delta Y/ \Delta Z) \cdot Z = 0.23 + 2.25
\cdot Z$ (\S \ref{subsec:Hyades_chars}). The $\odot$ symbol, which
coincides with a solid dot, denotes the position of the Sun $(Y,
Z)_\odot = (0.273, 0.019)$. The open dot and corresponding gray box
labeled `P98' denote the P98 Hyades value and corresponding 1$\sigma$
uncertainty; the $\odot$ symbol labeled `P98' denotes P98's Solar
value $(Y, Z)_\odot = (0.2659, 0.0175)$. The open square at $(Y, Z) =
(0.285, 0.024)$ labeled `Hyades' is discussed in \S
\ref{subsec:Hyades_chars}.\label{fig:YZ} }
\end{figure}
}
%--- Figure 12 ----------------------------------------------------------------

%--- Figure 13 ----------------------------------------------------------------
\def\placefigureThirteen{
\begin{figure}[t!]
\psfig{file=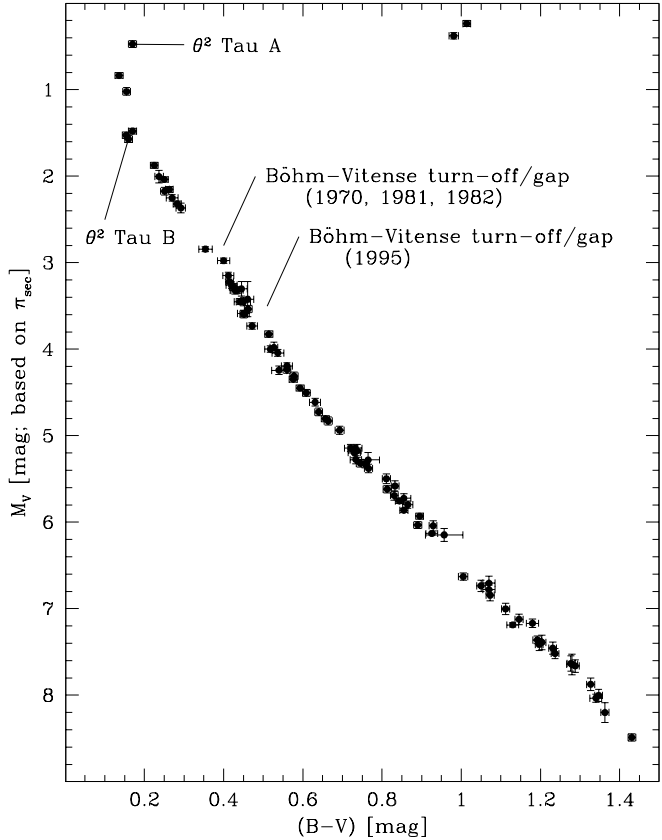,width=8.7truecm,silent=}
\caption[]{Colour-absolute magnitude diagram for 92 high-fidelity
members (\S \ref{subsec:HRD_stars}). This sample excludes all members
beyond 40~pc from the cluster center, multiple stars, and stars with
suspect secular parallaxes. The absolute magnitudes have been computed
using the observed $V$-band magnitudes (Hipparcos field H5) and
secular parallaxes (\S \ref{sec:sec_pars};
Table~\ref{tab:data_1}). The $(B-V)$ colours were directly taken from
the Hipparcos Catalogue (field H37). The gaps between $(B-V) = 0.15$
and $0.20$, between $(B-V) = 0.30$ and $0.35$, and the gap around
$(B-V) = 0.95$~mag are caused by the suppression of double, multiple,
and peculiar stars from our sample (cf.\ figure 21 in P98); the region
between $(B-V) = 0.30$ and $(B-V) = 0.35$~mag, e.g., is occupied by
Am-type stars, which have a high incidence of duplicity (\S
\ref{subsec:HRD_2}). The lines point at two conspicuous features in
the main sequence, the so-called B\"ohm--Vitense gaps (\S
\ref{subsec:HRD_2}). The `gaps' and corresponding `turn-offs' are most
likely caused by sudden changes in the properties of convective
atmospheres (cf.\ de Bruijne et al.\ 2000). Figure~\ref{fig:HRD_3}(a)
presents the corresponding theoretical HR diagram ($\log L$ versus
$\log T_{\rm eff}$) for the same sample of stars.\label{fig:HRD_4} }
\end{figure}
}
%--- Figure 13 ----------------------------------------------------------------

%--- Figure 14 ----------------------------------------------------------------
\def\placefigureFourteen{
\begin{figure}[t!]
\centerline{\psfig{file=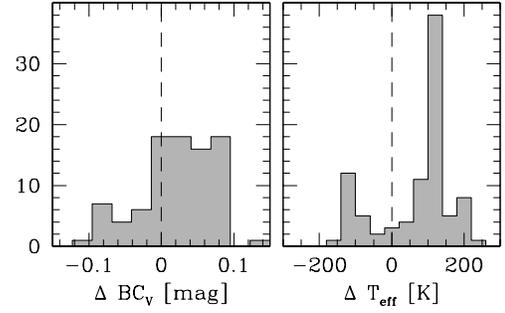,width=6.5truecm,silent=}}
\caption[]{The difference between bolometric corrections ({\it left})
and effective temperatures ({\it right}) derived using calibration (1)
(Bessell et al.\ (1998) plus an Alonso et al.\ (1996) metallicity
correction to $[{\rm Fe/H}] = +0.14$) and calibration (2) (Lejeune et
al.\ (1998) for $[{\rm Fe/H}] = +0.14$) for 92 high-fidelity members
(\S \ref{subsec:HRD_stars}), excluding the giants $\epsilon$ and
$\gamma$ Tau; $\Delta$ is defined as calibration (1) minus (2). Most
stars with $\Delta T_{\rm eff} < 0$~K have $(B-V) \geq 1.0$~mag,
indicating that calibrations (1) and (2) differ systematically with
effective temperature itself.\label{fig:Teff_rev} }
\end{figure}
}
%--- Figure 14 ----------------------------------------------------------------

%--- Figure 15 ----------------------------------------------------------------
\def\placefigureFifteen{
\begin{figure*}[t!]
\centerline{\psfig{file=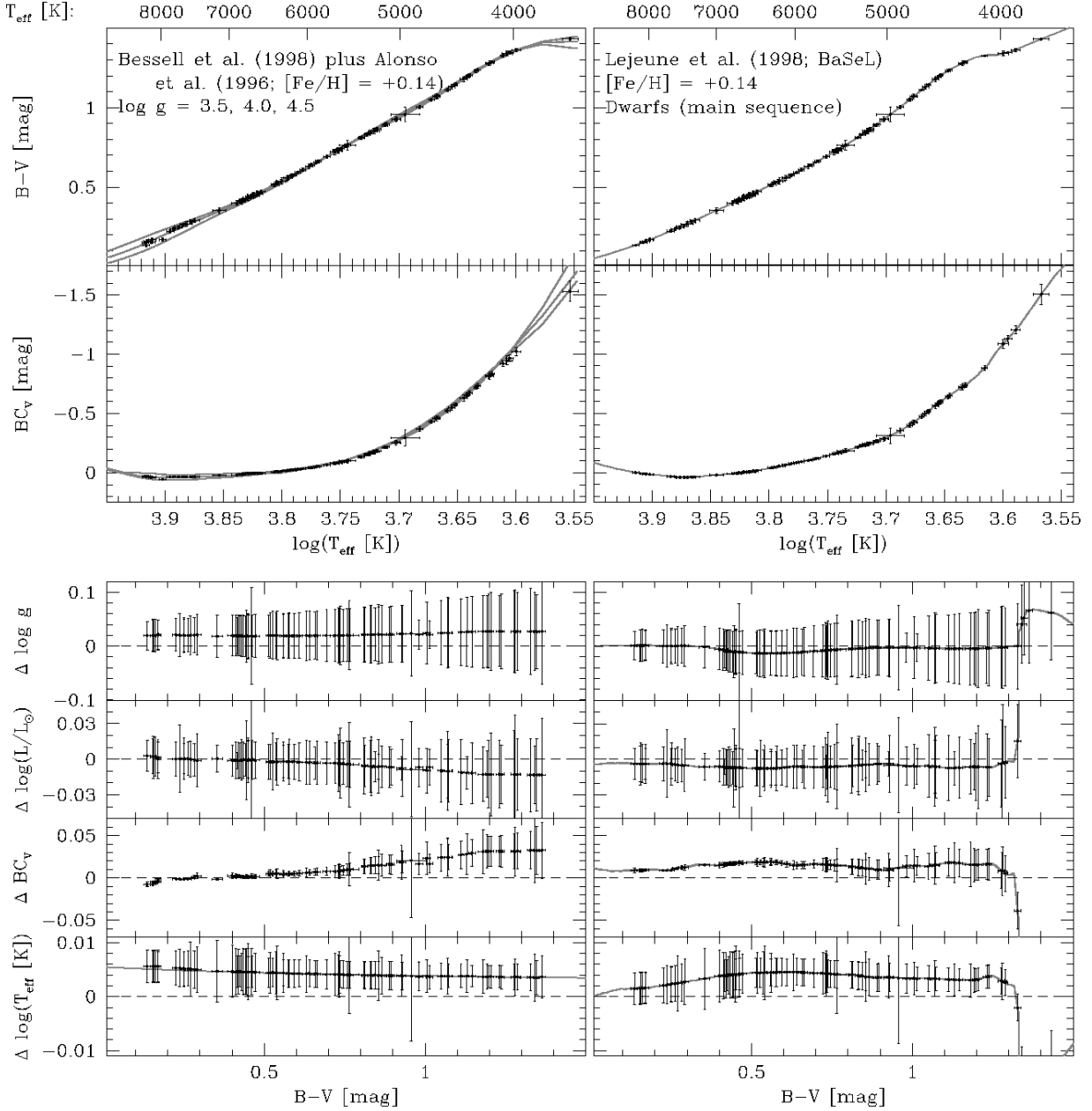,width=0.90\textwidth,silent=}}
\caption[]{Theoretical quantities $\log T_{\rm eff}$, ${\rm BC}_V$,
$\log(L/L_\odot)$, and $\log g$ derived from $(B-V)$ and $M_V$ (the
latter based on Hipparcos secular parallaxes) using Bessell et al.\
(1998) plus Alonso et al.\ (1996; {\it left column\/}) and Lejeune et
al.\ (1998; {\it right column\/}; \S\S
\ref{subsubsec:trans_cali1}--\ref{subsubsec:trans_cali2}). {\it Left
column:\/} the gray curves in the top two panels show the Bessell et
al.\ relation for $[{\rm Fe/H}] = 0$ and $\log g = 3.5, 4.0$, and
$4.5$ (from bottom to top along the vertical line $\log(T_{\rm eff}
[{\rm K}]) = 3.9$, respectively); the (tiny) dots and crosses show the
(metallicity corrected) values and uncertainties for the 92
high-fidelity Hyades (\S \ref{subsec:HRD_stars}), excluding the giants
$\epsilon$ and $\gamma$ Tau. The bottom four panels show the effect of
correcting the Bessell et al.\ calibration for the non-Solar Hyades
metallicity $[{\rm Fe/H}] = +0.14$ on (from the top panel down) $\log
g$, $\log(L/L_\odot)$, ${\rm BC}_V$, and $\log T_{\rm eff}$. The
difference $\Delta x$ is defined as $\Delta x \equiv x_{[{\rm Fe/H}] =
+0.14} - x_{[{\rm Fe/H}] = 0}$ for $x = \log g$, etc. The faint M
dwarf HIP 15720 ($(B-V) = 1.431 \pm 0.004$~mag) has no Bessell et al.\
solution for $[{\rm Fe/H}] = 0$. The gray line in the bottom panel
follows Alonso's eq.~(1). {\it Right column:\/} as left column, but
using the Lejeune et al.\ relation for dwarfs with $[{\rm Fe/H}] =
+0.14$. The bottom four panels show the effect of going from $[{\rm
Fe/H}] = 0$ to $[{\rm Fe/H}] = +0.14$ by means of interpolation. The
apparent discontinuities of the gray lines in the lower panels are
caused by the peculiar behaviour of the Lejeune et al.\ $[{\rm Fe/H}]
= +0.50$ relation (their table 9) around $(B-V) = 1.30$~mag. Several
stars with $(B-V) > 1.30$~mag fall outside the plotted ranges in the
bottom panels.\label{fig:B+L} }
\end{figure*}
}
%--- Figure 15 ----------------------------------------------------------------

%--- Figure 16 ----------------------------------------------------------------
\def\placefigureSixteen{
\begin{figure*}[t!]
\centerline{\psfig{file=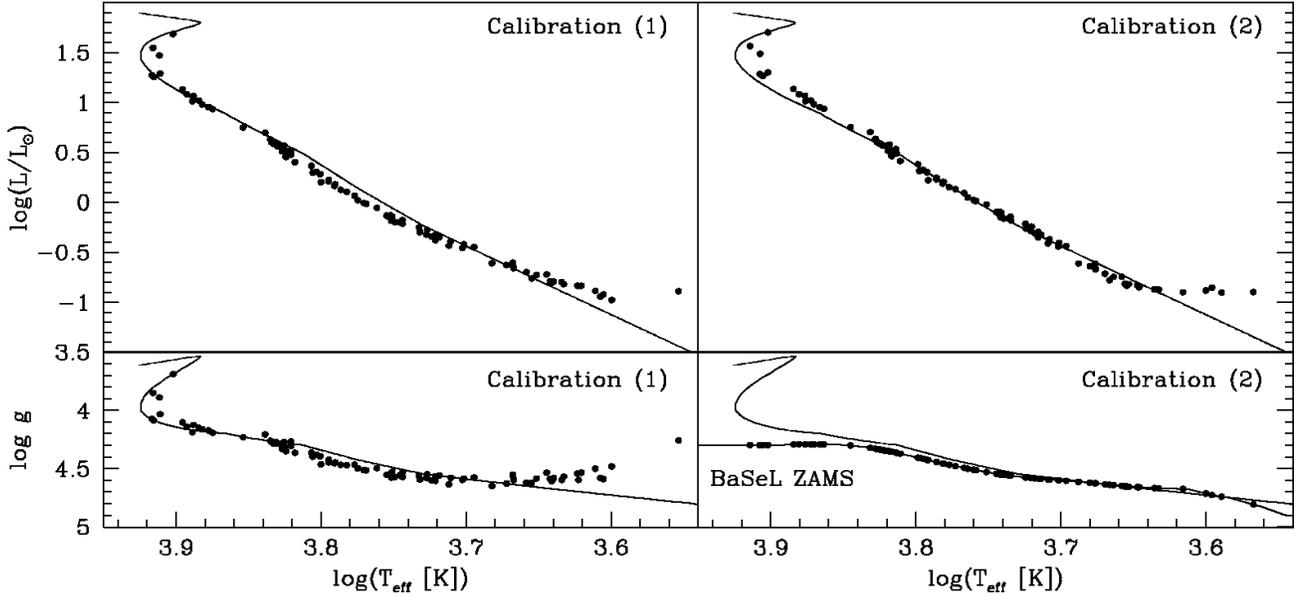,width=0.95\textwidth,silent=}}
\caption[]{The $\log T_{\rm eff}$--$\log(L/L_\odot)$ ({\it top\/}) and
$\log T_{\rm eff}$--$\log g$ ({\it bottom\/}) diagrams for the same 92
stars as in Figure~\ref{fig:HRD_3}(a) and (b), excluding the giants
$\gamma$ and $\epsilon$ Tau, but using calibration~(1) (\S
\ref{subsubsec:trans_cali1}; {\it left\/}) and (2) (\S
\ref{subsubsec:trans_cali2}; {\it right\/}) for all stars. The curves
denote the 625~Myr CESAM isochrones.\label{fig:HRD_5} }
\end{figure*}
}
%--- Figure 16 ----------------------------------------------------------------

%--- Figure 17 ----------------------------------------------------------------
\def\placefigureSeventeen{
\begin{figure}[t!]
\psfig{file=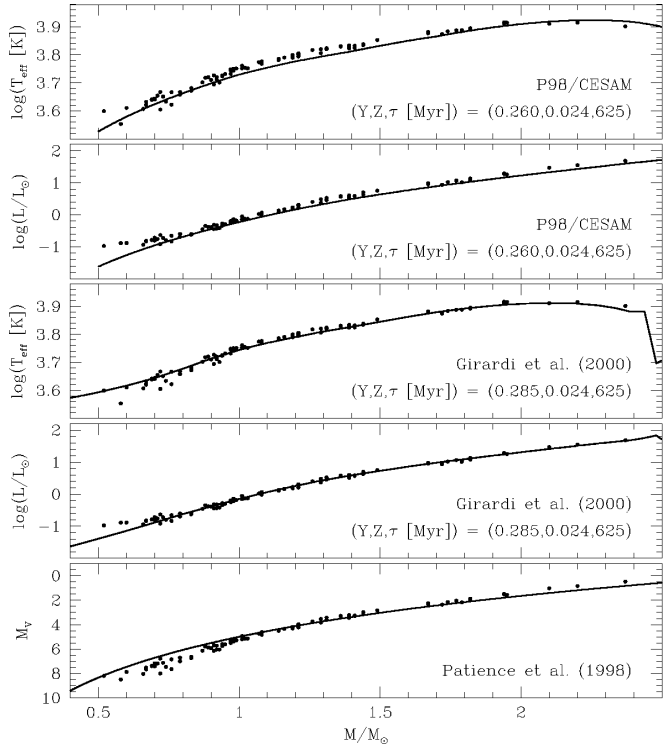,width=8.7truecm,silent=}
\caption[]{Masses of 92 high-fidelity members (\S
\ref{subsubsec:stellar_masses}), excluding the giants $\epsilon$ and
$\gamma$ Tau. The most massive stars in this figure are (in order of
decreasing mass): HIP 20894 A ($\theta^2$ Tau A; Appendix
\ref{subsec:trans_theta2}), 20635, and 23497. {\it Top two panels:\/}
the solid lines denote the effective temperature-- and
luminosity--stellar mass relations of the $(Y, Z)$$=$$(0.260, 0.024)$
$625$~Myr CESAM isochrone (P98; \S \ref{subsec:Hyades_chars}). {\it
Third and fourth panels:\/} as the top two panels, but for the $(Y,
Z)$$=$$(0.285, 0.024)$ $625$~Myr Padova isochrone (Girardi et al.\
2000a). {\it Bottom panel:\/} the empirical absolute magnitude--mass
relation of Patience et al.\ (1998) compared with the P98 stellar
masses. The absolute magnitudes are based on Hipparcos secular
parallaxes~(\S \ref{sec:sec_pars}).\label{fig:mass} }
\end{figure}
}
%--- Figure 17 ----------------------------------------------------------------

%--- Figure 18a ---------------------------------------------------------------
\def\placefigureEighteenA{
\begin{figure*}[t!]
\psfig{file=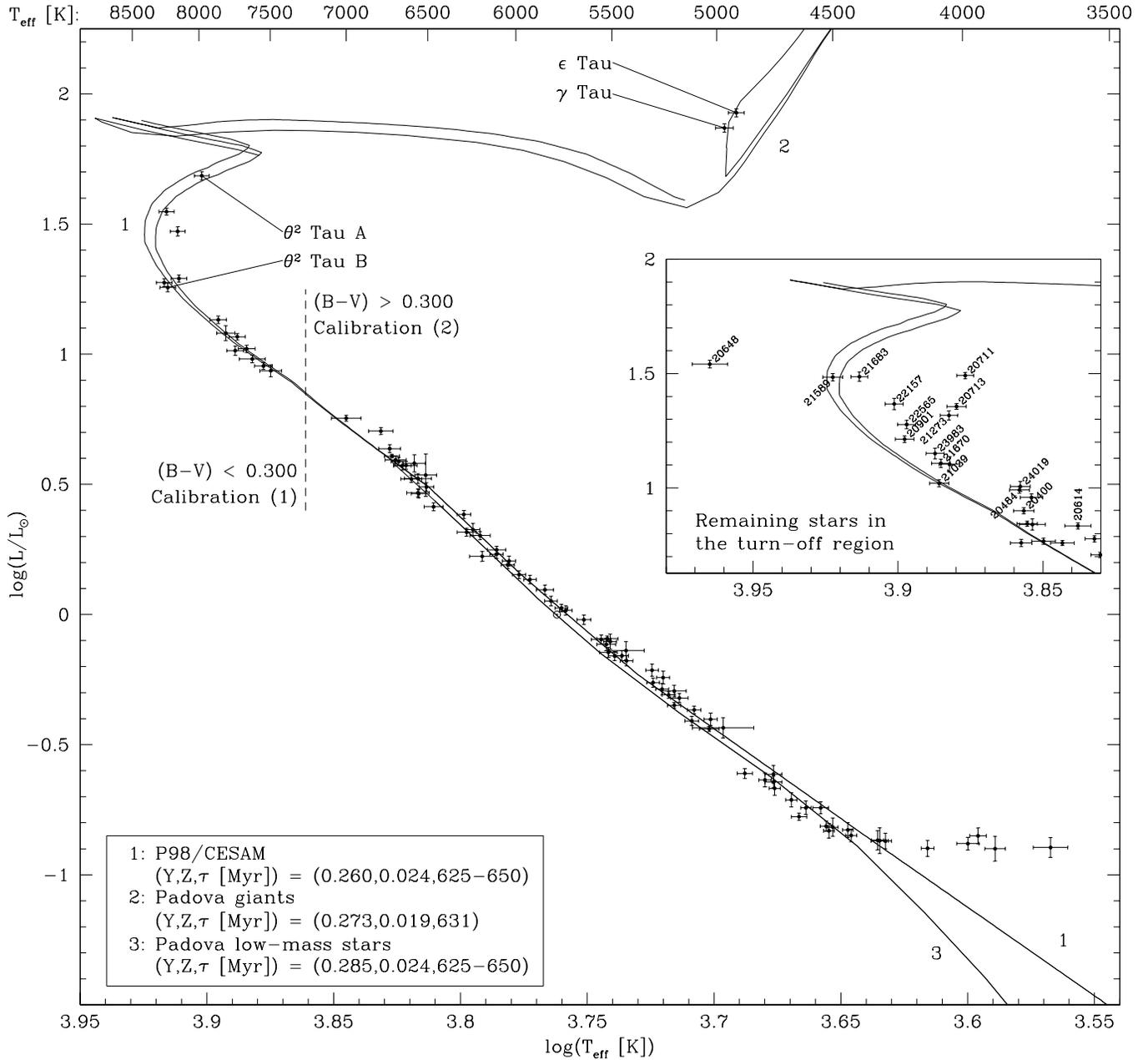,width=\textwidth,silent=}
\caption[]{{\bf panel (a).\/} HR diagram for 92 high-fidelity Hyades
(\S \ref{subsec:HRD_stars}). The stellar luminosities and effective
temperatures were derived from secular parallax-based absolute
magnitudes $M_V$ and $(B-V)$ colours using calibration (1) for $(B-V)
\leq 0.300$~mag ($T_{\rm eff} \geq 7250$~K; 14 stars; \S
\ref{subsubsec:trans_cali1}) and calibration (2) for $(B-V) >
0.300$~mag ($T_{\rm eff} < 7250$~K; 76 dwarfs; \S
\ref{subsubsec:trans_cali2}). The giants $\gamma$ and $\epsilon$ Tau
(HIP 20205 and 20889) are discussed in Appendix
\ref{subsec:trans_giants}; the components of the `resolved
spectroscopic binary' $\theta^2$ Tau (HIP 20894) are discussed in
Appendix \ref{subsec:trans_theta2}. The lines labeled `1', `2', and
`3' are CESAM and Padova isochrones for the Hyades with Helium content
$Y$, metal content $Z$, and age $\tau$ as indicated in the box (\S
\ref{subsec:Hyades_chars}). The position of the Sun is indicated by
$\odot$. The deviant locations of cool stars ($T_{\rm eff} \la
5000$~K) are a result of systematic errors in the $(B-V)$--$T_{\rm
eff}$ calibrations (\S \ref{subsec:HRD_1}; cf.\ Figure~\ref{fig:B+L}).
The {\it inset\/} shows the remaining members in the upper main
sequence and turn-off region of the cluster (excluding HIP 21459,
which has large photometric errors). HIP 20648 is discussed in \S
\ref{subsec:HRD_3}. Deviant locations can be attributed to
multiplicity, peculiar spectra, rotation, and/or suspect secular
parallaxes ($g_{\rm Hip} > 9$;
Table~\ref{tab:turn-off}).\label{fig:HRD_3} }
\end{figure*}
}
%--- Figure 18a ---------------------------------------------------------------

%--- Figure 18b ---------------------------------------------------------------
\def\placefigureEighteenB{
\begin{figure*}[t!]
\addtocounter{figure}{-1}
\psfig{file=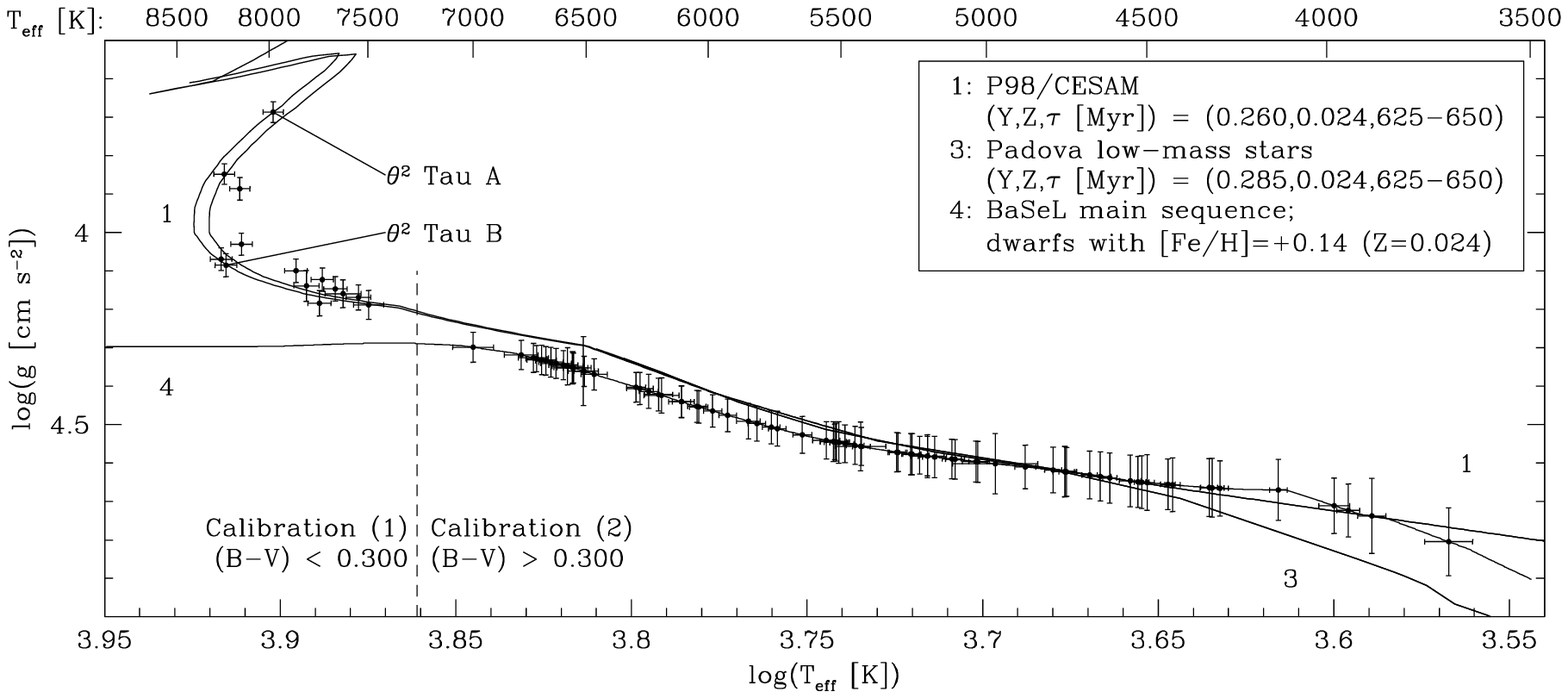,width=\textwidth,silent=}
\caption[]{{\bf panel (b).\/} The $\log T_{\rm eff}$--$\log g$ diagram for
the same 92 stars as in panel~(a), but excluding the giants $\gamma$
and $\epsilon$ Tau. The line labeled `4' denotes the BaSeL calibration
main sequence for $[{\rm Fe/H}] = +0.14$ ($Z = 0.024$; \S
\ref{subsubsec:trans_cali2}). Bessell et al.'s (1998) calibration (\S
\ref{subsubsec:trans_cali1}) is differently organized and cannot be
plotted in this diagram.
}
\end{figure*}
}
%--- Figure 18b ---------------------------------------------------------------

%--- Figure 19 ----------------------------------------------------------------
\def\placefigureNineteen{
\begin{figure*}[t!]
\centerline{\psfig{file=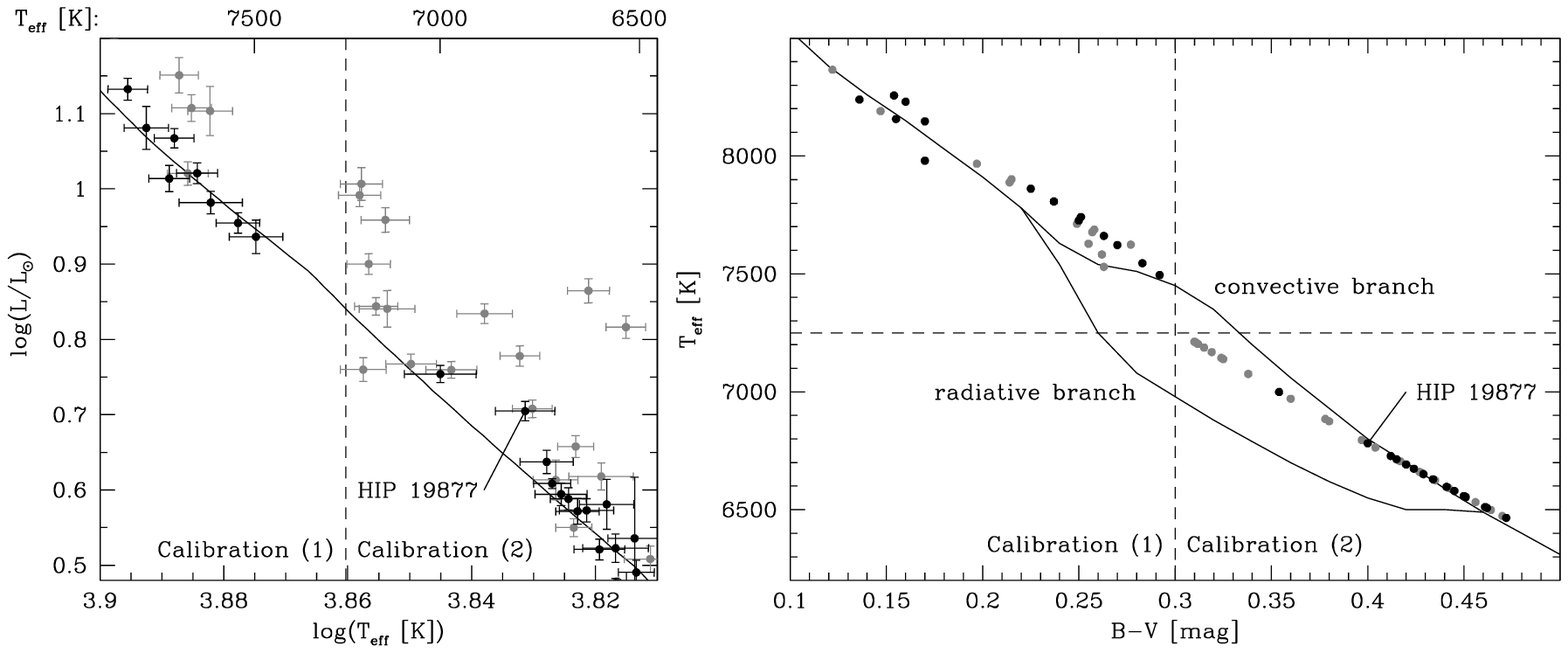,width=0.90\textwidth,silent=}}
\caption[]{{\it Left:\/} HR diagram of the region of the onset of
surface convection (the B\"ohm--Vitense gap; \S
\ref{subsec:HRD_2}). The solid line denotes the 625~Myr CESAM
isochrone. Black symbols show high-fidelity members; gray symbols
denote remaining members (excluding HIP 21459, which has large
photometric errors). Although the rotational velocity of the deviant
object HIP 19877 (F5var) is unknown, its detection in X-rays (H\"unsch
et al.\ 1998) suggests the star has a convective envelope. Patience et
al.\ (1998) did not detect a speckle companion. Its location above the
main sequence is hence possibly related to stellar variability
($\Delta V \sim 0.07$~mag) or activity. The dashed line denotes $(B-V)
= 0.30$~mag ($T_{\rm eff} \sim 7250$~K), which marks the boundary
between effective temperatures derived using calibrations (1) and (2)
(\S\S \ref{subsubsec:trans_cali1}--\ref{subsubsec:trans_cali2}). {\it
Right:\/} effective temperatures versus $(B-V)$ colours for the same
set of stars as shown in the left panel. The dashed horizontal line
denotes $T_{\rm eff} = 7250$~K, and is shown for reference. The solid
line shows the B\"ohm--Vitense (1981) Solar-metallicity
$(B-V)$--$T_{\rm eff}$ relation; the bifurcation into radiative and
convective branches is discussed in \S
\ref{subsec:HRD_2}.\label{fig:BVgap} }
\end{figure*}
}
%--- Figure 19 ----------------------------------------------------------------

%--- Figure 20 ----------------------------------------------------------------
\def\placefigureTwenty{
\begin{figure}[t!]
\psfig{file=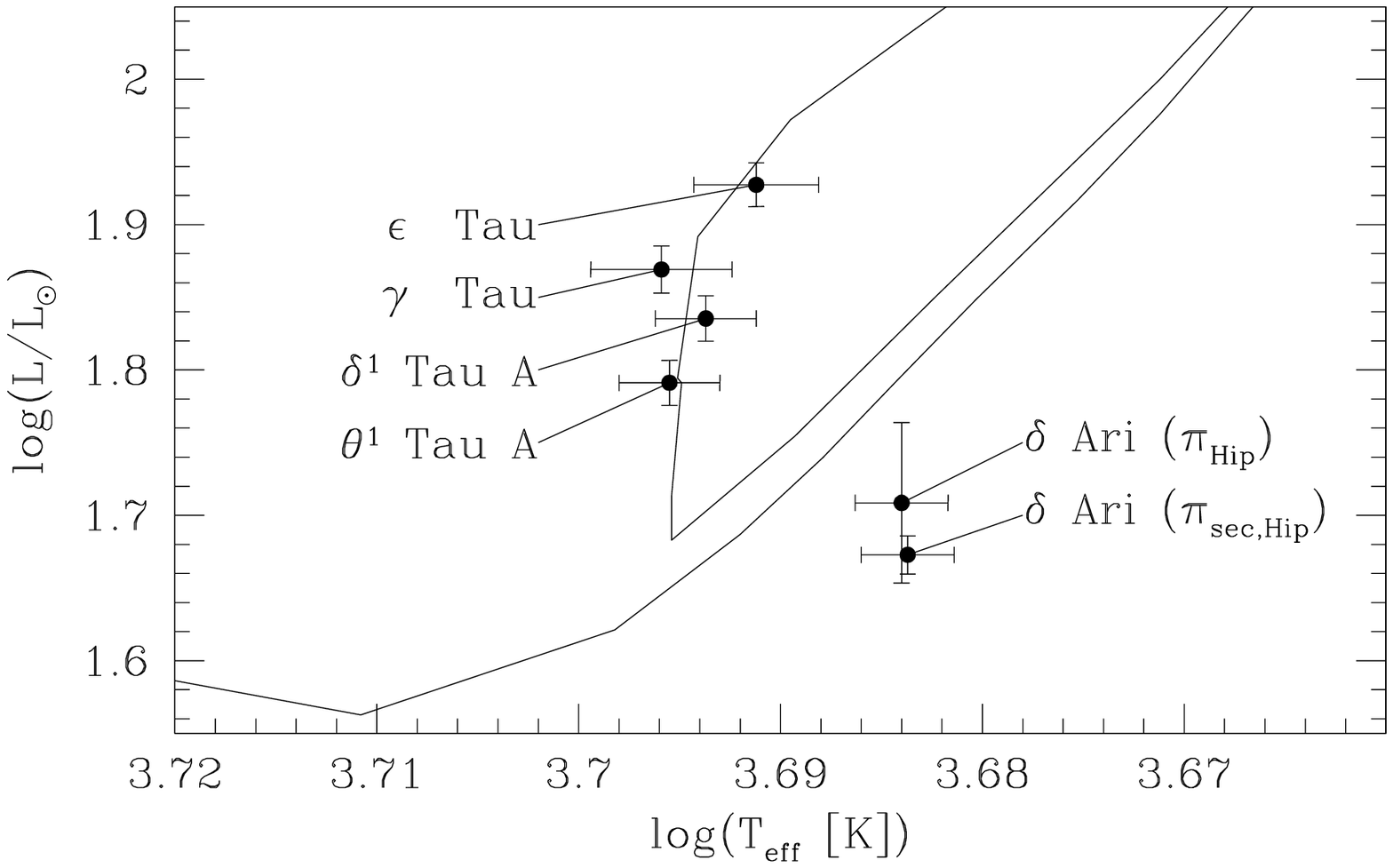,width=8.7truecm,silent=}
\caption[]{The Hyades red giant clump based on Hipparcos secular
parallaxes (Appendices
\ref{subsec:trans_giants}--\ref{subsec:trans_other_giants}). The solid
line denotes Girardi et al.'s (2000a) $(Y, Z) = (0.273, 0.019)$
631~Myr isochrone. The objects $\delta^1$ and $\theta^1$ Tau are
spectroscopic binaries; their secular parallaxes and luminosities
should be treated with care (\S \ref{subsec:HRD_4}). The `suspected
non-single' K2III giant $\delta$ Ari is not coeval with the Hyades;
this conclusion does not depend on the suspect secular parallax of the
star (\S \ref{subsec:HRD_4}).\label{fig:HRD_giants} }
\end{figure}
}
%--- Figure 20 ----------------------------------------------------------------

\section{Introduction}
\label{sec:intro}

The Hyades open cluster has for most of the past century been an
important calibrator of many astrophysical relations, e.g., the
absolute magnitude-spectral type and the mass-luminosity relation. The
cluster has been the subject of numerous investigations (e.g., van
Bueren 1952; Pels et al.\ 1975; Reid 1993; Perryman et al.\ 1998)
addressing, e.g., cluster dynamics and evolution, the distance scale
in the Universe (e.g., Hodge \& Wallerstein 1966; van den Bergh 1977),
and the calibration of spectroscopic radial velocities (e.g., Petrie
1963; Dravins et al.\ 1999).

The significance of the Hyades is nowadays mainly limited to the broad
field of stellar structure and evolution. Open clusters in general
form an ideal laboratory to study star formation, structure, and
evolution theories, as their members are thought to have been formed
simultaneously from the same molecular cloud material. As a result,
they have (1) the same age, at least to within a few Myr, (2) the same
distance, neglecting the intrinsic size which is at maximum 10--20~pc,
(3) the same initial element abundances, and (4) the same space
motion, neglecting the internal velocity dispersion which is similar
to the velocity dispersion within the parental molecular cloud
(typically several tenths of a km~s$^{-1}$). The Hyades open cluster
in particular is especially suitable and primarily has been used for
detailed astrophysical studies because of its (1) proximity (mean
distance $\sim$45~pc), also giving rise to several other advantages
such as relatively bright stars and negligible interstellar reddening
and extinction, (2) large proper motion ($\mu \sim 111$~mas~yr$^{-1}$)
and peculiar space motion ($\sim$35~km~s$^{-1}$ with respect to its
own local standard of rest), greatly facilitating both proper motion-
and radial velocity-based membership determinations, and (3) varied
stellar content ($\sim$400 known members, among which are white
dwarfs, red giants, mid-A stars in the turn-off region, and numerous
main sequence stars, at least down to $\sim$$0.10~M_\odot$ M dwarfs).
Its proximity, however, has also always complicated astrophysical
research: the tidal radius of $\sim$10~pc results in a significant
extension of the cluster on the sky ($\sim$20$^\circ\!$) and, more
importantly, a significant depth along the line of sight. As a result,
the precise definition and location of the main sequence and turn-off
region in the Hertzsprung--Russell (HR) diagram, and thereby, e.g.,
accurate knowledge of the Helium content and age of the cluster, has
always been limited by the accuracy and reliability of distances to
individual stars. Unfortunately, the distance of the Hyades is such
that ground-based parallax measurements, such as the Yale programme
(e.g., van Altena et al.\ 1995), have never been able to settle `the
Hyades distance problem' definitively.

The above situation improved dramatically with the publication of the
Hipparcos and Tycho Catalogues (ESA 1997). In 1998, Perryman and
collaborators (hereafter P98) published a seminal paper in which they
presented the Hipparcos view of the Hyades. P98 studied the
three-dimensional spatial and velocity distribution of the members,
the dynamical properties of the cluster, including its overall
potential and density distribution, and its HR diagram and age. At the
mean distance of the cluster ($D \sim 45$~pc), a typical Hipparcos
parallax uncertainty of 1~mas translates into a distance uncertainty
of $D^2/1000 \sim 2$~pc. Because this uncertainty compares favorably
with the tidal radius of the Hyades ($\sim$10~pc), the Hipparcos
distance resolution is sufficient to study details such as mass
segregation (\S 7.1 in P98). Uncertainties in absolute magnitudes, on
the other hand, are still dominated by Hipparcos parallax errors
($\ga$0.10~mag) and not by photometric errors ($\la$0.01~mag; \S 9.0
in P98).

Kinematic modelling of collective stellar motions in moving groups can
yield improved parallaxes for individual stars from the Hipparcos
proper motions (e.g., Dra\-vins et al.\ 1997, 1999; de Bruijne 1999b;
hereafter B99b). Such parallaxes, called `secular parallaxes', are
more precise than Hipparcos trigonometric parallaxes for individual
Hyades as the relative proper motion accuracy is effectively $\sim$3
times larger than the relative Hipparcos parallax accuracy. P98
discuss secular parallaxes for the Hyades (their \S 6.1 and
figures~10--11), but only in view of their statistical consistency
with the trigonometric parallaxes. Improved HR diagrams, based on
Hipparcos secular parallaxes, have been published on several
occasions, but these diagrams merely served as external verification
of the quality and superiority of the secular parallaxes (e.g., Madsen
1999; B99b). Narayanan \& Gould (1999a, b) derived secular parallaxes
for the Hyades, but used these only to study the possible presence and
size of systematic errors in the Hipparcos data. Although narrow main
sequences are readily observable for distant clusters, the absolute
calibration of the HR diagram of such groups is often uncertain due to
their poorly determined distances and the effects of interstellar
reddening and extinction. The latter problems are alleviated
significantly for nearby clusters, but a considerable spread in the
location of individual members in the HR diagram is introduced as a
result of their resolved intrinsic depths.

Hipparcos secular parallaxes for Hyades members provide a unique
opportunity to simultaneously obtain a well-defined and absolutely
calibrated HR diagram. In this paper we derive secular parallaxes for
the Hyades using a slightly modified version of the procedure
described by B99b (\S \ref{sec:method}). Sections
\ref{sec:space_motion} and \ref{sec:additional_members} discuss the
space motion and velocity dispersion of the Hyades, as well as
membership of the cluster, respectively. The secular parallaxes are
derived and validated in \S\S \ref{sec:sec_pars} and
\ref{sec:acc_and_prec}. The validation includes a detailed
investigation of the velocity structure of the cluster and of the
presence of small-angular-scale correlations in the Hipparcos
data. The three-dimensional spatial structure of the Hyades, based on
secular parallaxes, is discussed briefly in \S
\ref{sec:spatial_structure}. Readers who are primarily interested in
the secular parallax-based colour-absolute magnitude and HR diagrams
can turn directly to \S \ref{sec:cmd}; we analyze these diagrams in
detail, and also address related issues such as the transformation
from observed colours and magnitudes to effective temperatures and
luminosities, in \S\S \ref{sec:cmd}--\ref{sec:HRD}. Section
\ref{sec:disc} summarizes and discusses our findings. Appendices A and
B present observational data and discuss details of the derivation of
fundamental stellar parameters for the Hyades red giants and for the
$\delta$ Scuti pulsator $\theta^2$~Tau.

\section{History, outline, and revision of the method}
\label{sec:method}

\subsection{History}
\label{subsec:history}

We define a moving group (or cluster) as a set of stars which share a
common space motion ${\mathbf v}$ to within the internal velocity
dispersion $\sigma_v$. The canonical formula, based on the classical
moving cluster/convergent point method, to determine secular
parallaxes $\pi_{\rm sec}$ from proper motion vectors ${\mathbf \mu}$,
neglecting the internal velocity dispersion, reads (e.g., Bertiau
1958):
\begin{equation}
\pi_{\rm sec} = {{A |{\mathbf \mu}|} \over {|{\mathbf v}\,| \sin\lambda}},
\label{eq:canonical}
\end{equation}
where $A \equiv 4.740\,470\,446~{\rm km}~{\rm yr}~{\rm s}^{-1}$ is the
ratio of one astronomical unit in kilometers to the number of seconds
in one Julian year (ESA 1997, Vol.\ 1, table~1.2.2) and $\lambda$ is
the angular distance between a star and the cluster apex. We express
parallaxes in units of mas (milli-arcsec) and proper motions in units
of mas~yr$^{-1}$.

The derivation of secular parallaxes for Hyades (or Taurus cluster)
members from proper motions and/or radial velocities using
eq.~(\ref{eq:canonical}) dates back to at least Boss (1908; cf.\ Smart
1939; see Turner et al.\ 1994 for a review). Several authors have
criticized such studies for various reasons (e.g., Seares 1945; Brown
1950; Upton 1970; Hanson 1975; cf.\ Cooke \& Eichhorn 1997); the ideal
method {\it simultaneously\/} determines the individual parallaxes and
the cluster bulk motion, as well as the corresponding velocity
dispersion tensor, from the observational data. Murray \& Harvey
(1976) developed such a procedure, and applied it to subsets of Hyades
members. Zhao \& Chen (1994) presented a maximum likelihood method for
the simultaneous determination of the mean distance (and dispersion)
and kinematic parameters (bulk motion and velocity dispersion) of
moving clusters, and also applied it to the Hyades. Cooke \& Eichhorn
(1997) presented a method for the simultaneous determination of the
distances to Hyades and the cluster bulk motion.

\subsection{Outline}
\label{subsec:outline}

The Hipparcos data recently raised renewed interest in secular
parallaxes. Dravins et al.\ (1997) developed a maximum likelihood
method to determine secular parallaxes\footnote{
The aim of Dravins' investigations is the derivation of astrometric
radial velocities. Spectroscopic radial velocities are therefore not
used in their~modelling.
} based on Hipparcos positions, trigonometric parallaxes, and proper
motions, taking into account the correlations between the astrometric
parameters (cf.\ \S 2 in B99b). The algorithm assumes that the space
velocities of the $n$ cluster members follow a three-dimensional
Gaussian distribution with mean ${\mathbf v}$, the cluster space
motion, and standard deviation $\sigma_v$, the (isotropic)
one-dimensional internal velocity dispersion; the three components of
${\mathbf v}$ correspond to the ICRS equatorial Cartesian coordinates
$x$, $y$, and $z$ (ESA 1997, Vol.\ 1, \S 1.5.7). The model has $3 + 1
+ n$ unknown parameters: ${\mathbf v}$, $\sigma_v$, and the $n$ secular
parallaxes.

After maximizing the likelihood function, one can determine the
non-negative mod\-el-observation discrepancy parameter $g$ for each
star (eq.~8 in B99b). As the $g$'s are approximately distributed as
$\chi^2$ with 2 degrees of freedom (Dravins et al.\ 1997; B99b; cf.\
Lindegren et al.\ 2000; Makarov et al.\ 2000), $g > 9$ is a suitable
criterion for detecting outliers. Therefore, the procedure can be
iterated by rejecting deviant stars (e.g., undetected close binaries
or non-members) and subsequently re-computing the maximum likelihood
solution, until convergence is achieved in the sense that all
remaining stars have $g \leq 9$. By definition, the maximum likelihood
estimate of the velocity dispersion $\sigma_v$ decreases during this
process. Monte Carlo simulations show that this iterative procedure
can lead to underestimated values of $\sigma_v$ by as much as
$0.15$~km~s$^{-1}$ for Hyades-like groups (B99b).

\subsection{Revision}
\label{subsec:revision}

\placefigureOne

A wealth of ground-based radial velocity information was
accumulated for the Hyades over the last century. The cluster has an
extent on the sky that is large enough to allow an accurate derivation
of its three-dimensional space motion based on proper motion data
exclusively (e.g., de Bruijne 1999a, b; Hoogerwerf \& Aguilar
1999). We nonetheless decided to modify the maximum likelihood
procedure (\S \ref{subsec:outline}) so as to enforce global
consistency between the maximum likelihood estimate of the radial
component of the cluster space motion and the spectroscopic radial
velocity data as a set. Therefore, we multiplied the astrometric
likelihood merit function ${\cal L}$ (eq.~6 in B99b) by a radial
velocity penalty factor:\null\vskip-0.55truecm
\begin{equation}
{\cal L}
\quad\longrightarrow\quad
{\cal L} \cdot \exp{\left({{\Delta^2} \over {2~\sigma_\Delta^2}}\right)},
\label{eq:penalty}
\end{equation}
where $\Delta$ is the median value of $(v_{\rm rad} - v_{\rm rad,
pred})/$ $({\sigma_{v_{\rm rad}}^2 + \sigma_v^2})^{1/2}$, computed
over all stars with a (reliable) radial velocity (\S
\ref{subsec:space_motion}), where $v_{\rm rad, pred} = v_x \cos\alpha
\cos\delta + v_y \sin\alpha \cos\delta + v_z \sin\delta$, and
$(\alpha, \delta)$ denote the equatorial coordinates of a star. The
quantity $\sigma_\Delta$ is the allowed inconsistency in $\Delta$. We
choose $\sigma_\Delta = 0.5$, which corresponds to
$\sim$$0.25$~km~s$^{-1}$ when expressed in terms of the median
effective radial velocity uncertainty $(\sigma_{v_{\rm rad}}^2 +
\sigma_v^2)^{1/2} \sim 0.50$~km~s$^{-1}$ (\S
\ref{subsec:space_motion}).

In order to work around the $\sigma_v$ bias mentioned in \S
\ref{subsec:outline}, we decided to introduce a second modification of
the original procedure. This change involves the decoupling of the
determination of the cluster motion plus velocity dispersion (\S
\ref{sec:space_motion}) from the determination of the secular
parallaxes and goodness-of-fit parameters $g$ given the cluster
velocity and dispersion (\S \ref{sec:sec_pars}; cf.\ Narayanan \&
Gould 1999a, b). We thus reduce the dimensionality of the problem from
$3 + 1 + n$ to $n$ (cf.\ \S \ref{sec:space_motion}). However, the
$n$-dimensional maximum likelihood problem of finding $n$ secular
parallaxes and goodness-of-fit parameters $g$ for a given space motion
and velocity dispersion simplifies directly to $n$ independent
one-dimensional problems (\S 2 in B99b). This leads to three
additional advantages: (1) it reduces the computational complexity of
the problem; (2) it allows an a posteriori decision on the $g$
rejection limit (\S \ref{subsec:outline}); and (3) it allows a
treatment on the same footing of stars lacking trigonometric parallax
information, such as Hyades selected from the Tycho--2 catalogue (\S
\ref{subsec:faint_mems}). In practice, the analysis of such stars only
requires the reduction of the dimensionality of the vector of
observables ${\mathbf a}_{\rm Hip}$ and the corresponding vector
${\mathbf c}$ and matrices {\bf C}$_{\rm Hip}$ and {\bf D} from 3 to 2
and $3 \times 3$ to $2 \times 2$, respectively, through suppression of
the first component (see \S 2 in B99b).

\section{Space motion and velocity dispersion}
\label{sec:space_motion}

The procedure outlined in \S\S
\ref{subsec:outline}--\ref{subsec:revision} is based on two basic
assumptions, namely that (1) the cluster velocity field is known
accurately, and (2) the astrometric measurements of the stars
correctly reflect their true space motions. Assumption (2) can be
safely met by excluding from the analysis (close) multiple systems,
spectroscopic binaries, etc.\ (\S \ref{subsec:faint_mems}). Assumption
(1) requires a careful analysis, to which we will return in \S
\ref{subsec:vel_field}. For the moment, it suffices to say that there
exist neither conclusive observational evidence nor theoretical
predictions that the velocity field of, at least the central region
of, the Hyades cluster is non-Gaussian with an anisotropic velocity
dispersion.

We opt to define the space motion and velocity dispersion of the
Hyades based on a well-defined set of secure members. P98 used the
Hipparcos positions, parallaxes, and proper motions, complemented with
ground-based radial velocities if available, to derive velocities for
individual stars in order to establish membership based on the
assumption of a common space motion. P98 identified 218 candidate
members, all of which are listed in their table~2 (i.e., column (x) is
`1' or `?'). We take these stars, and exclude all objects without
radial velocity information (i.e., column (p) is `$\ast$') and all
(candidate) close binaries (i.e., either column (s) is `SB'
(spectroscopic binary) or column (u) is one of
`G$\,$O$\,$V$\,$X$\,$S'; see ESA 1997 for the definition of the
Hipparcos astrometric multiplicity fields H59 and H61). This leaves
131 secure single members with high-quality astrometric and radial
velocity information\footnote{
Contrary to what had been communicated, the 26 `new Coravel radial
velocities' in P98's table~2 (column (r) is `24') do {\it not\/}
include the standard zero-point correction of $+0.40$~km~s$^{-1}$ (\S
3.2 in P98; finding confirmed by J.--C. Mermilliod through private
communication).}$^,$\footnote{
As described in P98, the Griffin et al.\ radial velocities (column (r)
is `1') for {\it main sequence\/} stars in their table 2 have to be
corrected according to Gunn et al.'s (1988) eq.~(12), but accounting
for a sign error (Gunn's eq.~12 should read: $v_{\rm meas} - v_{\rm
true} = \ldots$ instead of $v_{\rm true} - v_{\rm meas} = \ldots$):
$v_{\rm rad, corrected} = v_{\rm rad} - q(V) - 0.5$~km~s$^{-1}$ for $V
> 6.0$~mag and $v_{\rm rad, corrected} = v_{\rm rad} -
0.5$~km~s$^{-1}$ for $V \leq 6.0$~mag, where $q(V) = 0.44 - 700 \cdot
10^{-0.4 \cdot V}$. The seemingly large discontinuity
($\sim$$2.3$~km~s$^{-1}$) at $V = 6$~mag in this correction is
academic for the Hyades as there are no {\it main sequence\/} members
brighter than this magnitude.
}.

\begin{table*}[t]
\caption[]{Comparison of the Hyades space motion of different studies:
this study ([$\sigma_v$, $\sigma_\Delta$], where $\sigma_v = 0.20,
0.30$, or $0.40$~km~s$^{-1}$ and $\sigma_\Delta = 0.1$ or $0.5$; the
default is [0.30, 0.50]), Perryman et al.\ (1998; P98 [134/180] for
the 134/180 stars within 10/20~pc of the cluster center), Narayanan \&
Gould (1999a; NG99a), Dravins et al.\ (1997; D97), and Lindegren et
al.\ (2000; LMD00). The components of the cluster space motion
${\mathbf v}$ in the equatorial Cartesian ICRS coordinate system are
$v_x$, $v_y$, and $v_z$ (in km~s$^{-1}$). The coordinate system $(v_r,
v_\perp, v_\parallel)$ is described in \S \ref{subsec:space_motion}.
The convergent point $(\alpha, \delta)_{\rm cp}$ is given in
equatorial coordinates (in the ICRS system in degrees). The Hyades
convergent point coordinates are strongly correlated; a typical value
for the correlation coefficient $\rho$ between $\alpha_{\rm cp}$ and
$\delta_{\rm cp}$ is $\rho \sim -0.8$ (\S 4.1 and figure 4 in de
Bruijne 1999a).}
\renewcommand{\arraystretch}{0.9}
\renewcommand{\tabcolsep}{10.0pt}
\begin{center}
\begin{tabular}{lrrrrrrrrrrr}
\noalign{\vskip -0.40truecm}
\noalign{\vskip 0.10truecm}
\hline
\hline
\noalign{\vskip 0.07truecm}
Study &
\multicolumn{1}{c}{$v_x$} &
\multicolumn{1}{c}{$\sigma_{v_x}$} &
\multicolumn{1}{c}{$v_y$} &
\multicolumn{1}{c}{$\sigma_{v_y}$} &
\multicolumn{1}{c}{$v_z$} &
\multicolumn{1}{c}{$\sigma_{v_z}$} &
\multicolumn{1}{c}{$v_r$} &
\multicolumn{1}{c}{$v_\perp$} &
\multicolumn{1}{c}{$v_\parallel$} &
\multicolumn{1}{c}{$\alpha_{\rm cp}$} &
\multicolumn{1}{c}{$\delta_{\rm cp}$}\\
\noalign{\vskip 0.07truecm}
\hline
\noalign{\vskip 0.07truecm}
$[0.30, 0.5]$ & $-5.84$ & 0.26 & 45.68 & 0.11 & 5.54 & 0.07 & 39.48 & $ 0.00$ & 24.35 & 97.29 & 6.86\\
$[0.30, 0.1]$ & $-5.74$ & 0.10 & 45.67 & 0.08 & 5.58 & 0.04 & 39.52 & $-0.00$ & 24.25 & 97.16 & 6.91\\
$[0.20, 0.5]$ & $-5.77$ & 0.10 & 45.64 & 0.05 & 5.59 & 0.07 & 39.49 & $-0.03$ & 24.25 & 97.21 & 6.93\\
$[0.40, 0.5]$ & $-5.99$ & 0.18 & 45.73 & 0.09 & 5.52 & 0.05 & 39.46 & $-0.03$ & 24.50 & 97.46 & 6.83\\
P98 [134]     & $-6.28$ & 0.20 & 45.19 & 0.20 & 5.31 & 0.20 & 38.82 & $-0.02$ & 24.56 & 97.91 & 6.66\\
P98 [180]     & $-6.32$ & 0.20 & 45.24 & 0.20 & 5.30 & 0.20 & 38.84 & $-0.02$ & 24.62 & 97.96 & 6.61\\
NG99a         & $-5.70$ & 0.20 & 45.62 & 0.11 & 5.65 & 0.08 & 39.51 & $-0.06$ & 24.17 & 97.12 & 7.01\\
D97           & $-6.07$ & 0.13 & 45.77 & 0.36 & 5.53 & 0.11 & 39.47 & $-0.06$ & 24.59 & 97.55 & 6.83\\
LMD00         & $-5.90$ & 0.13 & 45.65 & 0.34 & 5.56 & 0.10 & 39.44 & $-0.05$ & 24.38 & 97.36 & 6.89\\
\hline\hline
\end{tabular}
\label{tab:space_mot_comp}
\end{center}
\end{table*}

\subsection{Velocity dispersion}
\label{subsec:vel_disp}

We start the unrevised procedure (\S \ref{subsec:outline}) for
the above-described sample of 131 stars using a $g$ rejection limit of
9. Figure~\ref{fig:velocity} (panel d) shows the evolution of the
maximum likelihood value derived for $\sigma_v$ while rejecting
stars. The estimated velocity dispersion decreases rapidly,
more-or-less linearly, from $\sim$1~km~s$^{-1}$ initially to
$\sim$$0.30$~km~s$^{-1}$ in the first $\sim$11 steps. Previous studies
of the Hyades cluster show that its physical (one-dimensional)
velocity dispersion is $\sim$$0.30$~km~s$^{-1}$ (Gunn et al.\ 1988:
$0.23 \pm 0.05$~km~s$^{-1}$; Zhao \& Chen 1994: $0.37 \pm
0.04$~km~s$^{-1}$; Dravins et al.\ 1997: $0.25 \pm 0.04$~km~s$^{-1}$;
P98: 0.20--0.40~km~s$^{-1}$; Narayanan \& Gould 1999a, b: $0.32 \pm
0.04$~km~s$^{-1}$; Lindegren et al.\ 2000: $0.31 \pm
0.02$~km~s$^{-1}$; Makarov et al.\ 2000: $\sim$0.32~km~s$^{-1}$).
From step $\sim$12 onwards, the maximum likelihood dispersion estimate
decreases, again more-or-less linearly but much more gradually, to
$\sim$0~km~s$^{-1}$ in step $\sim$60--70. Not surprisingly, the
corresponding evolution of the space motion shows an unwanted trend
beyond step $\sim$12 (not shown): the unrevised method is forced to
search for a maximum likelihood solution which has a velocity
dispersion that is smaller than the physical value.

Given the Hyades space motion (or convergent point; \S
\ref{subsec:space_motion}), a semi-independent\footnote{
Kinematic member selection requires an a priori estimate of the
expected velocity dispersion in the cluster. The stars used in this
analysis were selected as members by P98 under the assumption that the
cluster velocity dispersion is small compared to the typical
measurement error of a stellar velocity.
} estimate of the internal velocity dispersion $\sigma_v$ can be
derived through a so-called $\mu_\perp$-component analysis (\S 20 in
Blaauw 1946; \S 7.2 in P98; \S 4.2 in B99b; Lindegren et al.\
2000). The $\mu_\perp$ proper motion components are directed
perpendicular to the great circle joining a star and the apex, and as
such, by definition, exclusively represent peculiar motions
($\Delta_{v, \perp}$; one-dimensional, in km~s$^{-1}$) and
observational errors ($\Delta_{\mu_\perp}$; in mas~yr$^{-1}$):
\begin{equation}
\mu_\perp = A^{-1} \pi \Delta_{v,\perp} + \Delta_{\mu_\perp}
\Longrightarrow
\overline{\mu_\perp^2} =
   \overline{(A^{-1} \pi \Delta_{v,\perp})^2} +
   \overline{\Delta_{\mu_\perp}^2},
\label{eq:sigma_v_perp}
\end{equation}
where the step follows from the statistical independence of the
peculiar motions and the observed proper motion errors. Upon using
$\pi = 21.58$~mas\footnote{
Individual secular parallaxes (\S \ref{sec:sec_pars}) give identical
results.
} ($D = 46.34$~pc; P98), and calculating $\mu_\perp$ and
$\Delta_{\mu_\perp}$ from the Hipparcos positions and proper motions
using the maximum likelihood apex $(\alpha_{\rm cp}, \delta_{\rm cp})
= (97\fdg29, 6\fdg86)$ (Table~\ref{tab:space_mot_comp}; \S
\ref{subsec:space_motion}), it follows that $({\overline{\Delta_{v,
\perp}^2}})^{1/2} \sim 0.20$--$0.40$~km~s$^{-1}$, where the precise
value of this quantity depends on the~details of the selection and
subdivision of the stellar sample (Table~\ref{tab:mu_perp}). The
abovementioned range is consistent with our assumed value of
$0.30$~km~s$^{-1}$. We therefore decided to take $\sigma_v =
0.30$~km~s$^{-1}$ fixed in the remainder of this study, i.e., we
reduce the dimensionality of the problem from $3 + 1 + n$ to $3 + n$
(\S \ref{subsec:revision}).

\subsection{Space motion}
\label{subsec:space_motion}

Our next step is to start the revised procedure (\S
\ref{subsec:revision}) for the same sample of 131 stars, but take
$\sigma_v = 0.30$~km~s$^{-1}$ and $\sigma_\Delta = 0.5$. We exclude
multiple systems without a known systemic (or center-of-mass or
$\gamma$-) velocity, as well as objects with a variable radial
velocity, in the calculation of the penalty factor
(eq.~\ref{eq:penalty}; i.e., all stars with a {\#}-sign preceding
column (q) in P98's table 2). Figure~\ref{fig:velocity} shows the
evolution of the maximum likelihood estimates of the space motion
components while rejecting stars; we derive $(v_x, v_y, v_z) = (-5.84
\pm 0.26, 45.68 \pm 0.11, 5.54 \pm 0.07)$~km~s$^{-1}$.
Table~\ref{tab:space_mot_comp} shows the results of varying
$\sigma_\Delta$ and $\sigma_v$. Changing $\sigma_\Delta$, for example,
from 0.5 to 0.1 at fixed $\sigma_v = 0.30$~km~s$^{-1}$ yields a set of
secular parallaxes (\S \ref{sec:sec_pars}) which differ systematically
in the sense $\langle \pi_{{\rm sec},\sigma_\Delta = 0.1} - \pi_{{\rm
sec},\sigma_\Delta = 0.5} \rangle \sim +0.08$~mas, independent of
visual magnitude. We take $(v_x, v_y, v_z) = (-5.84, 45.68,
5.54)$~km~s$^{-1}$ fixed in the remainder of this study, i.e., we
reduce the dimensionality of the problem, now from $3 + n$ to $n$.

\placefigureTwo

Table~\ref{tab:space_mot_comp} compares the space motions found by us
with results derived by P98\footnote{
Whereas our space cluster motion(s) and the values listed by Dravins
et al.\ and Narayanan \& Gould correspond to the arithmetic mean value
of individual motions of (a given set of) stars, P98 lists
mass-weighted mean values of individual velocities. We investigated
the effect of weighing the individual motions by stellar mass, and
found the difference between the final cluster space motions to be
generally less than $0.10$~km~s$^{-1}$ in each coordinate; we
therefore conservatively assume that quoting a $0.20$ instead of a
$0.10$~km~s$^{-1}$ error on the P98 space motion components `absorbs'
this uncertainty.
}, Dravins et al.\ (1997), Narayanan \& Gould (1999a), and Lindegren
et al.\ (2000) from Hipparcos data. We refrain from comparing our
values to pre-Hipparcos results (e.g., Schwan 1991; Zhao \& Chen 1994;
Cooke \& Eichhorn 1997), as these are (possibly) influenced by
fundamental uncertainties in the pre-Hipparcos proper motion reference
frames (the Hipparcos positions and proper motions are absolute, and
are given in the Hipparcos ICRS inertial reference frame; cf.\ \S 4 in
P98). Table~\ref{tab:space_mot_comp} also compares the different space
motions in the $(v_r, v_\perp, v_\parallel)$ coordinate system, which
is oriented such that the $v_r$-axis is along the radial direction of
the cluster center, which is (arbitrarily) defined as $(\alpha,
\delta)_{\rm center} = (4^{\rm h}\, 26^{\prime}\, 32^{\prime\prime},
17^\circ\!{~}13{\farcm}3)$ (J2000.0), the $v_\perp$-axis is along the
direction perpendicular to the cluster proper motion in the plane of
the sky, and the $v_\parallel$-axis is parallel to the cluster proper
motion in the plane of the sky. We conclude that our space motion is
consistent with the values derived by Dravins et al., P98, Narayanan
\& Gould, and Lindegren et al.; our radial motion agrees very well
with the Dravins et al., Narayanan \& Gould, and Lindegren et al.\
values, whereas our tangential motion perfectly agrees with Lindegren
et al.\ and lies between the Narayanan \& Gould value on the one hand
and the P98 and Dravins et al.\ values on the other hand. The radial
components of the P98 space motions ($38.82$--$38.84$~km~s$^{-1}$)
deviate significantly (at the level of $\sim$0.70~km~s$^{-1}$) from
all other values in Table~\ref{tab:space_mot_comp} (cf.\ \S 4.2 in
Narayanan \& Gould 1999a). Unfortunately, the mean spectroscopically
determined radial velocity of the Hyades cluster is not well defined;
Detweiler et al.\ (1984), for example, find $39.1 \pm
0.2$~km~s$^{-1}$, but their table~1 gives an overview of previous
determinations which show a discouragingly large spread (cf.\ Gunn et
al.\ 1988).  Figure~\ref{fig:v_rad} shows, for the 131 secure single
members, the distribution of observed radial velocities (left), the
distribution of observed minus predicted radial velocities (\S
\ref{subsec:revision}) given the cluster space motion (middle), and
the properly normalized distribution of observed minus predicted
radial velocities (right; taking into account a velocity dispersion of
$\sigma_v = 0.30$~km~s$^{-1}$). The distribution of observed radial
velocities is not symmetric but skewed towards lower $v_{\rm rad,
obs}$ values; the median $v_{\rm rad, obs}$ value
($38.60$~km~s$^{-1}$) is $0.66$~km~s$^{-1}$ larger than the straight
mean of the observed $v_{\rm rad, obs}$ values
($37.94$~km~s$^{-1}$). The large spread and skewness in the
distribution of observed radial velocities are caused by the
perspective effect, which is significant for the Hyades due to its
large extent on the sky (\S \ref{sec:intro}). The perspective effect
has been removed in the middle and right panels of
Figure~\ref{fig:v_rad}. The middle panel shows that the radial
component of our space motion ($39.48$~km~s$^{-1}$) yields an
acceptable $v_{\rm rad, obs} - v_{\rm rad, pred}$ distribution. The
deviation between the predicted zero-mean unit-variance Gaussian and
the observed histogram in the right panel is possibly caused by (1)
the presence of a few non-members (and possibly some not-detected
close binaries), (2) a slightly underestimated cluster velocity
dispersion, and/or (3) underestimated $v_{\rm rad, obs}$ errors. The
histogram and Gaussian prediction would agree, given the $v_{\rm rad,
obs}$ errors, if $\sigma_v$ is increased to
$\sim$0.80--0.90~km~s$^{-1}$, or, given $\sigma_v = 0.30$~km~s$^{-1}$,
if the individual random $v_{\rm rad, obs}$ errors are increased by an
amount of $\sim$0.50--0.60~km~s$^{-1}$. While the first possibility
seems highly unlikely (\S \ref{subsec:vel_disp}; cf.\ Gunn et al.\
1988), the required `extra radial velocity uncertainty' is not
unreasonable, given it is of the same order of magnitude as the
(poorly determined) non-physical zero-point shifts usually adopted in
radial velocity studies (e.g., Gunn et al.\ 1988; cf.\ \S\S 3.2 and
7.2 in P98).

\section{Membership}
\label{sec:additional_members}

Having determined the Hyades space motion and velocity dispersion, we
are in a position to discuss membership.

\subsection{Hipparcos: Perryman et al.\ (1998) candidates}
\label{subsec:mem_P98}

\begin{table}[t]
\centering
\caption[]{Statistics of the $\mu_\perp$ proper motion components for
the 63 brightest (${\rm spectral~type = SpT} \leq {\rm G5}$)
high-fidelity ($g \equiv g_{\rm Hip} \leq 9$; \S \ref{sec:sec_pars})
single P98 members (column~(s) in P98's table 2 is not `SB'), using
$(\alpha_{\rm cp}, \delta_{\rm cp}) = (97\fdg29, 6\fdg86)$ (see
eq.~\ref{eq:sigma_v_perp} and \S \ref{subsec:vel_disp}). Hyades main
sequence members with spectral types later than $\sim$G5 have modest
Hipparcos proper motion accuracies due to their faint magnitudes ($V
\mathrel{{\hbox to 0pt{\lower 3pt\hbox{$\sim$}\hss}} \raise
2.0pt\hbox{$>$}} 8.5$~mag; e.g., figure 1 in Hoogerwerf 2000). Results
are tabulated for four ranges in spectral type ($n$ stars from ${\rm
SpT}_{-}$ trough ${\rm SpT}_{+}$); the spectral-type averaged value of
{\vtop to 0pt {\hsize=35pt $({\overline{\Delta_{v,
\perp}^2}})^{1/2}$}}~\ for these 63 stars is $\sim$0.25~km~s$^{-1}$.}
\renewcommand{\arraystretch}{0.9}
\renewcommand{\tabcolsep}{6.0pt}
\begin{tabular}{cccccc}
\noalign{\vskip -0.40truecm}
\noalign{\vskip 0.10truecm}
\hline
\hline
\noalign{\vskip 0.07truecm}
${\rm SpT}_{-}$ &
${\rm SpT}_{+}$ &
$n$ &
$(\overline{{\mu_\perp^2}})^{1/2}$ &
$(\overline{{\Delta_{\mu_\perp}^2}})^{1/2}$ &
$({\overline{\Delta_{v, \perp}^2}})^{1/2}$\\
 & & & mas & mas & km~s$^{-1}$\\
\noalign{\vskip 0.07truecm}
\hline
\noalign{\vskip 0.07truecm}
A2 & F0 & 16 & 1.35 & 0.89 & 0.23\\
F0 & F5 & 16 & 1.71 & 0.88 & 0.31\\
F5 & F8 & 16 & 1.33 & 0.96 & 0.20\\
F9 & G5 & 15 & 1.71 & 1.13 & 0.28\\
\hline\hline
\end{tabular}
\label{tab:mu_perp}
\end{table}

The Hipparcos Catalogue contains 118$\,$218 entries which are
homogeneously distribut\-ed over the sky. The catalogue is complete to
$V \sim 7.3$~mag, and has a limiting magnitude of $V \sim
12.4$~mag. In the case of the Hyades, special care was taken to
optimize the number of candidate members in the Hipparcos target
list. As a result, the Hipparcos Input Catalogue (Turon et al.\ 1992)
contains $\sim$240 candidate Hyades members in the field $2^{\rm
h}\,15^{\rm m} < \alpha < 6^{\rm h}\,5^{\rm m}$ and $-2^\circ\! <
\delta < 35^\circ\!$ (\S 3.1 in P98). P98 considered all 5$\,$490
Hipparcos entries in this field for membership, and ended up with 218
members. The P98 member selection is generous: only very few genuine
members, contained in both the Hipparcos Catalogue and the selected
field on the sky, have probably not been selected, whereas a number of
field stars (interlopers) are likely to be present in their list. P98
distinguish members (197 stars) and possible members (21 stars); the
latter do not have measured radial velocities (column~(x) = `?' in
their table~2).

P98 divide the Hyades into four components ($r$ is the
three-dimensional distance to the cluster center): (1) a spherical
`core' with a $2.7$~pc radius and a half-mass radius of $5.7$~pc; (2)
a `corona' extending out to the tidal radius $r_{\rm t} \sim 10~{\rm
pc}$ (134 stars in core and corona); (3) a `halo' consisting of stars
with $r_{\rm t} \la r \la 2 r_{\rm t}$ which are still dynamically
bound to the cluster (45 stars; e.g., Pels et al.\ 1975); and (4) a
`moving group population' of stars, possibly former members, with $r
\ga 2 r_{\rm t}$ which have similar kinematics to the bound members in
the central parts of the cluster (39 stars; e.g., Asiain et al.\ 1999;
cf.\ \S 7 in~P98).

The fact that P98 restricted their search to a pre-defined field on
the sky limits knowledge on and completeness of membership, especially
in the outer regions of the cluster: the 10~pc tidal radius translates
to a cluster diameter of $\sim$$25^\circ\!$, whereas the P98 field
measures $57\fdg5$ in $\alpha$ and $37\fdg0$ in $\delta$. Although
this problem seems minor at first sight, suggesting a solution in the
form of simply searching the entire Hipparcos Catalogue for additional
(moving group) members, it is daunting in practice: thousands of
Hipparcos stars all over the sky have proper motions directed towards
the Hyades convergent point (\S 4.2 in de Bruijne 1999a). Whereas this
in principle means that these stars, in projection at least, are
`co-moving' with the Hyades, the identification of {\it physical
members\/} of a moving group (or `supercluster') population is not
trivial, and requires additional observational data (cf.\ \S\S7--8 and
table 6 in P98; \S \ref{subsubsec:vel_field_Hip_P98}). We therefore
restrict ourselves to the P98 field (\S \ref{subsec:addi_mems}).
Section \ref{subsec:faint_mems} discusses the possibility to extend
membership down to fainter magnitudes using the Tycho--2 astrometric
catalogue.

\subsection{Hipparcos: additional candidates}
\label{subsec:addi_mems}

De Bruijne (1999a) and Hoogerwerf \& Aguilar (1999) re-analyzed Hyades
membership, based on the refurbished convergent point and Spaghetti
method. These studies used Hipparcos data but excluded radial velocity
information. The convergent point method uses proper motion data only,
confirms membership for 203 of the 218 P98 members (cf.\
Table~\ref{tab:data_1}), and selects $12$ new candidates within 20~pc
of the cluster center. The Spaghetti method, using proper motion and
parallax data, selects six new candidate members, three of which are
in common with the $12$ proper motion candidates mentioned above. The
Spaghetti method does not confirm 56 P98 members (cf.\
Table~\ref{tab:data_1}); however, 49 of these are low-probability
(i.e., `1--3$\sigma$') P98 members. Table~\ref{tab:additional_members}
lists the 15 new candidates. We defer the discussion of these stars to
\S \ref{subsec:Hip_additional_members}.

\subsection{Tycho--2: bright binaries and faint candidates}
\label{subsec:faint_mems}

The Tycho(--1) Catalogue (ESA 1997), which is based on measurements of
Hipparcos' star\-mapper, contains astrometric data for $\sim$1 million
stellar systems with a $\sim$10--20 times smaller precision than
Hipparcos. Its completeness limit is $V \sim 10.5$~mag. Despite the
`inferior astrometric precision', the Tycho positions as a set are
superior to similar measurements in {\it any\/} other catalogue of
comparable size. The Tycho measurements (median epoch 1991.25) have
therefore been used as second epoch material in the construction of a
long time-baseline proper motion project, culminating in the Tycho--2
catalogue (H{\o}g et al.\ 2000a, b; cf.\ Urban et al.\ 1998a; Kuzmin
et al.\ 1999). This project uses the Astrographic Catalogue positions,
as well as data from 143 other ground-based astrometric catalogues, as
first epoch material (median epoch $\sim$1904). The Astrographic
Catalogue (D\'ebarbat et al.\ 1988; Urban et al.\ 1998b) was part of
the `Carte du Ciel' project, which envisaged the imaging of the entire
sky on 22$\,$660 overlapping photographic plates by 20 observatories
in different `declination zones'. The Tycho--2 catalogue contains
absolute proper motions in the Hipparcos ICRS frame for $\sim$2.5
million stars with a median error of $\sim$2.5~mas~yr$^{-1}$. Its
completeness limit is $V \sim 11.0$~mag. Tycho--2 contains proper
motions for 208 of the 218 P98 candidates; the entries HIP 20440,
20995, and 23205 are (photometrically) resolved binaries in Tycho--2.

\placefigureThree

The Tycho--2 proper motions can be used in two ways. First, as a
result of the $\sim$4~year time baseline over which Hipparcos obtained
its astrometric data, the proper motions of some multiple systems do
not properly reflect their true systemic motions as a result of
unrecognized orbital motion (e.g., de Zeeuw et al.\ 1999; Wielen et
al.\ 1999, 2000; these systems are called `$\Delta\mu$ binaries'). As
the long time-baseline Tycho--2 proper motions suffer from this effect
to a much smaller extent, they sometimes significantly better
represent the true motion of an object than the Hipparcos measurements
do. Second, the Tycho--2 catalogue can in principle be used to provide
membership information for stars beyond the Hipparcos completeness
limit, i.e., in the range $7.3 \la V \la 12.4$~mag (the Tycho--2
catalogue contains $\sim$90,000 stars in P98's Hyades field). The
search for new (faint) members is, however, non-trivial. The Hipparcos
Catalogue in the field of the Hyades is already relatively complete in
terms of (known) members of the cluster (cf.\ figure 2 and \S 3.1 in
P98), so that the majority of new members is necessarily quite faint
($V \ga 10.0$~mag). Moreover, most of the Tycho--2 entries have no
known radial velocity and/or parallax (although some stars in the
Tycho--2 catalogue have Tycho parallax measurements, the typical
associated random errors for Hyades fainter than $V \sim 7.3$~mag are
similar in size or larger than the expected parallaxes themselves,
$\pi \sim 22$~mas). We therefore cannot select (faint) Tycho--2 Hyades
members along the lines set out by P98, but must, e.g., follow
Hoogerwerf's (2000) method instead. This method, which is based on a
convergent point method, selects (faint) stars which (1) are
consistent with a given convergent point and the two-dimensional
proper motion distribution of a given set of (bright) Hipparcos
members, and (2) follow a given main sequence. However, the resulting
list of candidate members contains hundreds of falsely identified
objects (interlopers), especially in the faint magnitude regime (e.g.,
Hoogerwerf 2000; cf.\ de Bruijne 1999a). Reliable suppression of these
stars requires at least (yet unavailable) radial velocity and/or
parallax data. We therefore refrain from pursuing this route further
in this paper.

\section{Secular parallaxes}
\label{sec:sec_pars}

\begin{table*}[t]
\caption[]{Hipparcos and Tycho--2 secular parallaxes, and associated
goodness-of-fit parameters $g_{\rm Hip}$ and $g_{\rm Tycho-2}$, for
the binaries which have both trigonometric (ESA 1997) and orbital
parallaxes (Torres et al.\ 1997a, b, c).}
\renewcommand{\arraystretch}{0.9}
\renewcommand{\tabcolsep}{6.0pt}
\begin{center}
\begin{tabular}{lrrrrrrr}
\noalign{\vskip -0.40truecm}
\noalign{\vskip 0.10truecm}
\hline
\hline
\noalign{\vskip 0.07truecm}
\multicolumn{1}{c}{HIP} &
\multicolumn{1}{c}{TYC} &
\multicolumn{1}{c}{$\pi_{\rm Hip}$} &
\multicolumn{1}{c}{$\pi_{\rm sec, Hip}$} &
\multicolumn{1}{c}{$\pi_{\rm sec, Tycho-2}$} &
\multicolumn{1}{c}{$\pi_{\rm orb}$} &
\multicolumn{1}{c}{$g_{\rm Hip}$} &
\multicolumn{1}{c}{$g_{\rm Tycho-2}$}\\
&
&
\multicolumn{1}{c}{mas} &
\multicolumn{1}{c}{mas} &
\multicolumn{1}{c}{mas} &
\multicolumn{1}{c}{mas} &
& \\
\noalign{\vskip 0.07truecm}
\hline
\noalign{\vskip 0.07truecm}
20087$^{\rm a}$ & 1276~1622~1 & $18.25 \pm 0.82$ & $18.31 \pm 0.69$ & $18.70 \pm 0.29$ & $17.92 \pm 0.58$ & 0.19 & 0.00 \\
20661$^{\rm b}$ & 1265~1171~1 & $21.47 \pm 0.97$ & $21.29 \pm 0.37$ & $21.08 \pm 0.38$ & $21.44 \pm 0.67$ & 7.20 & 0.00 \\
20894$^{\rm c}$ & 1265~1172~1 & $21.89 \pm 0.83$ & $22.24 \pm 0.36$ & $22.35 \pm 0.36$ & $21.22 \pm 0.76$ & 0.26 & 0.22 \\
\hline\hline\\[-0.65truecm]
\end{tabular}
\label{tab:orbpar}
\end{center}
$^{\rm a}$: $\pi_{\rm trigonometric} = 19.4 \pm 1.1$~mas (Gatewood et
al.\ 1992) and $\pi_{\rm trigonometric} = 18.23 \pm 0.86$~mas
(S\"oderhjelm 1999).\\
$^{\rm b}$: $\pi_{\rm trigonometric} = 22.1 \pm 1.1$~mas (S\"oderhjelm
1999).\\
$^{\rm c}$: See \S\S \ref{subsec:HRD_3} and \ref{subsec:trans_theta2}
for a discussion of this system.\\
\end{table*}

We now determine secular parallaxes for the 218 P98 members and the 15
new candidates discussed in \S \ref{sec:additional_members}, using the
space motion and velocity dispersion found in \S
\ref{sec:space_motion} ($v_x, v_y, v_z, \sigma_v = -5.84, 45.68, 5.54,
0.30$~km~s$^{-1}$) as constants (\S \ref{subsec:revision}). This
provides, for each proper motion (either Hipparcos or Tycho--2), a
secular parallax ($\pi_{\rm sec, Hip}$ or $\pi_{\rm sec, Tycho-2}$)
and an associated random error ($\sigma_{\pi,{\rm sec, Hip}}$ or
$\sigma_{\pi, {\rm sec, Tycho-2}}$; \S \ref{subsec:accuracy}) and
goodness-of-fit parameter ($g_{\rm Hip}$ or $g_{\rm Tycho-2}$;
Appendix~A). As the Hipparcos and Tycho--2 proper motions are
independent measurements, the corresponding secular parallaxes can in
principle be averaged, taking the errors into account as weighting
factors. It is, however, less clear how to incorporate the
goodness-of-fit parameters $g_{\rm Hip}$ and $g_{\rm Tycho-2}$ in the
averaging process. We therefore provide both secular parallaxes for
all stars and refrain from giving any average value.

\subsection{Hipparcos: Perryman et al.\ (1998) candidates}
\label{subsec:Hip_P98_members}

Figure~\ref{fig:prlxs} shows a global comparison between the different
sets of parallaxes. The mean and/or median Hipparcos parallax is equal
to the mean and/or median secular parallax (either Hipparcos or
Tycho--2) to within $\la$0.10~mas. This implies that the secular
parallaxes are reliable and do not suffer from a significant
systematic component (cf.\ \S \ref{sec:acc_and_prec}). This conclusion
is supported by Table~\ref{tab:orbpar}, which compares trigonometric
and secular parallaxes for three Hyades binaries which also have
orbital parallaxes.

The goodness-of-fit parameter $g_{\rm Hip}$ allows a natural division
between high-fidelity kinematic members ($g_{\rm Hip} \leq 9$) and
kinematically deviant stars ($g_{\rm Hip} > 9$; \S
\ref{subsec:outline}; cf.\ Figure~\ref{fig:prlxs_and_gs}). The latter
are not necessarily non-members, but can also be (close) multiple
stars for which the Hipparcos proper motions do not properly reflect
the center-of-mass motion (\S \ref{subsec:faint_mems}). Fifty of the
197 P98 members with known radial velocities have $g_{\rm Hip} > 9$,
which leaves a number of high-fidelity members similar to that found
by Dravins et al.\ (1997; 133 stars), Madsen (1999; 136 stars), and
Narayanan \& Gould (1999b; 132 stars)(cf.\ table 3 in Lindegren et
al.\ 2000). Fourteen of the 21 {\it possible\/} P98 members (column
(x) = '?' in their table 2) have $g_{\rm Hip} > 9$. These stars do not
have measured radial velocities (\S \ref{subsec:mem_P98}), and P98
membership is based on proper motion data only. All but one of these
stars are rejected as Hyades members by de Bruijne (1999a) and/or
Hoogerwerf \& Aguilar (1999; Table~\ref{tab:data_1}; cf.\ \S
\ref{subsec:addi_mems}). The 14 suspect secular parallaxes thus most
likely indicate these objects are non-members.

\subsection{Hipparcos: additional candidates}
\label{subsec:Hip_additional_members}

\placefigureFour

Table~\ref{tab:additional_members} lists secular parallaxes for the 15
additional candidate members (\S \ref{subsec:addi_mems}). The
assumption that these stars share the same space motion as the Hyades
cluster (in other words: the assumption of membership) generally
results in both high values for the goodness-of-fit parameters $g_{\rm
Hip}$ and secular parallaxes which are inconsistent with the
trigonometric values. This means these stars are likely non-members
(cf.\ \S 4.2 in B99b); only three of them (HIP 19757, 21760, and
25730) have $g_{\rm Hip} \leq 9$. In retrospect, especially HIP 19757
is a likely new member:
it was selected as candidate both by de Bruijne (1999a) and Hoogerwerf
\& Aguilar (1999);
it has an uncertain trigonometric parallax ($\pi_{\rm Hip} = 16.56 \pm
4.48$~mas) due to its faint magnitude ($V = 11.09$~mag);
it has a Hipparcos secular parallax ($\pi_{\rm sec, Hip} = 20.19 \pm
1.04$~mas; $g_{\rm Hip} = 2.67$) which places it at only 7.15~pc from
the cluster center;
its Hipparcos secular parallax puts it on the Hyades main
sequence (\S \ref{sec:cmd}); and
it has an unknown radial velocity.

\subsection{Tycho--2: faint candidates}
\label{subsec:Tycho2}

The `Base de Donn\'ees des Amas ouverts' database (BDA; {\tt
http://obswww.unige.ch/webda/webda.html}) contains 23 photometric
Hyades which are not contained in the Hipparcos Catalogue but which
were observed by Tycho (cf.\ \S 3.1 and figure 2 in P98). The Tycho--2
secular parallaxes of most of these stars lie between 18 and 22~mas,
indicating they are located at the same distance as the bulk of the
bright Hyades. Only four of them have $g_{\rm Tycho-2} > 9$. Most of
these stars are thus likely members. We discuss their HR diagram
positions in \S \ref{sec:cmd}.

\subsection{Random secular parallax errors}
\label{subsec:accuracy}

Table~\ref{tab:data_1} contains random secular parallax errors
resulting from both uncertainties in the underlying proper motions and
the internal velocity dispersion in the cluster ($\sigma_v =
0.30$~km~s$^{-1}$; \S 4.1 in B99b). Hipparcos\-/Tycho--2 secular
parallax errors for Hyades are on average a factor $\sim$3.0 smaller
than the corresponding Hipparcos trigonometric parallax errors.

\section{Systematic secular parallax errors?}
\label{sec:acc_and_prec}

Although the secular parallaxes derived in \S \ref{sec:sec_pars} have
small random errors, they might suffer from significant systematic
errors. In this section, we investigate the influences of the maximum
likelihood method, the uncertainty of the tangential component of the
cluster space motion (\S \ref{subsec:acc_1}), the correlated Hipparcos
measurements (\S \ref{subsec:acc_2}), as well as possible unmodelled
patterns in the velocity field of the Hyades (\S
\ref{subsec:vel_field}). Section \ref{subsec:summary} summarizes our
results.

\subsection{Cluster space motion}
\label{subsec:acc_1}

Extensive Monte Carlo tests of the maximum likelihood procedure (\S\S
\ref{subsec:outline}--\ref{subsec:revision}) show that, given the
correct cluster space motion, the method is not expected to yield
systematic secular parallax errors larger than a few hundredths of a
mas (e.g., \S\S 3.2.2.1 and 3.2.2.3, table 3, and figure 5 in
B99b). It is possible, though, that a systematic error is introduced
through the use of an incorrect value for the tangential component of
the cluster space motion ($v_\parallel$; \S
\ref{subsec:space_motion}). We estimate $\sigma_{v_\parallel} \sim
0.15$~km~s$^{-1}$ from Table~\ref{tab:space_mot_comp}. This
uncertainty gives rise to {\it maximum\/} systematic secular parallax
errors of $\sim$0.14~mas ($\sigma_{v_\parallel} = 0.30$~km~s$^{-1}$
gives 0.28~mas; see \S 4.2 in B99b and use $v_\parallel =
24.35$~km~s$^{-1}$, $\overline\mu = 111.0$~mas~yr$^{-1}$,
$\sigma_{\overline\mu} = 0.15$~mas~yr$^{-1}$, and $\overline{\pi_{\rm
Hip}} = 1000.0/45.0$~mas). This value compares favorably to typical
random secular parallax errors for Hyades ($\sim$$0.50$~mas;
Table~\ref{tab:data_1}). It is, nonetheless, desirable to obtain a
more precise estimate of the tangential component of the cluster space
motion. This requires a better knowledge of the associated radial
space motion component and internal velocity dispersion and/or more
precise proper motion measurements (\S \ref{sec:disc}).

\subsection{Hipparcos correlations on small angular scales}
\label{subsec:acc_2}

The Hipparcos Catalogue contains absolute astrometric data. Absolute
in this sense should be interpreted as lacking global {\it
systematic\/} errors at the $\sim$0.10~mas~(yr$^{-1}$) level or larger
(ESA 1997; cf.\ Narayanan \& Gould 1999a, who quote an upper limit of
$0.47$~mas for the Hyades field). However, the measurement principle
of the satellite does allow for the existence of correlated
astrometric parameters on small angular scales
($\sim$$1^\circ\!$--$3^\circ\!$; e.g., Lindegren 1989; ESA 1997, Vol.\
3, p.\ 323 and 369). These correlations have been suggested to be
responsible for the so-called `Pleiades anomaly', i.e., the fact that
the mean distance of the Pleiades cluster as derived from the mean
Hipparcos trigonometric parallax differs from the value derived from
stellar evolutionary modelling (Pinsonneault et al.\ 1998; but see,
e.g., Robichon et al.\ 1999; van Leeuwen 1999).

\placefigureFive

The left panel of Figure~\ref{fig:NG99b_fig09} shows the
$1^\circ\!$-smoothed error-normalized difference field of the
Hipparcos trigonometric minus secular parallaxes for all stars with
non-suspect secular parallaxes ($g_{\rm Hip} \leq 9$) in the center of
the Hyades cluster ($170\fdg0 \leq \ell \leq 190\fdg0$ and $-32\fdg0
\leq b \leq -12\fdg0$; $\ell$ and $b$ denote Galactic coordinates). In
order to obtain this field, we convolved the sum of the discrete
quantity $s$:
\begin{equation}
s = s(\ell, b) \equiv \delta_{\rm D}(\ell, b) \cdot
{{\pi_{\rm Hip} - \pi_{\rm sec, Hip}} \over
{\sqrt(\sigma_{\pi, {\rm Hip}}^2 + \sigma_{\pi, {\rm sec, Hip}}^2)}},
\label{eq:signal}
\end{equation}
where $\delta_{\rm D}$ denotes the two-dimensional Dirac delta
function, for each star with the normalized two-dimensional Gaussian
smoothing kernel
\begin{equation}
G(\ell, b) \equiv
{{1}\over{2 \pi \sigma_{\rm s}^2}} \cdot \exp\left(-{{1}\over{2}}
\left[{{\ell^2 + b^2}\over{\sigma_{\rm s}^2}}\right]\right),
\label{eq:kernel}
\end{equation}
where $\sigma_{\rm s} = 1^\circ\!$ is the smoothing length. The
appearance of the difference field depends on the adopted smoothing
length, though not very sensitively. Taking a large smoothing length
returns a smooth field, whereas a small smoothing length gives a
`spiky distribution', reminiscent of the original delta function-type
field (eq.~\ref{eq:signal}). Given a Hyad, its closest neighbour on
the sky is typically found at an angular separation of
$\sim$$0\fdg5$. Our choice of the smoothing length ($1\fdg0$)
corresponds to the median value (for all stars) of the median angular
separation of the $\sim$3--4 nearest neighbours on the sky. We checked
that the smoothed difference field has the same overall appearance
when adopting smoothing lengths of $0\fdg5$ or $2\fdg0$.

The smoothed difference field shows several positive and negative
peaks with a full-width-at-half-maximum of a few degrees. These peaks
can be due to spatially correlated errors in the Hipparcos parallaxes
$\pi_{\rm Hip}$ on small angular scales, spatially correlated errors
in the Hipparcos secular parallaxes $\pi_{\rm sec, Hip}$ on small
angular scales, or both. From the fact that the peaks are not evident
in the smoothed difference field of the Hipparcos secular parallaxes
and the mean cluster parallax (not shown), whereas they are present in
the smoothed difference field of the Hipparcos trigonometric
parallaxes and the mean parallax (not shown), we conclude that they
are mainly caused by correlated Hipparcos measurements, notably the
trigonometric parallaxes (cf.\ Narayanan \& Gould 1999b). As the
relative precision of the Hipparcos proper motions is $\sim$5 times
higher than the relative precision of the Hipparcos trigonometric
parallaxes (cf.\ \S \ref{sec:intro}), the Hipparcos secular
parallaxes, which have more-or-less been directly derived from the
Hipparcos proper motions, are correlated as well, though with smaller
`amplitudes'.

The quantity $s$ (eq.~\ref{eq:signal}) denotes, for a given star, the
dimensionless significance (in terms of the effective Gaussian
standard deviation $\sigma \equiv (\sigma_{\pi, {\rm Hip}}^2 +
\sigma_{\pi, {\rm sec, Hip}}^2)^{1/2} \sim 1.5$~mas) of the parallax
difference $\pi_{\rm Hip} - \pi_{\rm sec, Hip}$. As the smoothing
kernel $G$ (eq.~\ref{eq:kernel}) is properly normalized to unit area
in two dimensions, the smoothed difference field can be interpreted in
terms of net significance {\it per square degree\/}. The right panel
of Figure~\ref{fig:NG99b_fig09} shows the corresponding smoothed
difference field expressed in terms of net significance {\it per
star\/}. This field was obtained by dividing the smoothed signal field
by an identically smoothed stellar number density field $\rho(\ell, b)
\equiv \delta_{\rm D}(\ell, b)$ (middle panel of
Figure~\ref{fig:NG99b_fig09}). The smoothed difference field
expressing significance per star also contains patches of size a few
degrees with both negative and positive contributions, although there
is no large-scale trend (cf.\ Figure~\ref{fig:prlxs}). We thus
conclude that the Hipparcos trigonometric parallaxes towards the
Hyades cluster are spatially correlated over angular scales of a few
degrees. The maximum deviations in the central region of the cluster
($\rho \ga 0.5~{\rm star}~{\rm deg}^{-2}$), however, are generally
less than $\sim$$0.50$--$0.75 \sigma$ per star (i.e., $\la$
$0.75$--$1.00$~mas). The signal in the outer parts of the cluster is
statistically non-interpretable as it is severely influenced by the
contributions of individual stars.

Our conclusions are qualitatively consistent with the results of
Narayanan \& Gould (1999b, their figure 9 and \S 6.2; cf.\ Lindegren
et al.\ 2000). These authors, however, overestimate, by a factor of
$\sim$2, the strength of the correlation by claiming that `the
Hipparcos parallaxes toward the Hyades are spatially correlated over
angular scales of a few degrees, {\it with an amplitude of about
1--2~mas}'.

\subsection{Three-dimensional location within the cluster}
\label{subsec:3d_pos}

Figure~\ref{fig:prlxs} (\S \ref{sec:sec_pars}) shows that the secular
parallaxes as a set (i.e., averaged over all regions within the
cluster) are statistically identical to the Hipparcos trigonometric
parallaxes. Figure~\ref{fig:prlxs_and_gs} compares Hipparcos secular
and trigonometric parallaxes as function of the three-dimensional
distance $r$ to the cluster center and as function of spatial location
within the cluster according to an equal-volume pyramid division: we
divide a(n artificial) three-dimensional box containing all cluster
members in six adjacent equal-volume pyramids, all of them having
their top at the cluster center. This division, as viewed from the
Sun in Galactic coordinates, yields six distinct regions: `back',
`left' (i.e., towards smaller longitudes), `front', `right' (i.e.,
towards larger longitudes), `top' (i.e., towards larger latitudes),
and `bottom' (i.e., towards smaller latitudes). Although systematic
differences seem to be present in Figure~\ref{fig:prlxs_and_gs}, they
are smaller than a few tenths of the median effective parallax
uncertainty $({\sigma_{\pi, {\rm Hip}}}^2 + {\sigma_{\pi, {\rm sec,
Hip}}}^2)^{1/2} \sim 1.0$--$1.5$~mas. Figure~\ref{fig:prlxs_and_gs}
also shows that the distribution of the goodness-of-fit parameter
$g_{\rm Hip}$ does not show unwarranted dependencies on distance from
the cluster center. We thus conclude that secular parallaxes for stars
in the inner and outer regions of the cluster do not differ
significantly (i.e., at the $\sim$0.30~mas level or larger).

\subsection{Cluster velocity field}
\label{subsec:vel_field}

\placefigureSix

The method described in \S\S
\ref{subsec:outline}--\ref{subsec:revision} assumes that the
expectation values ${\rm E}({\mathbf v}_i)$ of the individual stellar
velocities ${\mathbf v}_i$ ($i = 1, \ldots, n$) equal the cluster
space motion ${\mathbf v}$ (cf.\ \S \ref{sec:space_motion}). A {\it
random\/} internal velocity dispersion in addition to this common
space motion is allowed and accounted for in the procedure, as random
motions do not affect ${\rm E}({\mathbf v}_i)$ by definition. However,
a {\it systematic\/} velocity pattern, such as expansion or
contraction, rotation, and shear, has not been taken into account in
the modelling. The application of the procedure to data subject to
velocity patterns is thus bound to lead to incorrect and/or biased
results.

\subsubsection{Pre-Hipparcos results}
\label{subsubsec:vel_field_pre_Hip}

\placefigureSeven

Many studies have been devoted to the detection or exclusion of
velocity structure in the Hyades (see P98 for an overview), although
$N$-body simulations of open clusters generally predict that, for
gravitationally bound groups like the Hyades, velocity patterns are
generally too small to be measured with present-day data (e.g.,
Dravins et al.\ 1997; \S 7.2 in P98). In view of the Hyades age ($\tau
= 625 \pm 50$~Myr; P98), shear is not likely to be present in the core
and corona: using $\sigma_v = 0.30$~km~s$^{-1}$ and a half-mass radius
of 5~pc (Pels et al.\ 1975), it follows that $\tau \sim 20$~crossing
times, which means that the central parts of the cluster are relaxed
(cf.\ Hanson 1975). The Hyades age also puts a rough upper limit on
the linear expansion\footnote{
The only astrometrically non-observable velocity pattern is an {\it
isotropic\/} expansion at a rate $K$ (appendix~A in Dravins et al.\
1999): such velocity structure cannot be disentangled from a bulk
motion in the radial direction based on proper motion data only ($K$
is the linear expansion coefficient in km~s$^{-1}$~pc$^{-1}$; cf.\
\S\S 3.2.3--3.2.4 in B99b). Neglecting a uniform expansion for a
cluster at a distance $D$~[pc] yields a bias in its mean radial
velocity of $-D \cdot K$~[km~s$^{-1}$] (e.g., eq.~13 in B99b).
} coefficient $K \la 0.0016~{\rm km}~{\rm s}^{-1}~{\rm pc}^{-1}$
(resulting in a bias in the {\it radial\/} component of the maximum
likelihood cluster space motion of $\la$$-0.07$~km~s$^{-1}$). A global
rotation of the cluster could be present, although Wayman (1967; cf.\
Wayman et al.\ 1965 and Hanson 1975) claimed that the cluster rotation
about three mutually perpendicular axes is consistent with zero to
within 0.05~km~s$^{-1}$~pc$^{-1}$; Gunn et al.\ (1988) present (weak)
evidence for rotation at the level of $\la$1~km~s$^{-1}$~radian$^{-1}$
(cf.\ O'Connor 1914).

\subsubsection{Hipparcos parallaxes: Perryman et al.\ (1998)}
\label{subsubsec:vel_field_Hip_P98}

Figure~8(b) in P98 displays the three-dimensional velocity
distribution $(v_u, v_v, v_w)$ of the 197 P98 members with known
radial velocities. Although the velocity residuals seem to show
evidence for shear and/or rotation, notably for stars in the outer
regions (figure~9 in P98), the systematic pattern can be explained by
a combination of the transformation of the observables $(\pi_{\rm
Hip}, \mu_{\alpha^\ast}, \mu_\delta, v_{\rm rad, obs})$ to the linear
velocity components $(v_u, v_v, v_w)$ and the presence of Hipparcos
data covariances: P98 show that the assumption of a common space
motion for all members with a one-dimensional internal velocity
dispersion of $0.30$~km~s$^{-1}$, which allows averaging of the
individual motions and associated covariance matrices for all stars,
translates into a mean motion and associated mean covariance matrix
(i.e., 1, 2, and 3$\sigma$ confidence regions) which adequately follow
the observed velocity residuals (\S 7.2 and figures 16--17 in P98;
cf.\ top row of Figure~\ref{fig:vel_field}). Therefore, P98 conclude
that the observed kinematic data of the Hyades cluster is consistent
with a common space motion plus a 0.30~km~s$^{-1}$ velocity
dispersion, without the need to invoke the presence of rotation,
expansion, or shear.

The motions of members beyond the tidal radius ($\sim$10~pc), as
opposed to the motions of gravitationally bound members in the central
parts of the cluster, are predominantly influenced by the Galactic
tidal field (e.g., Pels et al.\ 1975). These (evaporated) stars do
therefore not {\it necessarily\/} adhere to the strict pattern of a
common space motion which is present in the core and corona. The
systematic velocity distortions are hard to predict, analytically and
numerically, as they depend critically on the details of the
evaporation mechanism(s) (e.g., Terlevich 1987). They are hard to
observe as well due to both the sparse sampling of `members' and the
uncertain criteria for membership in the outer regions of the cluster.

\subsubsection{Hipparcos parallaxes: this study}
\label{subsubsec:vel_field_Hip}

Figure~\ref{fig:vel_sys} (top series of panels) shows the Hyades
velocity field, based on Hipparcos trigonometric parallaxes, for
different spatial regions of the cluster (\S \ref{subsec:3d_pos}). The
observed velocities are not identically distributed in each region but
show systematic effects, although these are restricted to `the
3$\sigma$ confidence regions'. Notably the `front' and `back' of the
cluster show differences, indicative of a coupling between position
and velocity, i.e., a velocity pattern. Explanations for this trend
include (1) a rotation of the cluster, (2) a shearing pattern with
respect to an axis, and (3) a correlation between the trigonometric
parallaxes $\pi_{\rm Hip}$ and their associated errors $\sigma_{\pi,
{\rm Hip}}$ (cf.\ \S 7.2 in P98).
\placefigureEight
\null\vskip-1.2truecm
\paragraph{(1) Rotation:} Given the observed velocity field,
we determine the Galactic coordinates of the best-fitting rotation
axis ($\ell_{\rm rot}$ and $b_{\rm rot}$) and the corresponding
rotation period ($P_{\rm rot}$) by minimizing the dispersion of the
velocity residuals with respect to the rotation axis, after adding a
rotation pattern to the mean space motion. This results in the
estimates $\ell_{\rm rot} \sim 131\fdg5$, $b_{\rm rot} \sim +60\fdg0$,
and $P_{\rm rot} = 68.0$~Myr (i.e., $\sim$0.10~km~s$^{-1}$~pc$^{-1}$).
\placefigureNine
\null\vskip-1.2truecm
\paragraph{(2) Shear:} A shear pattern with respect to an axis
pointing towards $\ell_{\rm shear}$ and $b_{\rm shear}$ is described
by a constant $P_{\rm shear}$ which expresses the strength of the
shear. A least-squares fit returns $\ell_{\rm shear} \sim 131\fdg5$,
$b_{\rm shear} \sim +60\fdg0$, and $P_{\rm shear} \sim
0.13$~km~s$^{-1}$~pc$^{-1}$ (as our proper motion, parallax, and
radial velocity data do not have enough discriminating power to reveal
the subtle differences between a rotation and shear pattern, our fit
returns a shear axis which is identical to the rotation axis listed
above).
\null\vskip-1.2truecm
\paragraph{(3) Correlated trigonometric parallaxes and errors:} The
lower series of panels in Figure~\ref{fig:vel_sys} show a velocity
field decomposition for a realistic Monte Carlo realization of the
Hyades (500 stars, 10~pc radius, including Hipparcos data covariances)
in which the stars share a common space motion exclusively. Despite
the absence of intrinsic velocity structure, the `front' and `back'
distributions do show a systematic pattern which resembles the
observed distribution (upper series of panels) remarkably well. P98
(their \S 7.2) did already argue that correlated velocity residuals
are a natural result of the presence of a correlation between the
Hipparcos parallaxes $\pi_{\rm Hip}$ and the corresponding
observational errors $\sigma_{\pi, {\rm Hip}}$ in a sample of Hyades
members (left panel of Figure~\ref{fig:pearson_r_etc}; we find a
correlation coefficient $\rho = +0.56$ between $\pi_{\rm Hip} -
\pi_{\rm sec, Hip} \approx \pi_{\rm Hip} - \pi_{\rm true} \equiv
\Delta_{\pi, {\rm Hip}}$ and $\pi_{\rm Hip}$). Although the individual
Hipparcos trigonometric parallaxes are not correlated with their
associated observational errors, the selection of a set of stars with
(nearly) equal true parallaxes, such as the members of an open
cluster, induces the presence of a correlation in the sample: Hyades
with large observed parallaxes are, in general, more likely to have
$\Delta_{\pi, {\rm Hip}} \equiv \pi_{\rm Hip} - \pi_{\rm true} > 0$
than $\sigma_{\pi, {\rm Hip}} < 0$ (and vice versa for Hyades with
small observed parallaxes). The strength of this correlation between
the (sign of the) parallax error and the observed parallax depends on
the intrinsic size of the cluster: a small cluster gives a small
spread in true parallaxes, which implies a large correlation. The
right panel of Figure~\ref{fig:pearson_r_etc} shows the mean
correlation coefficient derived from Monte Carlo realizations of a
Hyades-like cluster as function of the cluster radius $R$. The
observed correlation coefficient $\rho = +0.56$ implies $R \sim
10$--$15~{\rm pc} \sim 1.0$--$1.5~r_{\rm t}$, which is a very
reasonable definition for the size of the Hyades cluster.
\null\vskip-1.2truecm
\paragraph{Discussion:} The analysis presented above shows that the
systematic velocity pattern displayed in Figure~\ref{fig:vel_sys} can
be due to rotation, shear, and/or a correlation between $\pi_{\rm
Hip}$ and $\sigma_{\pi, {\rm Hip}}$. Both rotation and shear provide
an equally good representation of the observations, but imply a
significant systematic velocity of $\sim$1~km~s$^{-1}$ at the tidal
radius of the cluster ($r_{\rm t} \sim 10$~pc). Unmodelled systematic
velocities at the level of 1--2~km~s$^{-1}$ in the outer regions of
the cluster ($r_{\rm t} \la r \la 2 r_{\rm t}$) would lead to
systematic secular parallax errors as large as 0.9--1.8~mas. These
values, however, are a factor 3--6 larger than the observed upper
limit of 0.3~mas at $r \sim 20$~pc (Figure~\ref{fig:prlxs_and_gs}; \S
\ref{subsec:3d_pos}), which argues against an explanation of the
velocity pattern in terms of rotation or shear. There is, moreover,
also a direct argument in favour of the apparent velocity pattern not
being caused by rotation or shear, but by the correlation between the
observed parallaxes and the parallax errors instead, for this should
result in an apparent rotation or shear axis pointing towards $(b_u,
b_v, b_w) \times (v_u, v_v, v_w)$, i.e., $(\ell, b) = (116^\circ\!,
+48^\circ\!)$ ($[b_u, b_v, b_w]$ and $[v_u, v_v, v_w]$ denote,
respectively, the position and velocity vector of the cluster center
expressed in Galactic Cartesian coordinates). This axis coincides
within $15^\circ\!$ with the rotation and shear axes found above. We
therefore conclude that the observed correlation between the Hipparcos
trigonometric parallaxes and their associated random errors
(Figure~\ref{fig:pearson_r_etc}) is mainly responsible for the
(apparent) velocity structure of the Hyades (Figure~\ref{fig:vel_sys};
cf.\ P98).

\subsubsection{Secular parallaxes}
\label{subsubsec:vel_field_sec}

Studying the Hyades velocity field using secular parallaxes, which
were derived under the {\it assumption\/} of a specific velocity
field, is of limited scientific merit. We therefore restrict such an
analysis to the straightforward comparison of the input and output
velocity fields (bottom row of Figure~\ref{fig:vel_field}), which turn
out to be fully consistent. A systematic pattern as observed in the
trigonometric parallax velocity field (\S
\ref{subsubsec:vel_field_Hip}) is absent in the secular parallax
velocity field (not shown).

\subsection{Summary}
\label{subsec:summary}

Monte Carlo tests combined with the uncertainty of the tangential
component of the cluster space motion set the maximum expected
systematic Hipparcos secular parallax error at $\sim$0.30~mas (\S
\ref{subsec:acc_1}). This value is consistent with the facts that (1)
the secular parallaxes as a set are statistically consistent with the
Hipparcos trigonometric parallaxes within $\la$0.10~mas
(Figure~\ref{fig:prlxs}; \S \ref{sec:sec_pars}), and (2) secular
parallaxes for stars in the inner and outer regions of the Hyades do
not differ significantly, i.e., at the $\sim$0.30~mas level or larger
(Figure~\ref{fig:prlxs_and_gs}; \S \ref{subsec:3d_pos}). We conclude
that secular parallaxes for Hyades within at least $r \la 2 r_{\rm t}
\sim 20$~pc of the cluster center can be regarded as absolute, i.e.,
having systematic errors smaller than $\sim$0.30~mas.

The Hipparcos trigonometric parallax errors are correlated on angular
scales of a few degrees with `amplitudes' smaller than
$\sim$$0.75$--$1.00$~mas per star (\S \ref{subsec:acc_2}). The mean
trigonometric parallax of the Hyades, however, is accurate to within
$\la$0.10~mas, as regions with positive and negative contributions
cancel when averaging parallaxes over the large angular extent of the
cluster.

\placefigureTen

The observed lack of significant systematics in the secular parallaxes
puts an upper limit on the size of possible velocity patterns
(rotation or shear) of a few hundredths of a km~s$^{-1}$~pc$^{-1}$ (\S
\ref{subsubsec:vel_field_Hip}). This upper limit, in its turn,
strongly suggests that the observed systematics in the trigonometric
parallax-based velocity field (Figure~\ref{fig:vel_sys}) are due to
the presence of a correlation between the Hipparcos parallaxes and
their associated random errors in our sample of Hyades.

\section{Spatial structure}
\label{sec:spatial_structure}

At the mean distance of the Hyades, a parallax uncertainty of
$\sigma_\pi$ (mas) corresponds to a distance error of $\sigma_\pi D^2
/ 1000 \sim 2 \sigma_\pi$~pc ($D \sim 45$~pc). Typical Hipparcos
parallax errors are 1.0--1.5~mas, thus yielding a $\sim$2--3~pc
distance resolution. Typical Hipparcos secular parallaxes are $\sim$3
times more accurate than the trigonometric values (\S
\ref{subsec:accuracy}). However, because the Hipparcos resolution is
already sufficient to resolve the internal structure of the Hyades
(with core and tidal radii of 2.7 and 10~pc, respectively; \S
\ref{subsec:mem_P98}), secular parallaxes cannot fundamentally improve
upon the P98 results regarding, e.g., the three-dimensional spatial
distribution of stars in the cluster, including the shape of the core
and corona and flattening of the halo, `the Hyades distance'\footnote{
The statistical consistency between the Hipparcos trigonometric and
secular parallaxes as a set (e.g., Figure~\ref{fig:prlxs}) implies
that the (mean) Hyades distance derived by P98 cannot be improved
upon.
}, the density and mass distribution of stars in the cluster, its
gravitational potential, moments of inertia, etc.\ (\S\S 7--8 in P98;
we investigated all aforementioned examples using secular parallaxes,
but were unable to obtain results which had not already been derived
by P98). Figure~\ref{fig:spatial_distribution}, for example, shows the
three-dimensional distribution of the 218 P98 members. Although the
internal spatial structure of the Hyades is resolved by the Hipparcos
trigonometric parallaxes, the Hipparcos secular parallaxes do provide
a sharper view.

\section{Colour-absolute magnitude diagram}
\label{sec:cmd}

The colour-absolute magnitude and HR diagrams of the Hyades cluster
have been studied extensively, mainly owing to the small distance of
the cluster. Among the advantages of this proximity are the negligible
interstellar reddening and extinction (e.g., Crawford 1975; Taylor
1980; $E(B-V) = 0.003 \pm 0.002$~mag) and the possibility to probe the
cluster (main sequence) down to low masses relatively easily. As
mentioned in \S \ref{sec:intro}, the significant cluster depth along
the line of sight has always complicated pre-Hipparcos stellar
evolutionary modelling (cf.\ \S 9.0 in P98). Unfortunately, even
Hipparcos parallax uncertainties (typically 1.0--1.5~mas) translate
into absolute magnitude errors of $\ga$0.10~mag at the mean distance
of the cluster ($D \sim 45$~pc), whereas $V$-band photometric errors
only account for $\la$0.01~mag uncertainties for most members. The
Hipparcos secular parallaxes derived in \S \ref{sec:sec_pars} are on
average a factor $\sim$3 times more precise than the Hipparcos
trigonometric values (i.e., $\sigma_{\pi, {\rm sec, Hip}} \la 0.5~{\rm
mas} \sim 0.05~{\rm mag}$; \S \ref{subsec:accuracy}). The maximum
expected systematic error in the secular parallaxes is $\la$$0.3~{\rm
mas}$ or $\la$$0.03~{\rm mag}$ (\S \ref{subsec:summary}). Secular
parallaxes therefore allow the construction of a well-defined and
well-calibrated Hyades colour-absolute magnitude (and HR) diagram.

\placefigureEleven

Figure~\ref{fig:HRD_1} shows colour-absolute magnitude diagrams of the
Hyades based on Hipparcos trigonometric (left), Hipparcos secular
(middle), and Tycho--2 secular parallaxes (right). The Hipparcos
secular parallax diagram shows a narrow main sequence consisting of
kinematic members ($g_{\rm Hip} \leq 9$; filled symbols; cf.\
Figure~\ref{fig:HRD_4}). Kinematically deviant stars ($g_{\rm Hip} >
9$; open symbols) are likely either non-members and/or close multiple
stars (\S \ref{subsec:faint_mems}). Most of the 15 new Hipparcos
candidates (open triangular symbols; \S \ref{subsec:addi_mems}) do not
follow the main sequence. This is not surprising, as secular
parallaxes for most of these stars are inconsistent with their
trigonometric parallaxes, suggestive of non-membership. The three
candidates with $g_{\rm Hip} \leq 9$ (filled triangles) identified in
\S \ref{subsec:Hip_additional_members} are labeled. Only HIP 19757
lies close to the main sequence and is a likely new member (cf.\
Table~\ref{tab:additional_members}).

The right panel of Figure~\ref{fig:HRD_1} shows, besides a narrow
cluster main sequence consisting of kinematic members ($g_{\rm
Tycho-2} \leq 9$; filled circles), a well-defined binary sequence for
$0.45 \mathrel{{\hbox to 0pt{\lower 3pt\hbox{$\sim$}\hss}} \raise
2.0pt\hbox{$>$}} (B-V) \mathrel{{\hbox to 0pt{\lower
3pt\hbox{$\sim$}\hss}} \raise 2.0pt\hbox{$>$}} 0.70$~mag. Most of the
photometrically deviant stars are low-probability kinematic members
($g_{\rm Tycho-2} > 9$; open circles), most likely indicating
non-membership. The 23 photometric BDA members (\S
\ref{subsec:Tycho2}) are indicated by triangles (filled for $g_{\rm
Tycho-2} \leq 9$; open for $g_{\rm Tycho-2} > 9$). About half of them
do not follow the main sequence. Most of these objects, nonetheless,
seem secure kinematic members ($g_{\rm Tycho-2} \leq 9$; cf.\ \S
\ref{subsec:Tycho2}). These stars are possibly interlopers but most
likely they are members with inaccurate $(B-V)$ photometry: objects
lacking accurate ground-based photometric observations, most likely as
a result of being pre-Hipparcos non-members, generally have Hipparcos
$(B-V)$ values derived from Tycho photometry (Hipparcos field H39 =
`T'). Corresponding $(B-V)$ errors can reach several tenths of a
magnitude for stars fainter than $V \sim 8.5$~mag ($M_V \sim
5.2$~mag). Hipparcos $(B-V)$ values for faint pre-Hipparcos members
contained in the Hipparcos Catalogue, on the other hand, are often
carefully selected accurate ground-based measurements (field H39 =
`G'); this explains the presence of a well-defined main sequence down
to faint magnitudes ($V \sim 11$--$12$~mag).

Figure~\ref{fig:HRD_2} compares colour-absolute magnitude diagrams for
different regions within the cluster (core, corona, halo, and moving
group; \S \ref{subsec:mem_P98}). Secular parallaxes clearly improve
the definition of the main sequence as compared to Hipparcos
trigonometric parallaxes in the central parts of the cluster. They
also significantly narrow the main sequence for stars in the halo ($10
\leq r < 20$~pc). The relatively large spread in the Hipparcos secular
parallax panel for $20 \leq r < 40$~pc is probably due to uncertain
membership assignment (cf.\ \S \ref{subsubsec:vel_field_Hip_P98}),
combined with inaccurate photometry, stellar multiplicity, and/or
suspect secular parallaxes. The latter uncertainty is possibly related
to unmodelled low-amplitude velocity patterns in the very outer parts
of the cluster ($r \ga 2 r_{\rm t}$; but see \S\S
\ref{subsec:vel_field}--\ref{subsec:summary}).

\section{Constructing the Hertzsprung--Russell diagram}
\label{sec:HRD_construction}

The secular parallaxes derived in \S \ref{sec:sec_pars} constrain the
locations of stars in the colour-absolute magnitude diagram with
unprecedented precision (\S \ref{sec:cmd}). An interpretation of these
high-quality observations in terms of stellar evolutionary models with
appropriately chosen input physics (\S \ref{subsec:ThStEvMo}) can
provide a wealth of information on the fundamental properties of the
Hyades cluster itself, such as its age and metallicity (\S
\ref{subsec:Hyades_chars}), as well as on the characteristics of stars
and stellar evolution in general (\S\S
\ref{subsec:HRD_stars}--\ref{subsec:calibration}).

\subsection{Theoretical stellar evolutionary models}
\label{subsec:ThStEvMo}

Stellar evolutionary models have been highly successful in explaining
the structure and evolution of stars (e.g., Cox \& Giuli 1968;
Kippenhahn \& Weigert 1990). Numerical stellar evolutionary codes
suffer from daunting practical problems, owing to, e.g., the large
dynamical range for the various quantities of interest such as
temperature, pressure, and density (e.g., Schwarzschild 1958; Henyey
et al.\ 1959). Moreover, they suffer from major uncertainties in the
appropriate input physics (e.g., Lebreton et al.\ 1995, 2000; Kurucz
2000). The most important of these uncertainties are, for Hyades main
sequence stars, related to:
(1) stellar atmosphere models, including issues related to atomic and
    molecular opacities and internal structure--external boundary
    conditions (\S \ref{subsec:HRD_1}); and
(2) the treatment of turbulent convection\footnote{
Corresponding physical theories do not exist; two numerical
prescriptions are in wide-spread use: the Mixing-Length Theory (MLT;
e.g., B\"ohm--Vitense 1953, 1958) and the Full Spectrum of Turbulent
eddies convection model (FST; e.g., Canuto \& Mazzitelli 1991, 1992;
Canuto et al.\ 1996).
}, including issues related to convective core and envelope overshoot.
Besides these two major uncertainties, numerous other physical
phenomena are generally either not or only partially taken into
account: (transport processes due to) stellar rotation (\S
\ref{subsec:HRD_3}), variability and/or pulsational instability (\S
\ref{subsec:HRD_3}), chromospheric activity (\S \ref{subsec:HRD_1}),
mass loss (\S \ref{subsec:HRD_4}), binary evolution, the evolution
towards the zero-age main sequence, etc. Moreover, stellar
evolutionary models are generally calibrated by using the Sun as
benchmark. Although the phenomenological treatment of convection in
models is considered appropriate for Sun-like stars, this is not
necessarily the case for stars with other mass, metallicity, and/or
stellar evolutionary status. Related to the latter issue is the
important open question to what extent the characteristics of
convection vary with location in a convection zone. A detailed study
of the secular parallax-based locations of Hyades in the
colour-absolute magnitude and HR diagrams of the cluster has the
potential to shed light on some of the abovementioned issues (e.g.,
Lebreton 2000).

\subsection{Helium content, metallicity, and age of the Hyades}
\label{subsec:Hyades_chars}

\paragraph{CESAM:} P98 used the CESAM evolutionary code (Morel 1997)
to interpret the Hipparcos HR diagram of the Hyades (their \S 9). In
order to derive the cluster Hydrogen, Helium, and metal abundances by
mass ($X \equiv 1 - Y - Z$, $Y$, and $Z$, respectively), P98 fitted
model zero-age main sequences to the observed trigonometric
parallax-based zero-age main sequence positions of 19 single low-mass
members\footnote{
CESAM employs Solar-calibrated mixing-length theory
($\alpha_{\rm MLT} = 1.64$) and convective core overshoot. P98 use
$(X, Y,$ $Z)_\odot = (0.7143, 0.2659, 0.0175)$.
}. By treating $Y$ and $Z$ as free parameters with the boundary
condition that the inferred metal content of the Hyades be
consistent\footnote{
For a Solar mixture of heavy elements, metallicity $Z$ and
Iron-to-Hydrogen ratio $[{\rm Fe/H}]$ (relative to Solar) are related
by: $[{\rm Fe/H}] = \log(Z/X) - \log(Z/X)_\odot$. The quantity
$(Z/X)_\odot$ is usually taken from Grevesse \& Noels (1993a, b).
} with the mean spectroscopically determined metallicity $[{\rm Fe/H}]
= +0.14 \pm 0.05$, P98 found $(X, Y, Z) = (0.716, 0.260 \pm 0.020,
0.024 \pm 0.003)$. The P98 $[{\rm Fe/H}]$ value is a
well-established\footnote{
Some Hyades have quite deviant metallicities. E.g., the
chromospherically active spectroscopic binary HIP 20577 has $[{\rm
Fe/H}] = 0.00 \pm 0.03$ (Cayrel et al.\ 1985; Smith \& Ruck 1997).
} quantity (cf.\ $[{\rm Fe/H}] \ga +0.12 \pm 0.03$ Cayrel et al.\
1985; $+0.13 \pm 0.02$ Boesgaard 1989 and Boesgaard \& Friel
1990). Unfortunately, the uncertainty on the derived Helium content of
the cluster, combined with the existing uncertainty on the Helium
content of the Sun (e.g., Brun et al.\ 1998), still prevents a
definitive answer to the question whether the Helium content of the
Hyades is sub-Solar or not (e.g., Str\"om\-gren et al.\ 1982; Hardorp
1982; Dobson 1990; Swenson et al.\ 1994; Pinsonneault et al.\ 1998).

After establishing the chemical composition of the Hyades, P98 derived
a nuclear age $\tau = 625 \pm 50$--$100$~Myr by fitting CESAM
isochrones to the upper main sequence of the trigonometric
parallax-based colour-absolute magnitude diagram (their figures 21--23
and \S 9.2; cf.\ $\tau = 600 \pm 50$~Myr from Torres et al.\ 1997a,
c).
\null\vskip-0.80truecm\null
\paragraph{Padova:} For an interpretation of the HR diagram of
the Hyades, the latest Padova iso\-chrones\footnote{
The Padova code uses Solar-calibrated mixing-length theory
($\alpha_{\rm MLT} = 1.68$) and stellar-mass dependent convective core
overshoot. Girardi found $(X, Y, Z)_\odot = (0.708, 0.273, 0.019)$.
} (Girardi et al.\ 2000a) offer the advantage that they include (post)
red giant branch evolution.

\placefigureTwelve

The six $(Y, Z)$ pairs discussed by Girardi et al.\ were not randomly
chosen but follow a fixed $Y(Z)$ relation (Figure~\ref{fig:YZ}), which
is inspired by the understanding of the origin of Helium and metals in
the universe: $Y \sim Y_{\rm p} + (\Delta Y/ \Delta Z) \cdot Z$, where
$Y_{\rm p} = 0.23$ is the primordial Helium abundance, and $\Delta Y/
\Delta Z = 2.25$ is the stellar evolution Helium-to-metal enrichment
ratio (e.g., Faulkner 1967; Pagel \& Portinari 1998; Lebreton et al.\
1999). The $Y(Z)$ relation implies we cannot obtain Padova isochrones
which are consistent with both the Helium content and metallicity of
the Hyades as derived by P98. Taking the latter fixed at $Z = 0.024$,
and interpolating between the sets $(Y, Z) = (0.273, 0.019)$ and
$(0.300, 0.030)$, provides isochrones with $Y = 0.285$ (open square in
Figure~\ref{fig:YZ}). Although this value is inconsistent at the
1.25$\sigma$ level with P98's value, it cannot be considered
inappropriate for the Hyades (VandenBerg \& Bridges 1984).

As the interpolation between published isochrones is practically
impossible after crossing the Hertzsprung gap, we use Girardi's
Solar-metallicity $(Y, Z, \tau~[{\rm Myr}]) = (0.273, 0.019, 631)$
isochrone for a comparison with the Hyades giants (\S
\ref{subsec:HRD_4}).

\subsection{A high-fidelity stellar sample}
\label{subsec:HRD_stars}

\placefigureThirteen

Following P98, we restrict attention to a high-fidelity subset of
members for the study of the HR diagram. We do not consider suspect
kinematic members and stars which have deviant HR diagram positions
for known reasons. We exclude the 16 stars beyond 40~pc from the
cluster center and all (possible) close multiple systems (98
spectroscopic binaries, Hipparcos DMSA--`G$\,$O$\,$V$\,$X$\,$S' stars,
and stars with $g_{\rm Hip} > 9$). We furthermore reject 11 stars
which are variable (Hipparcos field H52 is one of
`D$\,$M$\,$P$\,$R$\,$U') or have large photometric errors
($\sigma_{(B-V)} > 0.05$~mag), as well as the suspect objects HIP
20901, 21670, and 20614 (\S 9.2 in P98; cf.\ Wielen et al.\ 2000).

The final sample contains 90 single members. These stars follow the
main sequence (cluster isochrone) from $(B-V) \sim 1.43$~mag
(late-K/early-M dwarfs; $M \ga 0.5~M_\odot$) to $(B-V) \sim 0.10$~mag
(A7IV stars; $M \sim 2.4~M_\odot$). Two of the stars are evolved red
giants ($\epsilon$ and $\gamma$ Tau; \S \ref{subsec:HRD_4} and
\ref{subsec:trans_giants}). The two components of the `resolved
spectroscopic binary' $\theta^2$~Tau, located in the turn-off region
of the cluster (\S \ref{subsec:HRD_3} and \ref{subsec:trans_theta2};
cf.\ P98), contribute significant resolving power for distinguishing
between different evolutionary models as well as between different
isochrones from one evolutionary code. We therefore add them as single
stars to our sample, bringing the total number~of objects to~92.

Figure~\ref{fig:HRD_4} shows, for these 92 stars, the colour versus
secular parallax-based absolute magnitude diagram (cf.\
Figures~\ref{fig:HRD_1}--\ref{fig:HRD_2}). It shows, besides a
well-defined and very narrow main sequence, turn-off region, and giant
clump, substructure in the form of two `gaps'/`turn-offs' in the main
sequence around $(B-V) \sim 0.30$ and $\sim$$0.45$~mag (cf.\ de
Bruijne et al.\ 2000). These features are also present in the Tycho--2
secular parallax-based diagram (right panel of
Figure~\ref{fig:HRD_1}), but are not clearly discernible in the lower
quality trigonometric parallax-based version (left panel of
Figure~\ref{fig:HRD_1}). In \S \ref{subsec:HRD_2}, we will identify
these `turn-offs' with so-called B\"ohm--Vitense gaps, which are most
likely related to convective atmospheres. Although the reality
of the turn-offs in the `cleaned' secular parallax colour-magnitude
diagram is hard to establish beyond all doubt, the simultaneous
existence of both a turn-off and an associated gap at a location which
coincides with predictions made by stellar structure models (see \S
\ref{subsec:HRD_2}) strongly argues in favour of them being real (cf.\
Kjeldsen \& Frandsen 1991, and references therein).

\subsection{$(B-V)$--$M_V$ $\longrightarrow$ $\log T_{\rm eff}$--$\log(L/L_\odot)$}
\label{subsec:calibration}

In order to compare the locations of Hyades in the HR diagram to
theoretical isochrones, we need to transform the observables $(B-V)$
and $M_V$ to the theoretical quantities $T_{\rm eff}$ and luminosity
$L$. The usual procedure is to derive $T_{\rm eff}$ from $(B-V)$ and
then to compute the bolometric correction in the $V$-passband, ${\rm
BC}_V$, from $T_{\rm eff}$; $\log(L/L_\odot)$ then follows from
$M_{\rm bol} - M_{\rm bol, \odot}$, where $M_{\rm bol, \odot} =
4.74$~mag (Bessell et al.\ 1998; cf.\ footnote~16; the IAU value for
the solar bolometric magnitude is 4.75~mag). Both transformations
depend on metallicity $[{\rm Fe/H}]$ and on surface gravity $\log g$,
i.e., stellar evolutionary status.

\subsubsection{Previous work}
\label{subsubsec:trans_previous}

\placefigureFourteen

Numerous empirical and theoretical calibrations have been proposed in
the past, each of which has its own validity in terms of $\log g$,
$T_{\rm eff}$, $(B-V)$, and/or $[{\rm Fe/H}]$ (e.g., Flower 1977,
1996; Buser \& Kurucz 1992; Gratton et al.\ 1996). There is a large
uncertainty in and systematic disagreement between the different
$(B-V)$--$T_{\rm eff}$ relations. This is partly caused by the
uncertain Solar photospheric abundances and ill-defined $(B-V)$ colour
of the Sun (e.g., appendix~C in Bessell et al.\ 1998), but also partly
by the specific choice of the $(B-V)$ index. This colour is
particularly sensitive to opacity problems, mainly related to metal
lines and molecular electronic transitions in the UV-blue-optical,
especially for cool stars ($T_{\rm eff} \la 4500$~K; e.g., Lejeune et
al.\ 1998; Bessell et al.\ 1998). The effects of model uncertainties,
related to opacities and the treatment of convection (\S
\ref{subsec:ThStEvMo}), on theoretical calibrations are often
non-negligible, especially for K/M giants and low-mass dwarfs (e.g.,
Blackwell et al.\ 1991; cf.\ table 4 in Houdashelt et al.\ 2000). The
main reason for our ignorance is the lack of a representative set of
stars with (model-)\-independently determined effective temperatures
(e.g., Code et al.\ 1976; Ridgway et al.\ 1980; Blackwell \&
Lynas--Gray 1994). Differences between effective temperature scales,
established by empirical or theoretical calibrations, are generally
smaller than $\sim$200--400~K (Castelli 1999; Gardiner et al.\ 1999).

We consider two calibrations: (1) Bessell et al.\ (1998) in
combination with Alonso et al.\ (1996; \S
\ref{subsubsec:trans_cali1}), roughly following P98; and (2) Lejeune
et al.\ (1998; \S \ref{subsubsec:trans_cali2}).

\subsubsection{Calibration (1)}
\label{subsubsec:trans_cali1}

\placefigureFifteen

Bessell et al.\ (1998) present broad-band colour, bolometric
correction, and effective temperature calibrations for O to M stars,
based on Kurucz's (1995) ATLAS9 model atmospheres. Their
tables\footnote{
Tables 4 and 5 are valid for giants, which are treated separately in
\S \ref{subsec:HRD_4}. Table 6 is based on NMARCS M-dwarf
models. Although this calibration is preferred over the ATLAS9-based
relation for stars with $T_{\rm eff} \la 4000$~K, the modeled range in
$(B-V)$ is incompatible with the colours for dwarfs in our
sample. Moreover, the NMARCS relations do not link smoothly to the
ATLAS9 results. We therefore decided not to use them.
} 1 and 3 relate $T_{\rm eff}$, $\log g$, $(B-V)$, and ${\rm BC}_V$,
based on Solar-metallicity models with convective core
overshoot\footnote{
This choice is consistent with the results derived by P98 (their
\S 9.2), although we note that Castelli et al.\ (1997) found that
effective temperatures measured by means of the infra-red flux method
(e.g., Blackwell et al.\ 1990) for stars with $T_{\rm eff} \ga T_{{\rm
eff}, \odot} \sim 5765$~K are in general better reproduced by
theoretical ATLAS9 atmosphere models (Kurucz 1995) with the envelope
overshoot option switched {\it off\/} than with the option switched
{\it on} (table 2 in Bessell et al.\ 1998).
} ($\alpha_{\rm MLT} = 1.25$).

\placefigureSixteen

We use an iterative scheme to determine $\log T_{\rm eff}$,
$\log(L/L_\odot)$, and $\log g$ from the measured $(B-V)$ and $M_V$
values, in which we correct for the non-Solar metallicity of the
Hyades according to Alonso et al.\ (1996; their eq.~1). Step 1
involves the determination of $\log T_{\rm eff}$ for $[{\rm Fe/H}] =
0$, from a given $\log g$ and measured $(B-V)$. After differentially
applying Alonso's metallicity correction for $[{\rm Fe/H}] = +0.14$ in
step 2, the corrected $\log T_{\rm eff}$ provides ${\rm BC}_V$, and
thus $\log(L/L_\odot)$. Step 3 involves the determination of $\log g =
\log (4 \pi G M \sigma_{\rm Boltzmann} T^{4}_{\rm eff} L^{-1})$ using
stellar masses $M$ from P98 (\S \ref{subsubsec:stellar_masses}). This
recipe is repeated (typically 3 times) until convergence is achieved
in the sense that $\log g$ remains constant; the final results do not
depend on the initial estimate $\log(g [{\rm cm~s}^{-2}]) = 4.5$.

First-order error analysis allows an estimation of the uncertainties
on the derived quantities. We assume that $\sigma_{\log(L/L_\odot)}$
is influenced by both $\sigma_{M_V}$ and $\sigma_{{\rm BC}_V}$, which
in their turn are influenced by $\sigma_{V}$ plus $\sigma_{\pi}$ and
$\sigma_{\log T_{\rm eff}}$, respectively. We assume that
$\sigma_{\log T_{\rm eff}}$ is influenced by both $\sigma_{(B-V)} =
\min(\sigma_{(B-V), {\rm observed}}, 0.010~{\rm mag})$ and $\sigma_M =
0.10~M_\odot$ (P98); we neglect the contribution of $\sigma_{\log g}$
to $\sigma_{\log T_{\rm eff}}$ as it is typically an order of
magnitude smaller than the other contributors (cf.\ Castelli et al.\
1997).

\subsubsection{Calibration (2)}
\label{subsubsec:trans_cali2}

Lejeune et al.\ (1998) present (semi-)empirical calibrations linking
$T_{\rm eff}$, $(B-V)$, $\log g$, and ${\rm BC}_V$ for $[{\rm Fe/H}]$
between $-3.5$ and $+1.0$ based on BaSeL spectral energy
distributions\footnote{
Lejeune uses $M_{V, \odot} = 4.854$~mag and ${\rm BC}_{V, \odot} =
-0.108$~mag; we have transformed Lejeune's data to conform with
Bessell's zero point (cf.\ appendices C--D in Bessell et al.\ 1998).
} (Basel Stellar Library version 2.0; their tables 1--10). In the
range of stellar parameters considered here, these calibrations use
Kurucz (1995) ATLAS9 model atmospheres. The presentation of the data,
which is relevant for dwarfs ($\log g \ga 3.75$), allows a direct
determination of $\log T_{\rm eff}$, $\log(L/L_\odot)$, and $\log g$
from the measured $(B-V)$ and $M_V$ for $[{\rm Fe/H}] = +0.14$ by
means of interpolation between table 1 ($[{\rm Fe/H}] = 0$) and table
9 ($[{\rm Fe/H}] = +0.50$), without relying on stellar mass
information. We derive $\sigma_{\log(L/L_\odot)}$ and $\sigma_{\log
T_{\rm eff}}$ as in \S
\ref{subsubsec:trans_cali1}.

\subsubsection{Results for dwarfs}
\label{subsubsec:trans_results}

We applied calibrations (1) and (2) to the set of 92 members described
in \S \ref{subsec:HRD_stars}, excluding the giants $\epsilon$ and
$\gamma$ Tau (Appendix \ref{subsec:trans_giants}), using absolute
magnitudes $M_V$ based on Hipparcos secular parallaxes (\S
\ref{sec:sec_pars}). Details for the spectroscopic binary $\theta^2$
Tau are given in Appendix \ref{subsec:trans_theta2}

For a given calibration, the bolometric corrections are relatively
well defined, except at lower masses (i.e., redder $(B-V)$, lower
$T_{\rm eff}$; Figures~\ref{fig:Teff_rev}--\ref{fig:B+L}). Effective
temperatures, on the other hand, are quite uncertain, especially at
higher masses. Moreover, a comparison of the effective temperatures
derived from calibrations (1) and (2) reveals significant systematic
differences (at the level of $\sim$100~K; Figure~\ref{fig:Teff_rev})
as a function of $T_{\rm eff}$ itself. Taking the non-Solar
metallicity of the Hyades into account significantly changes the
derived parameters, and notably the effective temperatures
(Figure~\ref{fig:B+L}).

In order to establish which calibration is to be preferred, we
compared the effective temperatures following from Bessell's and
Lejeune's relations (for $[{\rm Fe/H}] = 0$ as well as $[{\rm Fe/H}] =
+0.14$) with a number of previously established effective temperature
scales for the Hyades (Bessell et al.\ [1998; with and without
overshoot; see footnote~15] plus Alonso et al.\ [1996; $[{\rm Fe/H}] =
0$ and $+0.14$]; Lejeune et al.\ [1998; $[{\rm Fe/H}] = 0$ and
$+0.14$]; Allende Prieto \& Lambert [1999; table 1]; Varenne \& Monier
[1999; table 2]; P98 [table 8]; and Balachandran [1995; table 4]).
This analysis reveals effective temperature differences, which
sometimes vary systematically with effective temperature itself (cf.\
Figure~\ref{fig:Teff_rev}), up to $\sim$300~K. Unfortunately, none of
these scales is truly fundamental in the sense of having been
established completely model-independently. In fact, agreement between
different calibrations might even be artificial to some degree, as
several of them have ultimately been calibrated using Kurucz
atmosphere models (e.g., Gardiner et al.\ 1999). We therefore decided
to enforce consistency between the calibration used here and the
spectroscopic effective temperatures and CESAM isochrones provided by
P98.

Figure~\ref{fig:HRD_5} compares the $\log T_{\rm
eff}$--$\log(L/L_\odot)$ and $\log$ $T_{\rm eff}$--$\log g$ diagrams
after applying calibration (1) and calibration (2) to all objects in
the sample. Calibration (2) has the problem that evolved stars in the
turn-off region (i.e., luminosity classes IV--V) are given too large
surface gravities because the procedure assumes that all stars are
dwarfs (i.e., luminosity class V). Calibration (1) has the `problem'
that stars on the main sequence fall significantly below the 625~Myr
CESAM isochrone, whereas stars in the turn-off region of the cluster
follow this curve acceptably well. These facts suggest the use of
calibration (2) for dwarfs\footnote{
We find $T_{\rm eff, \odot} = 5793$~K, $\log(L/L_\odot) = 0.004$, and
$\log g = 4.501$ for $(B-V)_\odot = 0.628$~mag (Taylor 1998).
} and calibration (1) for the 14 stars with $(B-V) \leq 0.300$~mag
($T_{\rm eff} \geq 7250$~K; $\log(T_{\rm eff} [{\rm K}]) \geq
3.8603$). We acknowledge that this approach is `ad hoc', as it naively
assumes the CESAM isochrones are correct. A full understanding of the
discrepancies shown in Figure~\ref{fig:HRD_5} requires a set of new
isochrones {\it and\/} calibrations which are tailored to the Hyades
in terms of metallicity and Helium content. The construction of such
isochrones and calibrations, using the secular parallax-based
$(B-V)$--$M_V$ diagram presented in Figure~\ref{fig:HRD_4} as boundary
condition, is beyond the scope of this paper (cf.\ \S \ref{sec:disc}).

\subsubsection{Stellar masses}
\label{subsubsec:stellar_masses}

Figure~\ref{fig:mass} compares the masses (\S 5.3 in P98) of the 92
high-fidelity members and the Hipparcos secular parallax-based
absolute magnitudes with an empirical mass-absolu\-te magnitude
relation derived for the Hyades (eq.~A3 in Patience et al.\ 1998). The
empirical relation deviates significantly from the P98 masses for
low-mass dwarfs ($M \la 1~M_\odot$; cf.\ \S \ref{subsec:HRD_1}). The
CESAM and Padova isochrone results (\S \ref{subsec:Hyades_chars}), on
the other hand, give an acceptable `fit'. This confirms that the P98
masses, which are used in calibration (1) to derive $T_{\rm eff}$ and
$\log(L/L_\odot)$ from $(B-V)$ and $M_V$ (\S
\ref{subsubsec:trans_cali1}), are well-defined.

\placefigureSeventeen

\section{Hertzsprung--Russell diagram}
\label{sec:HRD}

Figure~\ref{fig:HRD_3} shows 92 high-fidelity Hyades (\S
\ref{subsec:HRD_stars}) with Pa\-do\-va and CESAM isochrones (\S
\ref{subsec:Hyades_chars}) in the theoretical HR diagram (panel~a) and
the $\log T_{\rm eff}$--$\log g$ diagram (panel~b). We used
calibration (1) for the 14 evolved stars in the cluster turn-off
region ($(B-V) \leq 0.30$~mag; $T_{\rm eff} \geq 7250$~K; \S
\ref{subsubsec:trans_cali1}) and calibration (2) for the remaining 76
dwarfs ($(B-V) > 0.30$~mag; $T_{\rm eff} < 7250$~K; \S\S
\ref{subsubsec:trans_cali2}--\ref{subsubsec:trans_results}).

\subsection{Low-mass main sequence ($M \mathrel{{\hbox to 0pt{\lower 3pt\hbox{$\sim$}\hss}} \raise 2.0pt\hbox{$<$}} 0.9~M_\odot$)}
\label{subsec:HRD_1}

\placefigureEighteenA

Figure~\ref{fig:HRD_3}(a) shows, besides an inconsistency between the
Padova and CESAM isochro\-nes themselves, a significant discrepancy
between the isochrones on the one hand and the inferred effective
temperatures and luminosities for stars with $T_{\rm eff} \la 5000$~K
(K/M dwarfs) on the other hand. As the `turn-off' of the lower masses
is absent in the colour-absolute magnitude diagram
(Figure~\ref{fig:HRD_4}), we conclude that the model-observation
discrepancies for these cool stars are caused by inappropriate
$(B-V)$--$T_{\rm eff}$ calibrations (e.g., Gratton et al.\ 1996; cf.\
\S \ref{subsubsec:trans_previous} and \ref{subsubsec:trans_results},
Figures~\ref{fig:mass}--\ref{fig:HRD_5}, and footnote~15). Reliable
transformations for low-mass dwarfs with cool envelopes (or completely
convective interiors) require non-gray non-LTE line-blanketed model
atmospheres with appropriate molecular opacities (e.g., TiO, H$_2$O,
and VO; e.g., Hauschildt et al.\ 1999). Theoretical models in
this mass range similarly require adequate boundary conditions
(Chabrier \& Baraffe 1997). These models, for $M \la 0.6~M_\odot$,
also suffer from uncertainties related to the equation of state, and,
for $4000 \la T_{\rm eff} \la 5000$~K, from uncertainties related to
the treatment of convection (e.g., Ludwig et al.\ 1999).

\placefigureEighteenB

The low-mass stars as a set show an enhanced spread about the main
sequence compared to the higher mass stars. This effect is partly
related to the degradation of secular parallax (i.e., Hipparcos proper
motion) accuracies with $V$ magnitude. Part of the scatter might also
be due to spectral peculiarities (e.g., emission lines as in HIP 20527
[K5.5Ve], 20605 [M0.5Ve], and 21138 [K5Ve]) or colour anomalies
related to chromospheric activity (e.g., Campbell 1984; Stauffer et
al.\ 1991, 1997); a 5~per cent starspot coverage yields
$\sim$0.015~mag shifts in $(B-V)$, either towards the blue or towards
the red, as well as low-level photometric variations (on the order of
a few hundredths of a magnitude).

\subsection{Intermediate-mass main sequence ($0.9 \mathrel{{\hbox to 0pt{\lower 3pt\hbox{$\sim$}\hss}} \raise 2.0pt\hbox{$<$}} M \mathrel{{\hbox to 0pt{\lower 3pt\hbox{$\sim$}\hss}} \raise 2.0pt\hbox{$<$}} 1.6~M_\odot$)}
\label{subsec:HRD_2}

This mass range corresponds to F and G stars ($7500 \ga T_{\rm eff}
\ga 5000$~K). We divide it into three regimes corresponding to
different interiors:
(1) Stars with $M \ga 1.5~M_\odot$ ($B-V \la 0.30$~mag; $T_{\rm eff}
\ga 7000$~K) have a convective core and radiative envelope.
(2) With decreasing mass in the range $1.5 \ga M \ga
1.3$--$1.1~M_\odot$ ($0.30 \la B-V \la 0.40$~mag; $7000 \ga T_{\rm
eff} \ga 6500$--$6000$~K), the convective core shrinks to become
radiative, while a convective envelope develops at the same time. This
envelope gives rise to the formation of a chromosphere and corona.
(3) Stars with $M \la 1.3$--$1.1~M_\odot$ ($B-V \ga 0.40$~mag; $T_{\rm
eff} \la 6500$--$6000$~K) have a radiative core and convective
envelope.
Theoretical modelling for these regimes suffers mainly from
uncertainties related to convection, notably overshoot of the core and
envelope, and the issue of the universality of the mixing-length
parameter $\alpha_{\rm MLT}$ (e.g., Ludwig et al.\ 1999; \S
\ref{subsec:ThStEvMo}).

\placefigureNineteen

Figure~\ref{fig:HRD_3}(a) shows that the effective temperatures and
luminosities closely follow the CESAM isochrone, except in the range
$7000 \ga T_{\rm eff} \ga 6500$~K ($0.30 \la (B-V) \la 0.40$~mag)
where the observations are suggestive of a `turn-off' of the main
sequence around spectral type $\sim$F5V. Figure~\ref{fig:BVgap} shows
an expanded view of this region, including the locations of all
non-high-fidelity members. The `turn-off' is also clearly visible in
the secular parallax-based colour-absolute magnitude diagrams
displayed in Figures~\ref{fig:HRD_2} and \ref{fig:HRD_4}. We therefore
conclude that it is not caused by the $(B-V)$--$T_{\rm eff}$ relation
adopted in this study (\S \ref{subsubsec:trans_results}; cf.\
Figure~\ref{fig:HRD_5}; we do note, however, that the `turn-off'
roughly coincides with the transition at $(B-V) = 0.30$~mag ($T_{\rm
eff} \sim 7250$~K) between calibration (1) (\S
\ref{subsubsec:trans_cali1}) and calibration (2) (\S
\ref{subsubsec:trans_cali2})). We suspect, as argued below, that the
`turn-off' is related to the onset of surface convection around $(B-V)
= 0.30$--$0.40$~mag (cf.\ de Bruijne et al.\ 2000). This is consistent
with the work of Rachford (1997, 1998), who detected a transition in
stellar activity\footnote{
The onset of surface convection is accompanied by X-ray and near-UV
emission from coronal and chromospheric gas, which in its turn is due
to magnetic fields produced by a stellar rotation-induced dynamo
(e.g., Pallavicini et al.\ 2000).
} parameters at $(B-V) \sim 0.29$~mag as well as a chromospherically
active Hyad at $(B-V) = 0.26$~mag (HIP 21036). These findings indicate
that the onset of convection (in the Hyades) occurs at $(B-V) \ga
0.25$--$0.30$~mag (cf.\ Wolff et al.\ 1986; Rachford \& Canterna
2000).

B\"ohm--Vitense (1970, 1981, 1982) first realized that convective
atmospheres have relatively low temperatures in the deeper layers
which contribute to the surface flux in the spectral regions of the
$U$ and $B$ filters (cf.\ Nelson 1980). As a result, convective
atmospheres have reddened $(B-V)$ colours (as compared to radiative
atmospheres of the same $T_{\rm eff}$) by amounts of
$\sim$0.07--0.12~mag. As the reddening of the atmosphere is not
accompanied by a significant change in luminosity, the onset of
surface convection can cause a $\sim$$0.10$~mag `gap' and/or
`turn-off' in the $(B-V)$ colour-absolute magnitude diagram, starting
around $(B-V) = 0.25$--$0.35$~mag, the so-called B\"ohm--Vitense
gap\footnote{
The location of this gap roughly coincides with the Am-type stars. As
these objects are often close multiples (e.g., Jaschek \& Jaschek
1987; Debernardi et al.\ 2000), most of them are not contained in the
high-fidelity sample (\S \ref{subsec:HRD_stars}), causing a reduced
sampling of the main sequence (Figures~\ref{fig:HRD_3}(a) and
\ref{fig:BVgap}).
} (cf.\ figure~2 in de Bruijne et al.\ 2000). Observational evidence
for the existence of the B\"ohm--Vitense gap is sparse (e.g.,
B\"ohm--Vitense \& Canterna 1974; Jasniewicz 1984; Rachford \&
Canterna 2000), and its reality has been disputed (e.g., Mazzei \&
Pigatto 1988; Simon \& Landsman 1997; Newberg \& Yanny 1998). Previous
claims for its existence were based on the presence of gaps in either
colour-colour diagrams or in the cumulative distribution of cluster
members in some photometric index (e.g., Aizenman et al.\ 1969),
instead of on the presence of `gaps' or `turn-offs' in colour-absolute
magnitude diagrams. The secular parallax-based colour-absolute
magnitude diagrams of the Hyades presented in Figures~\ref{fig:HRD_2}
and \ref{fig:HRD_4} provide, in fact, the first direct evidence in
favor of the existence of the B\"ohm--Vitense gap.

\begin{table*}[t]
\caption[]{Some remaining stars in the upper main sequence and
turn-off region of the Hyades. Spectroscopic binarity (`SB'; table 2
in P98), high rotational velocities ($v \sin i$ from Abt \& Morrell
1995; km~s$^{-1}$), suspect secular parallaxes ($g_{\rm Hip} > 9$),
and spectral peculiarities can give rise to deviant HR diagram
positions. Hipparcos field H52 is a photometric variability flag (`M'
for micro-variable). The symbol $\delta$ Sct stands for $\delta$ Scuti
pulsator.}
\renewcommand{\arraystretch}{0.9}
\renewcommand{\tabcolsep}{4.5pt}
\begin{center}
\begin{tabular}{lrrlclrrlclrrl}
\noalign{\vskip -0.40truecm}
\noalign{\vskip 0.10truecm}
\hline
\hline
\noalign{\vskip 0.07truecm}
HIP & SB & $v \sin i$ & Notes & & HIP & SB & $v \sin i$ & Notes & & HIP & SB & $v \sin i$ & Notes\\
\noalign{\vskip 0.07truecm}
\hline
\noalign{\vskip 0.07truecm}
20400           & SB&  25& Am; $\delta$ Sct                   &&2090&   &  93& Am&&21683$^{\rm c}$ &   & 115& $g_{\rm Hip} = 11.92$; A5Vn          \\
20484           & SB&  15& Am                                 &&2103& SB&  31& Am&&22157           & SB&  63& Am; $g_{\rm Hip} = 23.34$            \\
20614$^{\rm a}$ &   & 145&                                    &&2127& SB& 130&   &&22565           &   & 165& H52 = `M'; $\delta$ Sct              \\
20711           &   & 225& H52 = `M'; $\delta$ Sct; A8Vn      &&2158& SB&  78&   &&23983           & SB&  13& Am                                   \\
20713$^{\rm b}$ & SB& 205& $g_{\rm Hip} = 30.30$; $\delta$ Sct&&2167&   &  75& Am&&24019$^{\rm d}$ &   &  45& Am                                   \\
\hline\hline\\[-0.65truecm]
\end{tabular}
\label{tab:turn-off}
\end{center}
$^{\rm a}$: Eggen's (1992) `photometry indicates a possible binary,
which is not resolved by speckle observations'. Patience et al.\
(1998) did not detect a speckle companion. Wielen et al.\ (2000) list
the star as `$\Delta\mu$ binary'.\\
$^{\rm b}$: A chromospherically active long-period F0V+G4V binary
(Peterson et al.\ 1981); the object is the second-brightest X-ray
source in the Hyades (e.g., Stern et al.\ 1992, 1994).\\
$^{\rm c}$: Parallaxes differ significantly ($\pi_{\rm Hip} = 20.51
\pm 0.82$, $\pi_{\rm sec,Hip} = 18.33 \pm 0.42$, $\pi_{\rm
sec,Tycho-2} = 18.70 \pm 0.38$~mas; $g_{\rm Hip\-/Tycho-2} =$
$11.92\-/8.74$), possibly as a result of multiplicity. Patience et
al.\ (1998) did not detect a speckle companion.\\
$^{\rm d}$: Primary target of a two-pointing system (field H60 = `P');
the A component is a periodic variable. The secondary target is HIP
24020; the double star processing used a linear system (H60 = `L'):
the parallax of the B component is constrained to be identical to the
A-component parallax, whereas the proper motions were
independently determined. The solution is uncertain (H61 = `D'). HIP
24020 is a duplicity-induced variable (H52 = `D'); the B component
itself is a resolved Hipparcos~binary.
\end{table*}

A careful inspection of Figures~\ref{fig:HRD_2}, \ref{fig:HRD_4}, and
\ref{fig:HRD_3}(a) reveals that the region around $(B-V) \sim
0.45$~mag ($T_{\rm eff} \sim 6400$~K) also shows an abrupt increase in
the $(B-V)$ colours of stars by an amount of $\sim$0.05~mag. This
feature was already commented on by B\"ohm--Vitense (1995a, b), who
attributed it to `{\it a sudden increase in convection zone
depths}'. The position of the second B\"ohm--Vitense gap coincides
with both the so-called Lithium gap, which is generally thought to be
related to the rapid growth of the depth of the surface convection
zone with effective temperature decreasing from $\sim$7000~K to
$\sim$6400~K (e.g., Boesgaard \& Tripicco 1986; Michaud 1986; Swenson
et al.\ 1994; Balachandran 1995), and the onset of dynamo-induced
magnetic chromospheric activity (cf.\ Wolff et al.\ 1986;
Garc{\'{\i}}a Lop\'ez et al.\ 1993).

As stellar rotation is known to influence significantly the precise
conditions for the onset of surface convection (e.g., Chandrasekhar
1961), B\"ohm--Vitense suggested that a range of $v \sin i$ values
within a stellar cluster can lead to a bifurcation of the
$(B-V)$--$T_{\rm eff}$ relation in the range $6500 \la T_{\rm eff} \la
7500$~K (cf.\ Simon \& Landsman 1997). This means that a star with a
given $(B-V)$ colour can be either a slowly rotating weakly convective
star with a `high' effective temperature or a rapidly rotating
radiative star with a `lower' effective temperature, where the `high'
and `low' temperatures can differ by as much as 500~K. The righthand
panel of Figure~\ref{fig:BVgap} compares B\"ohm--Vitense's (1981; her
table 3 and figure 4) predictions of the two branches to the effective
temperatures derived using calibrations (1) and (2) (\S\S
\ref{subsubsec:trans_cali1}--\ref{subsubsec:trans_cali2}). The
calibration (1) and (2) effective temperatures do not show a
bifurcation, simply because they do not take stellar rotation into
account. Systematic differences outside the bifurcation region can be
due to the presently outdated Solar-metallicity atmosphere models used
by B\"ohm--Vitense in 1981. A natural next step would be to
determine, for each Hyades member in the region of the (first)
B\"ohm--Vitense gap, its rotational velocity, to determine its
effective temperature based on adequate stellar modelling, i.e.,
taking stellar rotation into account, and to compare the resulting
location in the $T_{\rm eff}$--$(B-V)$ diagram
(Figure~\ref{fig:BVgap}) with the predictions of B\"ohm--Vitense
(1981).

Naively, one would expect that the peculiar $(B-V)$ behaviour
resulting from (the onset of) surface convection is incorporated in
the (synthetic) $(B-V)$--$T_{\rm eff}$ relations adopted in this
paper. However, as the onset of surface convection is a subtle effect
when translated into the mass range at stake, its visibility in
the theoretical HR diagram (Figure~\ref{fig:HRD_3}(a)) is probably
the result of an inappropriate mass sampling in the construction of
the synthetic $(B-V)$--$T_{\rm eff}$ relations (\S\S
\ref{subsubsec:trans_cali1}--\ref{subsubsec:trans_cali2}). Only a
future investigation of the interplay between stellar rotation and
(the onset of) surface convection, and the corresponding effects on
the atmospheric parameters of mid-F stars (e.g., Hartmann \& Noyes
1987; Chaboyer et al.\ 1995), can shed more light on the issue of the
`turn-off(s)' of the Hyades main sequence and the bifurcation of the
$(B-V)$--$T_{\rm eff}$ relation.

\subsection{The upper main sequence and turn-off region ($M \mathrel{{\hbox to 0pt{\lower 3pt\hbox{$\sim$}\hss}} \raise 2.0pt\hbox{$>$}} 1.6~M_\odot$)}
\label{subsec:HRD_3}

Beyond $T_{\rm eff} \ga 7500$~K, we find the early-F and late-A stars
which are evolving towards the end of the core Hydrogen burning
phase. These objects are powered by the CNO-cycle and have convective
cores and radiative envelopes. As a result, the precise locations of
the isochrones depend strongly on the treatment of
rotationally-induced mixing (Maeder \& Meynet 2000) and the amount of
convective core overshoot included in the models (e.g., \S 9.2 in
P98). Given these uncertainties, Figure~\ref{fig:HRD_3} shows that the
stars on the upper main sequence of the Hyades follow the 625--650~Myr
CESAM isochrones (including convective core overshoot) remarkably
well, both in luminosity (panel~a) and in surface gravity (panel~b).

Many stars in the turn-off region of the Hyades rotate rapidly (e.g.,
table 11 in P98). Stellar rotation influences the observed colours and
magnitudes (and thus the inferred temperatures and luminosities) of
stars, the amounts depending on the rotation characteristics
(solid-body or differential rotation), the rate of rotation, the
orientation of the rotation axis with respect to the observer, and the
spectral type of the star (e.g., Kraft \& Wrubel 1965; Maeder \&
Peytremann 1970, 1972; Collins \& Smith 1985; P\'erez Hern\'andez et
al.\ 1999; Maeder \& Meynet 2000). Although the photometric effect of
rotation also depends on the specific filter system in use, rotating
stars generally become redder (typically a few hundredths of a
magnitude in $(B-V)$; $\sim$200--250~K in $T_{\rm eff}$) as well as
brighter or fainter (typically a few tenths of a magnitude in $V$). An
example of a rapidly rotating star in the Hyades turn-off region is
HIP 23497 ($v \sin i \sim 126$~km~s$^{-1}$; $\log(T_{\rm eff} [{\rm
K}]) = 3.9115$; $\log(L/L_\odot) = 1.472$); its deviant location in
the HR diagram is most likely due to rotation: although the object is
listed as equal-magnitude occultation double by Hoffleit \& Jaschek
(1991), Patience et al.\ (1998) did not detect a secondary.

The turn-off region of the Hyades coincides with the lower part of the
instability strip (e.g., Liu et al.\ 1997). Some of the stars
accordingly pulsate (e.g., Antonello \& Pasinetti Fracassini 1998). An
example is the primary component of the `resolved spectroscopic
binary' $\theta^2$ Tau (HIP 20894), which is a $\delta$ Scuti pulsator
(e.g., Breger et al.\ 1987, 1989; Kennelly et al.\ 1996; cf.\ Appendix
\ref{subsec:trans_theta2} and Table~\ref{tab:orbpar}). The A7III--IV
primary has an A5V secondary. The exact location of both stars in the
turn-off region of the HR diagram therefore puts a severe constraint
on stellar evolutionary models of the system as well as of the Hyades
cluster (e.g., Lastennet et al.\ 1999). Unfortunately, the precise
evolutionary status of the evolved component is unknown. Moreover,
observations as well as theoretical modelling are complicated by the
fact that both stars, which are roughly equally bright, are rapid
rotators ($[v \sin i]_{\rm primary} \sim 80$~km~$^{-1}$; $[v \sin
i]_{\rm secondary} \ga 90$~km~$^{-1}$), resulting in severely blended
spectra.

The secular parallax of the binary allows to position both components
in the HR diagram with unprecedented
precision. Figure~\ref{fig:HRD_3}(a) suggests that $\theta^2$ Tau A is
close to the end of the core Hydrogen burning phase, and is about to
undergo an overall gravitational contraction which will ignite thick
Hydrogen shell burning. This is consistent with the low amplitude of
the pulsations, indicative of main sequence evolution (e.g., Li \&
Michel 1999; large-amplitude pulsators (0.10~mag or higher) are
thought to be evolved stars in the Hydrogen shell burning
phase). These results confirm the conclusion of Torres et al.\
(1997c), which was based on the orbital parallax of the binary
(Table~\ref{tab:orbpar}), but are in conflict with Kr\'olikowska's
(1992) finding that $\theta^2$ Tau A currently burns Hydrogen in a
thick shell.

The bluest Hyad is HIP 20648 (A2IVm; $(B-V) = 0.049 \pm 0.007$~mag).
Being a Hipparcos duplicity-induced variable (field H52 is `D') and
component binary (H59 is `C'; $\Delta Hp = 4.02 \pm 0.09$~mag), the
object is not included in our high-fidelity sample. Nonetheless, the
object is a secure member of the Hyades: its secular parallax solution
(based on the Hipparcos A-component proper motion; H10 is `A') is of
high quality ($g_{\rm Hip} = 0.46$), and the observed radial velocity
and parallax (secular as well as trigonometric) are consistent with
membership (e.g., table 2 in P98). Applying calibration (1) to the
observed colour and secular parallax returns $\log(T_{\rm eff} [{\rm
K}]) = 3.9649 \pm 0.0061$ and $\log(L/L_\odot) = 1.541 \pm
0.017$. These parameters put the star on the Hyades main sequence, but
far beyond the turn-off (inset of Figure~\ref{fig:HRD_3}(a)). This
`blue straggler' nature is not readily explained by rapid rotation
(since $v \sin i = 15$~km~s$^{-1}$ and a pole-on orientation is
unlikely) or an inappropriate calibration (e.g., Burkhart \& Coupry
(1989) found $T_{\rm eff} = 9050 \pm 100$~K). The observation of a
large photospheric (and thus presumably also large interior) magnetic
field (Babcock 1958) suggests that the apparently prolonged hydrogen
core burning phase of this star, as compared to other members of
similar mass which have already evolved off the main sequence, can be
explained if the Helium produced in the core is continuously replaced
with fresh Hydrogen through large-scale magnetic mixing (e.g., Hubbard
\& Dearborn 1980; Abt~1985).

The inset of Figure~\ref{fig:HRD_3}(a) shows all objects in the
turn-off region of the Hyades which are not contained in the
objectively defined sample of high-fidelity members (cf.\
Table~\ref{tab:turn-off}). The deviant HR diagram locations of some
of these members are most likely caused by: (1) suspect secular
parallaxes (i.e., $g_{\rm Hip} > 9$); (2) stellar rotation; (3)
(spectroscopic) binarity; or (4) an inappropriate calibration (\S
\ref{subsec:calibration}). The latter option might be especially
relevant for the slowly rotating metallic-line A stars. Uncertainties
in (metal-)line-blanketing, line lists, and opacities in atmosphere
models, the treatment of the diffusion of chemical elements, and the
treatment of large-scale motions in the envelopes of these stars give
rise to significant uncertainties in the $T_{\rm eff}$-scale (e.g.,
Smalley \& Dworetsky 1993; Richer et al.\ 2000).

\subsection{The giant region}
\label{subsec:HRD_4}

The red clump of the Hyades contains four giants: $\theta^1$,
$\delta^1$, $\epsilon$, and $\gamma$ Tau; the determination of their
effective temperatures and secular parallax-based luminosities is
discussed in Appendices
\ref{subsec:trans_giants}--\ref{subsec:trans_other_giants} Whereas
$\epsilon$ and $\gamma$ Tau are single stars, $\theta^1$ and
$\delta^1$ Tau are spectroscopic binaries. The secular parallaxes of
these double stars should be treated with care as they could be based
on astrometric data which do not properly reflect their true space
motions (\S \ref{subsec:faint_mems}).

\placefigureTwenty

The location of isochrones in the giant region not only depends quite
sensitively on metallicity but also on the mass loss history on the
red giant branch and the adopted value of the mixing-length parameter
$\alpha_{\rm MLT}$. Nonetheless, all Hyades giants precisely follow
Girardi et al.'s (2000a) Solar-metallicity 631~Myr isochrone (\S
\ref{subsec:Hyades_chars}; Figure~\ref{fig:HRD_giants}), despite the
fact that Girardi simply accounted for mass loss by means of Reimers'
(1975) empirical formula (using a mass-loss efficiency parameter of
$0.4$; Renzini \& Fusi Pecci 1988). A natural next step would be to
investigate the variation of the location of the red giant clump with
metallicity, mass loss, and mixing-length parameter (cf.\ Girardi et
al.\ 2000b). Such an analysis would give, e.g., more insight into the
reliability of red clump giants as distance calibrators (e.g., Alves
2000; Udalski 2000).

The P98 member list contains one additional evolved star of spectral
type K2III. This object ($\delta$ Ari; HIP 14838; Appendix
\ref{subsec:trans_other_giants}) has several characteristics of a
non-member: it is located $15.34$~pc from the cluster center, it has a
near-Solar metallicity as opposed to the mean Hyades value $[{\rm
Fe/H}] = +0.14 \pm 0.05$, it has an unreliable secular parallax
solution, and it was rejected by de Bruijne (1999a) and Hoogerwerf \&
Aguilar (1999) as proper motion member of the Hyades
(Table~\ref{tab:data_1}). Although the latter two facts could be
spurious due to possible duplicity (Appendix
\ref{subsec:trans_other_giants}; Wielen et al.\ 2000),
Figure~\ref{fig:HRD_giants} shows that $\delta$ Ari is not coeval with
the classical giants in the cluster. We conclude that the star is most
likely a non-member.

\subsection{White dwarfs}
\label{subsec:HRD_5}

The Hyades contains a dozen\footnote{
The present-day luminosity function of the Hyades predicts the cluster
contains $\sim$25--30 white dwarfs (e.g., Chin \& Stothers 1971). The
discrepancy between the observed and predicted number of these objects
is possibly explained by evaporation from the cluster (e.g., Weidemann
et al.\ 1992; Eggen 1993; \S \ref{subsec:mem_P98}). One possible
example of an escaped white dwarf is the P98 candidate HIP 12031
(DAwe...). It is located beyond 40~pc from the cluster center, and is
possibly a kinematic member ($g_{\rm Hip} = 0.80$; $g_{\rm Tycho-2} =
12.61$).
} known white dwarfs (e.g., Humason \& Zwicky 1947; Luyten 1954, 1956;
B\"ohm--Vitense 1995). They typically have $V \ga 14$~mag, which means
they are too faint to be observed directly by Hipparcos. However,
several white dwarfs are contained in multiple systems (e.g., Lanning
\& Pesch 1981), the primary components of which were observed by
Hipparcos. We discuss two of such systems.

\paragraph{V471 Tau:} HIP 17962 is a post-common-envelope
detached eclipsing binary. It is composed of a DA white dwarf and a
coronally active K2V star. Hosting the hottest and youngest Hyades
white dwarf, V471 Tau has been studied extensively (e.g., Nelson \&
Young 1970, 1972; Guinan \& Sion 1984; Clemens et al.\ 1992; Shipman
et al.\ 1995; Marsh et al.\ 1997). The system is a pre-cataclysmic
variable: the K star does not fill its Roche lobe yet. The observed
periodic optical and X-ray variations are related to material from the
K-star wind being accreted onto the magnetic poles of the rotating
white dwarf (e.g., Jensen et al.\ 1986; Barstow et al.~1992).

\begin{table*}[t]
\caption[]{Fundamental parameters of two Hyades white dwarfs. {\it
V471 Tau:\/} effective temperatures and surface gravities determined
from ORFEUS/IUE spectra (Barstow et al.\ 1997; Werner \& Rauch
1997). The orbital elements of the eclipsing binary imply a mass
function of $0.174 \pm 0.002~M_\odot$; using $80^\circ\! \leq i \leq
90^\circ\!$ returns an astrometric white dwarf mass $M = 0.759 \pm
0.020~M_\odot$ for an assumed K-dwarf mass $M = 0.800~M_\odot$ (Young
1976; Bois et al.\ 1988). The white dwarf radius $R$ is independently
determined from model fluxes and the observed parallax and flux. The
inferred surface gravity then follows from combining this radius with
the astrometric mass. Barstow et al.\ used Wood's (1995) evolutionary
models to derive a spectroscopic mass $M/M_\odot =
0.61^{+0.14}_{-0.10}$ and radius $R/R_\odot = 0.014 \pm 0.003$. `This
study' combines the secular parallax of V471 Tau with Werner \&
Rauch's model. {\it HD 27483:\/} results from B\"ohm--Vitense (1993)
and Burleigh et al.\ (1998). The Burleigh et al. and `This study'
results were derived from an interpolation in Burleigh's table 5 using
the Hipparcos trigonometric and secular parallax,
respectively. Burleigh et al.\ themselves quote $\log g = 8.5$,
$T_{\rm eff} = 2\,2000$~K, and $M/M_\odot = 0.94$.}
\renewcommand{\arraystretch}{0.9}
\renewcommand{\tabcolsep}{15.0pt}
\begin{center}
\begin{tabular}{lrrr}
\noalign{\vskip -0.40truecm}
\noalign{\vskip 0.10truecm}
\hline
\hline
\noalign{\vskip 0.07truecm}
V471 Tau                 & Barstow et al.\      & Werner \& Rauch   & This study       \\
HIP 17962  & (1997; $\pi_{\rm Hip}$) & (1997; $\pi_{\rm Hip}$) & ($\pi_{\rm sec, Hip}$)\\
\noalign{\vskip 0.07truecm}
\hline\\[-0.30truecm]
%\noalign{\vskip 0.07truecm}
$T_{\rm eff} [{\rm K}]$ (spectroscopic)
                         &$32\,400^{+270}_{-800}$ & 35\,125$\pm$1275    &                  \\
$\log(g [{\rm cm~s}^{-2}])$ (spectroscopic)
                         &$8.16^{+0.18}_{-0.24}$& 8.21$\pm$0.23     &                  \\
$M/M_\odot$ (from orbit) & 0.76$\pm$0.02        & 0.759$\pm$0.020   & 0.759$\pm$0.020  \\
$R/R_\odot$ (from parallax)
                         & 0.0107$\pm$0.0009    & 0.0097$\pm$0.0013 & 0.0098$\pm$0.0011\\
$\log(g [{\rm cm~s}^{-2}])$ (inferred)
                         & 8.27$\pm$0.07        & 8.35$\pm$0.12     & 8.34$\pm$0.10    \\
\noalign{\vskip 0.07truecm}
\hline
\noalign{\vskip 0.07truecm}
HD 27483                 & B\"ohm--Vitense      & Burleigh et al.\  & This study       \\
HIP 20284 & (1993; $\pi_{\rm sec, Schwan~(1991)}$) & (1998; $\pi_{\rm Hip}$) & ($\pi_{\rm sec, Hip}$) \\
\noalign{\vskip 0.07truecm}
\hline
\noalign{\vskip 0.07truecm}
$T_{\rm eff} [{\rm K}]$  & $23\,800$                & $21\,815\pm178$     & $21\,555\pm84$     \\
$\log(g [{\rm cm~s}^{-2}])$
                         & 8.03                 & $8.34\pm0.16$     & $8.11\pm0.07$    \\
$M/M_\odot$              & 0.60                 & $0.84\pm0.10$     & $0.696\pm0.041$  \\
$R/R_\odot$              & 0.012                & $0.010\pm0.002$   & $0.0121\pm0.0006$\\
\hline\hline\\[-0.6truecm]
\end{tabular}
\label{tab:wd}
\end{center}
\end{table*}

The effective temperature and surface gravity of the white dwarf in
V471 Tau were recently determined by fitting synthetic spectra to
observed spectra of the Hydrogen Lyman lines (Barstow et al.\ 1997;
Werner \& Rauch 1997; Table~\ref{tab:wd}). Unfortunately, the
relatively large uncertainty in the best-fit $\log g$ values prevents
an accurate mass determination. It has therefore been common practice
to infer the surface gravity of the white dwarf from its astrometric
mass, obtained from the orbital elements of the binary, and its
radius, obtained from its observed and modelled flux combined with its
Hipparcos parallax (Table~\ref{tab:wd}).

A more precise estimate of the white dwarf radius is available through
its secular parallax. The latter is non-suspect ($g_{\rm Hip} =
0.06$), and fully consistent with the Hipparcos parallax ($\pi_{\rm
Hip} = 21.37 \pm 1.62$~mas; $\pi_{\rm sec, Hip} = 21.00 \pm
0.40$~mas). The long time-baseline Tycho--2 secular parallax, which
might be preferred over the Hipparcos secular parallax in view of the
binary nature of the system (although $P_{\rm orb} = 0.521$~days only;
Stefanik \& Latham 1992; \S \ref{subsec:faint_mems}), places the
object at a slightly, though not significantly, larger distance
($\pi_{\rm sec,Tycho-2} = 20.56 \pm 0.33$~mas with $g_{\rm Tycho-2} =
0.02$). The Hipparcos secular parallax fits Werner \& Rauch's (1997)
model for a radius of $R/R_\odot = 0.0098 \pm 0.0011$; the
corresponding surface gravity is $\log g = 8.34 \pm 0.10$ for
$M/M_\odot = 0.759 \pm 0.020$ (Table~\ref{tab:wd}). These values are
fully consistent with but (in principle) more precise than the results
of Werner \& Rauch (1997) and Barstow et al.\ (1997).
\smallskip

\noindent{\it HD 27483:\/} In 1993, B\"ohm--Vitense reported the
serendipitous discovery of a DA white dwarf companion around the close
F6V--plus--F6V binary HD 27483 (HIP 20284). B\"ohm--Vitense
interpreted her spectra using Wesemael et al.'s (1980) unblanketed
white dwarf models and Hamada \& Salpeter's (1961) mass--radius
relation, assuming a distance to the system of 47.6~pc (the secular
parallax derived by Schwan 1991; Table~\ref{tab:wd}). Recently,
Burleigh et al.\ (1998) presented an analysis of the object, based on
updated atmosphere and evolutionary models (Koester 1991; Wood 1995),
using its Hipparcos parallax ($\pi_{\rm Hip} = 21.80 \pm
0.85$~mas). The orbital motion of the binary ($P_{\rm orb} =
3.05$~days; Mayor \& Mazeh 1987) has not hampered the Hipparcos
measurements: its secular parallax ($\pi_{\rm sec, Hip} = 20.59 \pm
0.35$~mas) is well defined ($g_{\rm Hip} = 2.73$). We use cubic spline
interpolation in Burleigh's table 5 to derive $\log g$, $T_{\rm eff}$,
$M$, and $R$ for both the Hipparcos trigonometric and secular
parallaxes (Table~\ref{tab:wd}). The secular parallax-based white
dwarf mass ($M/M_\odot = 0.70 \pm 0.04$) is significantly smaller than
Burleigh et al.'s value ($M/M_\odot = 0.94$). The new mass estimate
resolves the problem (acknowledged by Burleigh et al.) that the sum of
the cooling age of a $M = 0.94~M_\odot$ white dwarf and the
evolutionary age of its progenitor is significantly shorter than the
nuclear age of the Hyades.

\section{Summary and discussion}
\label{sec:disc}

At $\sim$45~pc, the Hyades is the nearest open cluster to the Sun.
Its tidal radius of $\sim$10~pc translates to an angular extent of
several tens of degrees on the sky and to a significant line-of-sight
extension compared to its mean distance. Hipparcos trigonometric
parallaxes, with typical accuracies of $\sim$1.5~mas, constrain the
positions of members in the cluster to within a few parsec. Although
this precision is sufficient to study the three-dimensional structure
of the cluster, and related issues like mass segregation.
uncertainties in the `trigonometric distances' to individual stars
dominate the error budget when constructing the colour-absolute
magnitude diagram of the cluster. The considerable tangential velocity
of the cluster (relative to the Sun) opens the possibility to derive
parallaxes for individual stars from their measured proper motions (\S
\ref {sec:sec_pars}). Despite the presence of an internal velocity
dispersion in the cluster of $\sim$0.30~km~s$^{-1}$ (in one
dimension), the proper motion-based (secular) parallaxes are $\sim$3
times more precise than the Hipparcos trigonometric values (\S
\ref{subsec:accuracy}).

The secular parallaxes are derived under the assumption that all
cluster stars follow a three-dimensional Gaussian velocity
distribution with an isotropic dispersion (\S
\ref{sec:space_motion}). A careful analysis confirms the absence of
significant velocity structure in the form of expansion, rotation, or
shear (\S \ref{sec:acc_and_prec}). Numerical and theoretical work
suggests that members beyond the tidal radius of the cluster (the halo
and moving group population) (may) have a systematically different
velocity field from the main body of the cluster (the core and
corona). We show that the maximum expected systematic secular parallax
errors in the outer regions of the cluster are $\la$0.30~mas (\S
\ref{subsec:summary}), i.e., a factor $\sim$2 smaller than typical
random secular parallax errors. The Hipparcos trigonometric and
secular parallaxes as a set are statistically fully
consistent. Nonetheless, we do find evidence for the presence of
spatially correlated Hipparcos measurements on small angular scales
(i.e., a few degrees), consistent with the predictions of the
Hipparcos data reduction consortia (\S \ref{subsec:acc_2}). The
maximum `amplitude' of the correlation is $\la$$0.50$--$0.75 \sigma
\sim 0.75$--$1.00$~mas per star, which is a factor of $\sim$2 smaller
than the value quoted by Narayanan \& Gould (1999b).

Our list of Hyades candidate members contains 233 stars (\S
\ref{sec:additional_members}). These are the 218 candidates selected
by P98 and 15 new Hipparcos stars selected by de Bruijne's (1999a)
and/or Hoogerwerf \& Aguilar's (1999) methods. Only one of the latter
stars (HIP 19757) is a likely member (\S
\ref{subsec:Hip_additional_members}). Long time-baseline
Tycho--2 proper motions are available for (most of) the brighter
Hyades that were observed by Hipparcos; the corresponding secular
parallaxes are fully consistent with both the Hipparcos trigonometric
and the Hipparcos secular parallaxes.

The secular parallaxes for members of the Hyades allow the
construction of the most precise colour--absolute magnitude diagram of
the cluster to date (Figures~\ref{fig:HRD_1} and \ref{fig:HRD_4}; see
also, e.g., Madsen 1999). The main sequence is well defined, and shows
a few conspicuous but artificial gaps, e.g., around $(B-V) = 0.95$ and
between $(B-V) = 0.30$ and $(B-V) = 0.35$ mag, caused by the
suppression of (double and multiple) stars from our sample (cf.\
figure~21 of P98). The small gap between $(B-V) = 0.75$ and $0.80$~mag
is also present in the original sample and is therefore probably real
(cf.\ figure~2 of Mermilliod 1981). Somewhat further to the blue,
there are two conspicuous features (`gaps' and/or `turn-offs') around
$(B-V) \sim 0.30$ ($T_{\rm eff} \sim 7000$~K) and $\sim$0.45~mag
($\sim$$6400$ K). These features, which have never been observed this
clearly before, but the existence of which was predicted by E.\
B\"ohm--Vitense already $\sim$30 years ago, are related to the use of
the $(B-V)$ colour as temperature indicator. We suspect, following
B\"ohm--Vitense, that sudden changes in the properties of surface
convection zones in the atmospheres of stars with $(B-V) \sim 0.30$
and $\sim$0.45~mag significantly affect the emergent UV and
blue-optical fluxes, and thus the $(U-B)$ and $(B-V)$ colours (\S
\ref{subsec:HRD_2}).

As the Hipparcos members of the Hyades span a large range in mass and
occupy a number of different evolutionary states, their effective
temperatures and luminosities provide stringent constraints on both
the global characteristics of the Hyades (such as metallicity, Helium
content, and age) and stellar evolutionary modelling in general (cf.\
Lebreton 2000). We combine the secular parallaxes derived in this
study with two existing $(B-V) \longrightarrow \log T_{\rm eff}$ and
$M_V \longrightarrow \log(L/L_\odot)$ calibrations (\S
\ref{subsec:calibration}) to infer the fundamental properties of the
cluster as well as of a variety of members. The latter include, among
others, stars with surface convection zones, Am stars, a $\delta$
Scuti pulsator, red giants, and white dwarfs (\S \ref{sec:HRD}). We
show that neither the Bessell et al.\ (1998) nor the Lejeune et al.\
(1998) $(B-V)$--$T_{\rm eff}$ and $T_{\rm eff}$--${\rm BC}_V$
calibrations (whether correcting for the non-Solar metallicity of the
Hyades or not) are appropriate throughout the entire mass range
studied in this paper; only an ad-hoc combination of the two
calibrations provides an acceptable fit to the P98 CESAM isochrones.

An optimum exploitation of the secular parallax data requires several
future steps. An obvious need is the construction of a set of
isochrones that are based on homogeneous stellar evolutionary
modelling, tailored to the Hyades cluster in terms of chemical
composition and age, from the low-mass main sequence ($M \ga
0.50~M_\odot$) through the clump-giant region (\S
\ref{subsec:HRD_4}). The lack of this information prevents us from
validating the chemical composition and age of the cluster derived by
P98. The precision with which the secular parallaxes constrain the
locations of stars in the HR diagram brings out such detailed
structure in the main sequence (e.g., the B\"ohm--Vitense gaps) that
the construction of future isochrones will require a fine-tuned mass
sampling as well as, for stars near the B\"ohm--Vitense gap, inclusion
of stellar rotation (\S \ref{subsec:HRD_2}). This study has shown that
the most significant uncertainty in the secular parallax-based HR
diagram locations of stars is now set by systematic errors in the
available transformations of the observed optical broad-band colours
and absolute magnitudes to effective temperatures and luminosities
(cf.\ Nordstr\"om et al.\ 1997; Castellani et al.\ 2000). This issue
clearly requires future study, for which the secular parallaxes
provide stringent boundary conditions.

An application of the secular parallax method to other nearby
clusters, such as Coma Berenices, the Pleiades, and Praesepe, is
feasible. Improved parallaxes can be obtained for these groups (cf.\
Dravins et al.\ 1999), although several complicating factors exist (as
compared to the Hyades). Among these are their larger mean distances,
smaller fields on the sky, less well-defined (Hipparcos) membership
lists, and the presence of interstellar reddening and extinction. The
secular parallaxes for the Hyades cluster presented in this paper will
only be superseded by the measurements of a second generation of
astrometric satellites, such as FAME, DIVA, and notably GAIA ({\tt
http://aa.usno.navy.mil/fame/, www.-} {\tt aip.de/groups/DIVA/,
astro.estec.esa.nl/GAIA/}).

\begin{acknowledgements}

The authors wish to thank warmly Adriaan Blaauw for his continuous
interest in and impetus on this project. Stimulating discussions with
Francesca d'Antona, Anthony Brown, Jan Lub, and Michael Perryman, as
well as the constructive criticism of the referees, Lennart Lindegren
and Yveline Lebreton, are gratefully acknowledged. The Hipparcos
Hyades team (P98) kindly provided BDA Tycho membership information,
CESAM isochrones, and stellar masses. This research is based to a
large extent on data obtained by ESA's Hipparcos satellite, and has
made use of the ADS (NASA) and SIMBAD (CDS) services; it is partly
supported by NWO.

\end{acknowledgements}

\appendix

% Appendix A
\section{Data}
%\label{sec:data}

Tables~\ref{tab:data_1}--\ref{tab:additional_members} contain data for
the 218 Hyades candidates selected by P98 (\S \ref{subsec:mem_P98})
and the 15 new candidates selected by de Bruijne (1999a) and
Hoogerwerf \& Aguilar (1999; \S \ref{subsec:addi_mems}),
respectively. The fundamental stellar parameters listed in
columns~(10)--(13) of Table~\ref{tab:data_1} are based on the
Hipparcos secular parallaxes, and the $V$ and $(B-V)$ values listed in
the Hipparcos Catalogue (ESA 1997; fields H5 and H37, respectively;
cf.\ \S \ref{subsec:calibration}).

% Appendix B
\section{$(B-V)$--$M_V$ $\longrightarrow$ $\log T_{\rm eff}$--$\log(L/L_\odot)$}
%\label{sec:transformation}

\subsection{The giants $\epsilon$ and $\gamma$ Tau}
\label{subsec:trans_giants}

The Hyades cluster contains two single giants, $\epsilon$ Tau (HIP
20889) and $\gamma$ Tau\footnote{
Although $\gamma$ Tau was reported as $\rho \sim 0\farcs395$ speckle
double by Morgan et al.\ (1982), Mason et al.\ (1993) and Patience et
al.\ (1998) could not confirm this result. Griffin \& Holweger (1989)
summarize the confusing literature on (non-existent) radial velocity
variability, and conclude the object is single.
} (HIP 20205; \S\S \ref{subsec:HRD_stars} and
\ref{subsec:HRD_4}). Both stars are located in the giant clump, i.e.,
the core Helium burning phase. Table~\ref{tab:giants}\footnote{
Although Bessell et al.\ do provide separate relations for red giants
in their table 5, the modeled $(B-V)$ range is inconsistent with the
measured values for both stars; the same holds for Lejeune's giant
calibration in their tables 1 and 9.
} shows the results of applying calibration (1) (\S
\ref{subsubsec:trans_cali1}) to $\epsilon$ and $\gamma$ Tau for $[{\rm
Fe/H}] = 0$ and $+0.14$ (although it formally applies to dwarfs only).

The effective temperatures, surface gravities, metallicities, and
bolometric fluxes of $\epsilon$ and $\gamma$ Tau have recently been
determined through combining the infra-red flux method with modelling
of high-resolution spectra using MARCS model atmospheres
(Table~\ref{tab:giants}; Smith \& Ruck 1997; Smith 1999). Luminosities
for both stars can be obtained by combining their bolometric fluxes
with parallaxes; results for Hipparcos trigonometric and secular
parallaxes are listed in Table~\ref{tab:giants} under Smith \& Ruck
(1997) and Smith (1999), respectively. Although calibration (1) gives
very similar results, we use the infra-red flux method temperatures
and secular parallax-based luminosities.

\subsection{The giants $\delta^1$ and $\theta^1$ Tau and $\delta$ Ari}
\label{subsec:trans_other_giants}

\begin{table*}[t]
\caption[]{Fundamental properties of the single Hyades red giants
(cf.\ table 2 in Smith \& Ruck 1997). The luminosities
$\log(L/L_\odot)$ listed for Smith \& Ruck (1997) and Smith (1999)
were derived from combining integrated stellar fluxes above the
Earth's atmosphere determined using the infra-red flux method
($F_{\gamma~{\rm Tau}} = 1.14 \cdot 10^{-9}$~W~m$^{-2}$;
$F_{\epsilon~{\rm Tau}} = 1.28 \cdot 10^{-9}$~W~m$^{-2}$; Blackwell \&
Lynas--Gray 1998) with Hipparcos trigonometric and secular parallaxes,
respectively. The giant metallicities $[{\rm Fe/H}] = +0.14$ and
masses $M/M_\odot = 2.32 \pm 0.10$ used in calibration (1) were taken
from P98. These mass estimates are consistent with the clump giant
masses of $\sim$$2.4~M_\odot$ listed by Torres et al.\ (1997a, c).}
\renewcommand{\arraystretch}{0.9}
\renewcommand{\tabcolsep}{15.0pt}
\begin{center}
\begin{tabular}{lrrrr}
\noalign{\vskip-0.30truecm}
\hline
\hline
\noalign{\vskip 0.07truecm}
$\gamma$ Tau (K0III)  & Smith \& Ruck (1997) & Smith (1999)        & Calibration (1)     & Calibration (1)     \\
HIP 20205       & $L=L(\pi_{\rm Hip})$ & $L=L(\pi_{\rm sec,Hip})$ & $[{\rm Fe/H}] = +0.14$ & $[{\rm Fe/H}] = 0$\\
\noalign{\vskip 0.07truecm}
\hline
\noalign{\vskip 0.07truecm}
$\log(L/L_\odot)$      & $1.901  \pm 0.016 $ & $1.869  \pm 0.016 $ & $1.843  \pm 0.016 $ & $ 1.850  \pm 0.016 $\\
$\log(T_{\rm eff} [{\rm K}])$$^{\rm a}$ 
                       & $3.6959 \pm 0.0035$ &                     & $3.7004 \pm 0.0027$ & $ 3.6967 \pm 0.0027$\\
$\log(g [{\rm cm~s}^{-2}])$
                       & $2.65   \pm 0.20  $ & $2.63   \pm 0.03  $ & $2.714  \pm 0.027 $ & $ 2.692  \pm 0.027 $\\
$[{\rm Fe/H}]$$^{\rm c}$
                       & $+0.12  \pm 0.03  $ & $+0.150 \pm 0.029 $ & $+0.14            $ & $ +0.00            $\\
$M/M_\odot$            &                     & $2.30   \pm 0.10  $ & $ 2.32  \pm 0.10  $ & $ 2.32   \pm 0.10  $\\
\noalign{\vskip 0.07truecm}
\hline
\noalign{\vskip 0.07truecm}
$\epsilon$ Tau (K0III) &                     &                     &                     &                     \\
HIP 20889              &                     &                     &                     &                     \\
\noalign{\vskip 0.07truecm}
\hline
\noalign{\vskip 0.07truecm}
$\log(L/L_\odot)$      & $1.956  \pm 0.015 $ & $1.927  \pm 0.015 $ & $1.912  \pm 0.015 $ & $ 1.919  \pm 0.015 $\\
$\log(T_{\rm eff} [{\rm K}])$$^{\rm b}$
                       & $3.6912 \pm 0.0031$ &                     & $3.6938 \pm 0.0022$ & $ 3.6902 \pm 0.0022$\\
$\log(g [{\rm cm~s}^{-2}])$
                       & $2.45   \pm 0.20  $ & $2.57   \pm 0.03  $ & $2.619  \pm 0.026 $ & $ 2.598  \pm 0.026 $\\
$[{\rm Fe/H}]$$^{\rm c}$
                       & $+0.15  \pm 0.03  $ & $+0.163 \pm 0.030 $ & $+0.14            $ & $ +0.00            $\\
$M/M_\odot$            &                     & $2.30   \pm 0.10  $ & $ 2.32  \pm 0.10  $ & $ 2.32   \pm 0.10  $\\
\noalign{\vskip 0.07truecm}
\hline
\noalign{\vskip 0.07truecm}
$T_{{\rm eff},\gamma~{\rm Tau} - \epsilon~{\rm Tau}}$~[K]
                       & $   +54  \pm 53   $ & $   +80 \pm 20    $ & $   +76 \pm 40    $ & $    +74 \pm 40    $\\
\noalign{\vskip 0.07truecm}
\hline\hline\\[-0.60truecm]
\end{tabular}
\label{tab:giants}
\end{center}

$^{\rm a}$:  $\log(T_{\rm eff} [{\rm K}]) = 3.6955 \pm 0.0013$ (Taylor 1999).\\
$^{\rm b}$: $\log(T_{\rm eff} [{\rm K}]) = 3.6985 \pm 0.0013$ (Taylor 1999).\\
$^{\rm c}$: $[{\rm Fe/H}] = +0.104 \pm 0.009$ (Taylor 1999).
\end{table*}

The giant $\delta^1$ Tau (HIP 20455) forms a common-proper-motion pair
with the turn-off-region object $\delta^2$ Tau (HIP 20542) at a
separation of $\sim$$0\fdg05$. The giant itself is a single-lined
spectroscopic binary (Griffin \& Gunn 1977; $P = 530$~days, $K_1 =
3$~km~s$^{-1}$; $1 = {\rm A} = {\rm primary}$; $2 = {\rm B} = {\rm
secondary}$); speckle observations have allowed the detection of a
companion at $0\farcs273$ (Mason et al.\ 1993; but see Patience et
al.\ 1998). Despite the duplicity of $\delta^1$ Tau, its secular
parallax solution is well defined ($\pi_{\rm sec,Hip} = 21.16 \pm
0.37$~mas with $g_{\rm Hip} = 0.07$ and $\pi_{\rm sec,Tycho-2} = 21.14
\pm 0.28$~mas with $g_{\rm Tycho-2} = 0.36$; cf.\ $\pi_{\rm Hip} =
21.29 \pm 0.93$~mas). Table \ref{tab:other_giants} provides $\log
T_{\rm eff}$, $\log(L/L_\odot)$, and $\log g$ for the primary
component using calibration (1) (\S \ref{subsubsec:trans_cali1}) for
$[{\rm Fe/H}] = +0.14$.

\begin{table*}[t]
\caption[]{Fundamental parameters of the primary components of
$\delta^1$ and $\theta^1$ Tau and $\delta$ Ari. The McWilliam
temperature errors have been chosen as $0.0025$. Taylor's (1999)
catalogue lists (homogeneously determined) mean metallicities and
temperatures based on literature results. The giant masses and
metallicities used in calibration (1) were taken from P98.}
\renewcommand{\arraystretch}{0.9}
\renewcommand{\tabcolsep}{15.0pt}
\begin{center}
\begin{tabular}{lrrrr}
\noalign{\vskip-0.30truecm}
\hline
\hline
\noalign{\vskip 0.07truecm}
$\delta^1$ Tau A (K0III)& Blackwell \&        & McWilliam (1990)    & Taylor (1999)       & Calibration (1)     \\
HIP 20455               & Lynas--Gray (1998)  &                     &                     & $[{\rm Fe/H}] = +0.14$\\
\noalign{\vskip 0.07truecm}
\hline
\noalign{\vskip 0.07truecm}
$\log(L/L_\odot)$       &                     &                     &                     & $ 1.835 \pm 0.016 $ \\
$\log(T_{\rm eff} [{\rm K}])$
                        & $3.6927 \pm 0.0039$ & $3.6937\pm0.0025$   & $ 3.6973\pm 0.0013$ & $ 3.6999\pm 0.0022$ \\
$\log(g [{\rm cm~s}^{-2}])$
                        &                     & $2.85$              &                     & $ 2.716 \pm 0.026 $ \\
$[{\rm Fe/H}]$
                        &                     & $+0.00 \pm 0.07$    & $+0.104 \pm 0.009 $ & $+0.14            $ \\
$M/M_\odot$             &                     &                     &                     & $ 2.30  \pm 0.10  $ \\
\noalign{\vskip 0.07truecm}
\hline
\noalign{\vskip 0.07truecm}
$\theta^1$ Tau A (K0III)& Torres et al.       &                     &                     &                     \\
HIP 20885               & (1997c)             &                     &                     &                     \\
\noalign{\vskip 0.07truecm}
\hline
\noalign{\vskip 0.07truecm}
$\log(L/L_\odot)$       & $1.98$              &                     &                     & $ 1.791 \pm 0.016 $ \\
$\log(T_{\rm eff} [{\rm K}])$$^{\rm a}$
                        & $3.6902 \pm 0.0089$ & $3.6955\pm0.0025$   & $ 3.7042\pm 0.0013$ & $ 3.7062\pm 0.0022$ \\
$\log(g [{\rm cm~s}^{-2}])$
                        & $2.63 \pm 0.07$     & $3.17$              &                     & $ 2.789 \pm 0.026 $ \\
$[{\rm Fe/H}]$
                        &                     & $+0.04 \pm 0.10$    & $+0.104 \pm 0.009 $ & $+0.14            $ \\
$M/M_\odot$             & $2.91 \pm 0.88$     &                     &                     & $ 2.32  \pm 0.10  $ \\
\noalign{\vskip 0.07truecm}
\hline
\noalign{\vskip 0.07truecm}
$\delta$ Ari (K2IIIvar) & Calibration (1)     &                     &                     &                     \\
HIP 14838               & $[{\rm Fe/H}] = 0$  &                     &                     &                     \\
\noalign{\vskip 0.07truecm}
\hline
\noalign{\vskip 0.07truecm}
$\log(L/L_\odot)$       & $1.673 \pm  0.013$  &                     &                     & $ 1.665 \pm 0.013 $ \\
$\log(T_{\rm eff} [{\rm K}])$
                        & $3.6837\pm 0.0023$  & $3.6821\pm0.0025$   & $ 3.6817\pm 0.0025$ & $ 3.6872\pm 0.0024$ \\
$\log(g [{\rm cm~s}^{-2}])$
                        & $2.815 \pm  0.025$  & $2.93$              &                     & $ 2.837 \pm 0.025 $ \\
$[{\rm Fe/H}]$
                        & $+0.00$             & $-0.03 \pm 0.09$    & $-0.012 \pm 0.049 $ & $+0.14            $ \\
$M/M_\odot$             & $ 2.31  \pm 0.10  $ &                     &                     & $ 2.31  \pm 0.10  $ \\
\hline\hline\\[-0.65truecm]
\end{tabular}
\label{tab:other_giants}
\end{center}

$^{\rm a}$:  $\log(T_{\rm eff} [{\rm K}]) = 3.6803$ (Ridgway et al.\ 1980) and $\log(T_{\rm eff} [{\rm K}]) = 3.6665 \pm 0.0075$ (Peterson et al.\ 1981).
\end{table*}

\begin{table*}[t]
\caption[]{Fundamental parameters of the components of the
spectroscopic binary $\theta^2$ Tau (HIP 20894; cf.\ Table
\ref{tab:orbpar}). The symbol $q \equiv M_2 / M_1 \leq 1$ denotes the
mass ratio of the two components ($1 = {\rm A} = {\rm primary}$; $2 =
{\rm B} = {\rm secondary}$). Peterson et al.\ (1993) based their
results on spectral line modelling using Kurucz (1991) models; their
component masses are likely underestimated. Kr\'olikowska (1992) used
stellar evolutionary models lacking convective core overshoot. Her
results depend sensitively on the adopted evolutionary status and
chemical composition of the primary; values listed here assume a
primary in the thick Hydrogen shell burning phase, $(X, Y, Z) = (0.70,
0.27, 0.03)$, and $\log(\tau/{\rm yr}) \sim 8.73$ (cf.\
$\log(\tau/{\rm yr}) = 8.80^{+0.05}_{-0.11}$ for $Z =
0.027^{+0.023}_{-0.011}$ [Lastennet et al.\ 1999]; $\log(\tau/{\rm
yr}) = 8.80 \pm 0.04$ for $Z = 0.024 \pm 0.003$ [P98]). A \& PF stands
for Antonello \& Pasinetti Fracassini (1998). We adopt calibration (1)
for $[{\rm Fe/H}] = +0.14$ using $M_1/M_\odot = 2.37 \pm 0.10$ and
$M_2/M_\odot = 1.95 \pm 0.10$ (Lastennet et al.\ 1999).}
\null\vskip-0.65truecm \renewcommand{\arraystretch}{0.9}
\renewcommand{\tabcolsep}{15.0pt}
\begin{center}
\begin{tabular}{lrrrr}
\noalign{\vskip 0.07truecm}
\hline
\hline
\noalign{\vskip 0.07truecm}
$\theta^2$ Tau A  & Peterson et
                                   & Kr\'olikowska (1992)  & A \& PF + Tom-        & Calibration (1)    \\
(A7III--IV)       & al.\ (1993)  & + Breger et al.\ (1987) &    kin et al.\ (1995) & $[{\rm Fe/H}] = +0.14$\\
\noalign{\vskip 0.07truecm}
\hline
\noalign{\vskip 0.07truecm}
$\log(L/L_\odot)$ &                     & $ 1.75  \pm 0.11  $ & $ 1.72  \pm 0.04  $ & $ 1.686 \pm 0.016 $\\
$\log(T_{\rm eff} [{\rm K}])^{\rm a}$
                  & $3.9243 \pm 0.0052$ & $3.9191 \pm 0.0052$ & $ 3.9138\pm 0.0053$ & $ 3.9020\pm 0.0029$\\
$\log(g [{\rm cm~s}^{-2}])$
                  & $ 3.9   \pm 0.05  $ & $ 3.8             $ & $ 3.701 \pm 0.058 $ & $ 3.687 \pm 0.027 $\\
$M/M_\odot$       & $ 1.71  \pm 0.20  $ & $ 2.63  \pm 0.10  $ & $ 2.33  \pm 0.21  $ & $ 2.37  \pm 0.10  $\\
\noalign{\vskip 0.07truecm}
\hline
\noalign{\vskip 0.07truecm}
$\theta^2$ Tau B (A5V)  &               &                     &                     &                    \\
\noalign{\vskip 0.07truecm}
\hline
\noalign{\vskip 0.07truecm}
$\log(L/L_\odot)$ &                     & $ 1.31  \pm 0.11  $ &                     & $ 1.256 \pm 0.016 $\\
$\log(T_{\rm eff} [{\rm K}])^{\rm a}$
                  & $3.9243 \pm 0.0052$ & $3.9243 \pm 0.0052$ &                     & $ 3.9154\pm 0.0031$\\
$\log(g [{\rm cm~s}^{-2}])$
                  & $ 4.3   \pm 0.05  $ & $ 4.0             $ &                     & $ 4.085 \pm 0.030 $\\
$M/M_\odot$       & $ 1.61  \pm 0.12  $ & $ 2.23  \pm 0.10  $ & $ 1.80  \pm 0.31  $ & $ 1.95  \pm 0.10  $\\
\noalign{\vskip 0.07truecm}
\hline
\noalign{\vskip 0.07truecm}
$q^{\rm b}$       & $ 0.94  \pm 0.13  $ & $ 0.85  \pm 0.05  $ & $ 0.76  \pm 0.14  $ & $ 0.82  \pm 0.05  $\\
$\pi_{\theta^2~{\rm Tau}}$~[mas]$^{\rm c}$
                  & $ 23.58 \pm 0.83  $ & $ 19.72 \pm 1.36  $ & $ 22.68 \pm 1.13  $ & $ 22.24 \pm 0.36  $\\
\noalign{\vskip 0.07truecm}
\hline\hline\\[-0.65truecm]
\end{tabular}
\end{center}

         $^{\rm a}$:  $\log(T_{\rm eff,1} [\rm K]) = 3.9138 \pm 0.0053$;
                      $\log(T_{\rm eff,2} [\rm K]) = 3.9191 \pm 0.0052$
                                          (Breger et al.\ 1987).\\
         $^{\rm b}$: $q = 0.82 \pm 0.05$ and $q \sim 0.88$ (Peterson 1991);\\
\phantom{$^{\rm b}$:} $q = 0.76 \pm 0.14$ ($M_1/M_\odot = 2.10 \pm 0.30$;
                                           $M_2/M_\odot = 1.60 \pm 0.20$;
                                           Tomkin et al.\ 1995;
                                           P98);\\
\phantom{$^{\rm b}$:} $q = 0.87 \pm 0.13$ ($M_1/M_\odot = 2.42 \pm 0.30$;
                                           $M_2/M_\odot = 2.11 \pm 0.17$;
                                           Torres et al.\ 1997c).\\
\phantom{}$^{\rm c}$: $\pi_{\theta^2~{\rm Tau}} = 22.88 \pm 0.99$~mas
                                           (Pan et al.\ 1992);\\
\phantom{$^{\rm c}$:} $\pi_{\theta^2~{\rm Tau}} = 22.68 \pm 0.87$~mas
                                           (Torres et al.\ 1997a);\\
\phantom{$^{\rm c}$:} $\pi_{\theta^2~{\rm Tau}} = 21.22 \pm 0.76$~mas
                                           (Torres et al.\ 1997c);\\
\phantom{$^{\rm c}$:} $\pi_{\theta^2~{\rm Tau}} = 21.89 \pm 0.83$~mas
                                           (ESA 1997).
\label{tab:theta2}
\end{table*}

The giant $\theta^1$ Tau (HIP 20885) forms a common-proper-motion pair
with the spectroscopic binary $\theta^2$ Tau (Appendix
\ref{subsec:trans_theta2}) at a separation of $\sim$$0\fdg10$
($\theta^1$ and $\theta^2$ Tau were independently observed by
Hipparcos as `single pointings'; field H58 = `1' for both systems).
The object $\theta^1$ Tau itself is a `resolved single-lined
spectroscopic binary' (e.g., Griffin \& Gunn 1977; Torres et al.\
1997c; $P = 5939$~days; $K_1 = 7.17$~km~s$^{-1}$). Peterson et al.\
(1981) derived a magnitude difference $\Delta V \sim 3.5$~mag by using
Lunar occultations; Mason et al.\ (1993) detected a speckle companion
at $0\farcs048$. Torres et al.\ (1997c) constructed an
astrometric-spectroscopic orbital solution for the binary; the
fundamental parameters of the primary component were derived by
assuming a distance of $47.6 \pm 1.9$~pc (following from the orbital
parallax of $\theta^2$ Tau). This distance, which corresponds to $\pi
= 21.01 \pm 0.84$~mas, is fully consistent with both the Hipparcos
trigonometric parallax ($\pi_{\rm Hip} = 20.66 \pm 0.85$~mas) and the
Hipparcos secular parallax ($\pi_{\rm sec,Hip} = 21.29 \pm 0.37$~mas),
although the latter is suspect, presumably as a result of duplicity
($g_{\rm Hip} = 48.26$; S\"oderhjelm (1999) lists $\pi_{\rm
trigonometric} = 21.3 \pm 1.0$~mas). Fundamental parameters for
$\theta^1$ Tau A are listed in Table~\ref{tab:other_giants}. A
comparison with literature values shows that the effective
temperatures for both $\delta^1$ and $\theta^1$ Tau are uncertain; we
decided to use the McWilliam (1990) effective temperatures and secular
parallax-based luminosities.

The giants $\theta^1$, $\delta^1$, $\epsilon$, and $\gamma$ Tau are
all located within $\sim$2.5~pc of the cluster center (i.e., in the
core region). The P98 member list contains one other red giant (HIP
14838; $\delta$ Ari), which is, however, located at $15.34$~pc from
the cluster center (e.g., Eggen 1983). Although its proper motion,
parallax, and radial velocity are consistent with membership ($c =
8.05$ in P98's table 2), the object was not confirmed as member by de
Bruijne (1999a) and Hoogerwerf \& Aguilar (1999) based on its proper
motion and parallax (\S \ref{subsec:addi_mems}; cf.\
Table~\ref{tab:data_1}). The secular parallax solution derived in \S
\ref{sec:sec_pars} is suspect ($g_{\rm Hip} = 66.87$; $\pi_{\rm Hip} =
19.44 \pm 1.23$ and $\pi_{\rm sec,Hip} = 20.28 \pm 0.28$~mas), which
is not surprising {\it if\/} the object is a non-member. However,
$\delta$ Ari is a `suspected non-single' star (i.e., Hipparcos field
H61 = `S') and a `$\Delta\mu$ binary' (Wielen et al.\ 2000), which
implies the Hipparcos astrometric parameters, and thus the secular
parallax $\pi_{\rm sec,Hip}$ and corresponding goodness-of-fit
parameter $g_{\rm Hip}$ as well as de Bruijne's conclusion on
membership, could be erroneous due to unmodelled orbital motion.
The goodness-of-fit parameter of the Tycho--2 secular parallax
($g_{\rm Tycho-2} = 74.87$), nonetheless, indicates that the long
time-baseline Tycho--2 proper motion is inconsistent with the mean
cluster motion. Although the WEB radial velocity catalogue (Duflot et
al.\ 1995) lists the star as single, de Medeiros \& Mayor (1999)
estimate the `probability that the radial velocity of the star is
constant' at 0.494 (based on 2 observations separated by 265 days); no
speckle measurements have been reported for $\delta$ Ari. The
metallicity of the star has not been determined unambiguously (e.g.,
$[{\rm Fe/H}] = -0.03 \pm 0.09$ [McWilliam 1990]; $-0.012 \pm 0.049$
[Taylor 1999]), although published values are more consistent with
sub-Solar than with the mean value of the Hyades ($[{\rm Fe/H}] =
+0.14$; \S \ref{subsec:Hyades_chars}). Table~\ref{tab:other_giants}
lists the parameters of $\delta$ Ari. In view of its near-Solar
metallicity, we also applied calibration (1) (\S
\ref{subsubsec:trans_cali1}) for $[{\rm Fe/H}] = 0$; we use these
parameters. The star is likely a non-member (\S \ref{subsec:HRD_4}).

\subsection{The spectroscopic binary $\theta^2$ Tau}
\label{subsec:trans_theta2}

Table~\ref{tab:theta2} summarizes results on the fundamental stellar
parameters of the `resolved double-lined spectroscopic binary'
$\theta^2$ Tau (HIP 20894; \S \ref{subsec:HRD_3}). The uncertain
semi-amplitude of the secondary velocity curve translates into a wide
range of published mass ratios $q$ (e.g., Peterson 1991, 1993; Torres
et al.\ 1997c). We applied calibration (1) (\S
\ref{subsubsec:trans_cali1}) for $ [{\rm Fe/H}] = +0.14$ to both
components separately (Table~\ref{tab:theta2}). Lunar occultations
have provided magnitude and colour differences: $\Delta V \equiv V_2 -
V_1 = 1.10 \pm 0.05$~mag; $\Delta(B-V) \equiv (B-V)_2 - (B-V)_1 =
-0.006 \pm 0.005$~mag ($1 = {\rm A} = {\rm primary}$; $2 = {\rm B} =
{\rm secondary}$; Peterson et al.\ 1981, 1993). The combined magnitude
$V = 3.40 \pm 0.02$~mag and colour $(B-V) = 0.179 \pm 0.004$~mag
therefore yield $V_1 = 3.736 \pm 0.020$~mag, $V_2 = 4.836 \pm
0.020$~mag, $(B-V)_1 = 0.170 \pm 0.010$~mag, and $(B-V)_2 = 0.160 \pm
0.010$~mag. We adopt the calibration (1) values for both components.

% Dirty solution; these are tables in Appendix A
\addtocounter{section}{-1}
\addtocounter{table}{-3}

% Make some phantom definitions
\def\s1{\phantom{}}
\def\r1{\phantom{0}}
\def\q1{\phantom{00}}
\def\p1{\phantom{000}}

\begin{landscape}
\setlength{\textwidth}{56pc}
\setlength{\textheight}{177.6mm}
\begin{table*}[t]
\caption[]{The 218 P98 Hyades candidates. Columns:
(1) Hipparcos number;
(2) Tycho identifier;
(3) Hipparcos trigonometric parallax $\pi_{\rm Hip}$ and associated error (mas);
(4) Hipparcos secular parallax $\pi_{\rm sec, Hip}$ and error (mas);
(5) Hipparcos goodness-of-fit parameter $g_{\rm Hip}$;
(6) Tycho--2 secular parallax $\pi_{\rm sec, Tycho-2}$ and error (mas);
(7) Tycho--2 goodness-of-fit parameter $g_{\rm Tycho-2}$;
(8) mass $M$ (in $M_\odot$; from \S 5.3 in P98; cf.\ \S \ref{subsubsec:stellar_masses});
(9) logarithm of effective temperature $T_{\rm eff}$ (in K) and associated error;
(10) bolometric correction ${\rm BC}_{V}$ and error (mag);
(11) logarithm of luminosity $\log(L/L_\odot)$ and error;
(12) logarithm of surface gravity $\log(g [{\rm cm}~{\rm s}^{-2}])$ and error;
(13) miscellaneous flags:
(a) `1' if an optical companion was not counted in the mass;
(b) `1' if the mass is particularly uncertain (from P98);
(c) `1' = spectroscopic binary (`SB'), `2' = radial velocity variable (`RV') (cf.\ column~(s) in P98's table~2);
(d) Hipparcos duplicity flag H56 (`I' = 1, `M' = 2, `H' = 3; cf.\ column~(t) in P98's table~2);
(e) Hipparcos duplicity flags H59 and H61 (`C' = 1, `G' = 2, `O' = 3, `V' = 4, `X' = 5, `S' = 6, `X/S' = 7; cf.\ column~(u) in P98's
     table~2);
(f) Hipparcos photometric variability flag H52 (`C' or `empty' = 0, `D' = 1, `M' = 2, `P' = 3, `R' = 4, `U' = 5);
(g) `1' or `2' if columns (10)--(13) were derived using calibration (1) or (2) (\S\S \ref{subsubsec:trans_cali1} and
     \ref{subsubsec:trans_cali2});
(h) P98 membership (column~(x) in their table~2;
`1' if based on proper motion and radial velocity,
`2' if based on proper motion only,
`3'/`6' if `1'/`2' but rejected by de Bruijne (1999a),
`4'/`7' if `1'/`2' but rejected by Hoogerwerf \& Aguilar (1999),
`5'/`8' if `1'/`2' but rejected by de Bruijne and Hoogerwerf \& Aguilar);
(i) index to notes (`1' if note is present).}
\null\vskip-0.60truecm
\begin{center}
{\scriptsize
\renewcommand{\arraystretch}{0.9}
\renewcommand{\tabcolsep}{5.7pt}
\begin{tabular}{lrrrrrrrrrrrrc}
\noalign{\vskip 0.07truecm}
\hline
\hline
\noalign{\vskip 0.07truecm}
\multicolumn{1}{c}{HIP}& \multicolumn{1}{c}{TYC}& \multicolumn{1}{c}{$\pi_{\rm Hip}$}& \multicolumn{2}{c}{$\pi_{\rm sec}$ \& $g$ [Hip]}& \multicolumn{2}{c}{$\pi_{\rm sec}$ \& $g$ [Tycho--2]}& \multicolumn{1}{c}{$M$}& \multicolumn{1}{c}{$\log T_{\rm eff}$}& \multicolumn{1}{c}{${\rm BC}_{V}$}& \multicolumn{1}{c}{$\log(L/L_\odot)$}& \multicolumn{1}{c}{$\log g$}& \multicolumn{1}{c}{a b c d e f g h i}\\[0.05truecm]
\noalign{\vskip 0.07truecm}
\hline
\noalign{\vskip 0.07truecm}
10672 &    44\s1  1162 1& $15.37\pm 1.29$& $ 16.98\pm0.28$&  12.95& $ 16.94\pm  0.24$&   5.52& 1.14& $3.7838\pm0.0054$& $-0.073\pm0.006$& $ 0.048\pm0.015$& $4.446\pm0.046$& 0 0 0 0 0 0 2 1 0\\
12031 &    49\r1   886 1& $13.44\pm 3.62$& $ 10.33\pm0.63$&   0.80& $ 10.00\pm  0.34$&  12.61& 1.00&                  &                 &                 &                & 0 1 0 0 0 5 0 7 1\\
12709 &  1226\r1   710 1& $53.89\pm 1.27$& $ 53.98\pm0.96$&  13.62& $ 52.88\pm  0.26$&  46.82& 0.77& $3.6759\pm0.0022$& $-0.430\pm0.015$& $-0.686\pm0.017$& $4.623\pm0.059$& 0 1 1 0 3 0 2 3 0\\
13042 &   643\r1   529 1& $11.18\pm17.11$& $  6.98\pm3.89$&   3.52& $  7.78\pm  0.37$& 405.66& 0.91&                  &                 &                 &                & 1 0 0 1 1 1 0 8 0\\
13117 &      \s1        & $29.67\pm 9.34$& $ 61.80\pm1.76$&  18.66&                  &       & 0.99&                  &                 &                 &                & 0 0 0 3 1 1 0 4 0\\
13600 &  1227\s1  1301 1& $18.89\pm 1.29$& $ 15.32\pm0.31$&  12.16& $ 15.05\pm  0.26$&   7.60& 1.00& $3.7486\pm0.0056$& $-0.147\pm0.010$& $ 0.055\pm0.018$& $4.532\pm0.052$& 0 0 0 0 0 0 2 4 0\\
13806 &  1794\r1   435 1& $25.77\pm 1.39$& $ 24.45\pm0.28$&   0.93& $ 24.36\pm  0.28$&   0.40& 0.89& $3.7157\pm0.0026$& $-0.244\pm0.010$& $-0.349\pm0.011$& $4.582\pm0.051$& 0 0 0 0 0 0 2 1 0\\
13834 &  1230\s1  1425 1& $31.41\pm 0.84$& $ 30.51\pm0.23$&   1.97& $ 30.38\pm  0.21$&   1.18& 1.39& $3.8270\pm0.0030$& $ 0.003\pm0.002$& $ 0.608\pm0.007$& $4.328\pm0.034$& 0 0 0 0 0 0 2 1 0\\
13976 &   641\r1   305 1& $42.66\pm 1.22$& $ 42.87\pm0.31$&   2.38& $ 43.05\pm  0.33$&   1.65& 0.87& $3.7019\pm0.0039$& $-0.286\pm0.018$& $-0.439\pm0.010$& $4.596\pm0.053$& 0 0 0 0 0 0 2 1 0\\
14838 &  1228\s1  2019 1& $19.44\pm 1.23$& $ 20.28\pm0.28$&  66.87& $ 20.84\pm  0.19$&  74.87& 2.31&                  &                 &                 &                & 0 0 0 0 6 0 0 5 1\\
14976 &  2344\s1  1916 1& $23.73\pm 1.18$& $ 25.05\pm0.27$&   1.46& $ 25.04\pm  0.27$&   0.32& 1.03& $3.7421\pm0.0042$& $-0.166\pm0.008$& $-0.093\pm0.010$& $4.545\pm0.046$& 0 0 0 0 0 0 2 1 0\\
15300 &  1792\r1   913 3& $29.49\pm 4.70$& $ 25.49\pm0.94$&   0.90& $ 26.11\pm  0.47$&   3.83& 1.04& $3.5749\pm0.0042$& $-1.399\pm0.039$& $-0.799\pm0.036$& $4.778\pm0.057$& 0 0 0 1 1 1 2 1 0\\
15304 &   649\r1   766 1& $20.20\pm 1.18$& $ 23.12\pm0.32$&  75.59& $ 23.19\pm  0.33$&  41.92& 1.76& $3.7838\pm0.0026$& $-0.073\pm0.003$& $ 0.247\pm0.012$& $4.446\pm0.029$& 1 0 0 1 0 1 2 5 0\\
15310 &   649\s1  1241 1& $21.64\pm 1.33$& $ 23.31\pm0.36$&  61.36& $ 23.62\pm  0.33$&  60.85& 1.14& $3.7704\pm0.0025$& $-0.098\pm0.003$& $ 0.091\pm0.013$& $4.481\pm0.042$& 1 0 0 1 0 0 2 5 0\\
15368 &   649\r1   815 3& $13.76\pm 5.62$& $  8.28\pm1.30$&   1.76& $  7.12\pm  0.54$&   0.00& 1.31&                  &                 &                 &                & 0 0 0 3 1 1 0 7 0\\
15374 &      \p1        & $24.54\pm 3.95$& $ 16.58\pm0.85$&  16.40&                  &       & 0.55& $3.5993\pm0.0030$& $-1.094\pm0.026$& $-0.867\pm0.046$& $4.712\pm0.092$& 0 0 0 0 0 0 2 7 0\\
15563 &   649\p1     5 1& $34.18\pm 1.70$& $ 32.17\pm0.38$&   2.91& $ 31.42\pm  0.40$&   0.09& 0.71& $3.6665\pm0.0030$& $-0.499\pm0.023$& $-0.777\pm0.014$& $4.635\pm0.064$& 0 0 0 0 0 0 2 1 0\\
15720 &  1805\r1   247 1& $29.75\pm 2.73$& $ 31.02\pm0.62$&   0.50& $ 30.78\pm  0.44$&   2.27& 0.58& $3.5673\pm0.0068$& $-1.505\pm0.085$& $-0.895\pm0.038$& $4.804\pm0.088$& 0 0 0 0 0 0 2 1 1\\
16377 &  4716\r1   414 1& $10.48\pm 1.61$& $  8.59\pm0.41$&  30.62& $  8.88\pm  0.31$&  75.54& 0.99&                  &                 &                 &                & 0 0 0 0 0 0 0 7 0\\
16529 &  1798\s1  1193 1& $22.78\pm 1.26$& $ 23.71\pm0.33$&   0.78& $ 23.90\pm  0.27$&   2.09& 0.94& $3.7179\pm0.0026$& $-0.237\pm0.010$& $-0.309\pm0.013$& $4.579\pm0.049$& 0 0 0 0 0 0 2 1 0\\
16548 &    67\r1   844 1& $17.20\pm 3.36$& $ 18.32\pm0.80$&  10.07& $ 16.55\pm  0.56$&  23.01& 0.62& $3.5847\pm0.0062$& $-1.260\pm0.052$& $-0.875\pm0.043$& $4.744\pm0.086$& 0 0 0 0 0 0 2 4 0\\
16908 &  1247\r1   684 1& $25.23\pm 1.58$& $ 21.26\pm0.38$&  11.80& $ 21.99\pm  0.26$&  16.68& 0.84& $3.7036\pm0.0026$& $-0.280\pm0.012$& $-0.385\pm0.016$& $4.594\pm0.055$& 0 0 0 0 0 0 2 4 0\\
17609 &  1804\s1  2036 1& $68.62\pm 1.78$& $ 66.40\pm0.43$& 190.48& $ 65.22\pm  0.42$& 173.34& 0.55& $3.5653\pm0.0084$& $-1.533\pm0.105$& $-0.977\pm0.042$& $4.811\pm0.096$& 0 0 0 0 0 0 2 5 0\\
17766 &    71\s1  1173 1& $24.02\pm 2.27$& $ 27.33\pm0.59$&   2.95& $ 27.62\pm  0.59$&   0.06& 0.66& $3.5999\pm0.0044$& $-1.087\pm0.038$& $-0.880\pm0.024$& $4.711\pm0.072$& 0 0 0 0 0 0 2 1 0\\
17950 &  4718\s1  1021 1& $22.22\pm 0.97$& $ 16.74\pm0.31$&  38.16& $ 16.77\pm  0.33$&   2.34& 2.13& $3.8211\pm0.0034$& $-0.004\pm0.003$& $ 0.865\pm0.016$& $4.341\pm0.029$& 0 0 0 1 1 1 2 7 0\\
17962 &  1252\r1   212 1& $21.37\pm 1.62$& $ 21.00\pm0.40$&   0.06& $ 20.56\pm  0.33$&   0.02& 0.87& $3.7309\pm0.0100$& $-0.198\pm0.028$& $-0.451\pm0.020$& $4.562\pm0.067$& 0 0 0 0 0 3 2 1 1\\
18018 &  1800\s1  1424 1& $24.72\pm 4.62$& $ 24.30\pm0.94$&   0.13&                  &       & 0.73& $3.6614\pm0.0081$& $-0.540\pm0.065$& $-0.725\pm0.043$& $4.642\pm0.080$& 0 1 0 0 5 0 2 1 1\\
18096 &  1253\r1   799 1& $11.19\pm 1.65$& $ 11.88\pm0.42$&  31.79& $ 11.85\pm  0.33$&  51.82& 1.11& $3.7718\pm0.0091$& $-0.096\pm0.011$& $ 0.047\pm0.031$& $4.478\pm0.062$& 0 0 0 0 0 0 2 4 0\\
18170 &  1253\r1   716 1& $24.14\pm 0.90$& $ 23.69\pm0.31$&   3.48& $ 23.74\pm  0.26$&   2.87& 1.49& $3.8451\pm0.0058$& $ 0.018\pm0.001$& $ 0.754\pm0.011$& $4.299\pm0.039$& 0 0 0 0 0 0 2 1 0\\
18322 &   662\r1   171 1& $26.49\pm 1.98$& $ 21.47\pm0.51$&   7.17& $ 21.34\pm  0.41$&   4.97& 0.72& $3.6766\pm0.0033$& $-0.425\pm0.022$& $-0.643\pm0.022$& $4.622\pm0.066$& 0 0 0 0 0 0 2 4 0\\
18327 &  1253\r1   868 1& $24.16\pm 1.40$& $ 24.43\pm0.41$&   0.92& $ 24.01\pm  0.35$&  11.01& 0.90& $3.7078\pm0.0027$& $-0.267\pm0.011$& $-0.366\pm0.015$& $4.590\pm0.052$& 0 0 0 0 0 0 2 1 0\\
18617 &  4718\r1   634 1& $10.38\pm 2.61$& $ 13.59\pm0.79$&  30.30& $ 12.46\pm  0.39$&  80.63& 1.68&                  &                 &                 &                & 0 0 0 2 1 0 0 8 0\\
18658 &   662\r1   633 1& $25.42\pm 1.05$& $ 23.56\pm0.71$&   6.22& $ 23.58\pm  0.40$&   0.40& 1.36& $3.8264\pm0.0036$& $ 0.002\pm0.003$& $ 0.613\pm0.026$& $4.329\pm0.044$& 0 1 0 0 2 0 2 1 0\\
18692 &   665\s1  1120 1& $10.93\pm 1.19$& $ 10.01\pm0.35$&  86.55& $ 10.11\pm  0.33$&  60.79& 1.34&                  &                 &                 &                & 0 0 0 0 0 0 0 4 0\\
18735 &  1254\r1   866 1& $21.99\pm 0.81$& $ 22.06\pm0.29$&   0.18& $ 22.19\pm  0.28$&   1.65& 1.57& $3.8554\pm0.0035$& $ 0.028\pm0.003$& $ 0.844\pm0.011$& $4.291\pm0.033$& 1 0 1 1 0 0 2 1 0\\
18946 &  1258\r1   768 1& $23.07\pm 2.12$& $ 20.98\pm0.55$&   2.19& $ 21.20\pm  0.42$&   0.01& 0.76& $3.6724\pm0.0262$& $-0.455\pm0.186$& $-0.611\pm0.078$& $4.627\pm0.142$& 0 0 0 0 0 0 2 1 0\\
19082 &  1258\r1   652 1& $14.56\pm 3.17$& $ 20.82\pm0.68$&   5.00& $ 21.58\pm  0.46$&   0.77& 0.72& $3.5959\pm0.0033$& $-1.131\pm0.028$& $-0.850\pm0.030$& $4.723\pm0.069$& 0 0 0 0 0 0 2 4 0\\
19098 &  1254\r1   178 1& $19.81\pm 1.39$& $ 22.10\pm0.42$&   2.92& $ 22.24\pm  0.34$&   0.03& 0.92& $3.7088\pm0.0027$& $-0.264\pm0.011$& $-0.409\pm0.017$& $4.589\pm0.051$& 0 0 0 0 0 0 2 1 0\\
19148 &  1250\r1   686 1& $21.41\pm 1.47$& $ 20.88\pm0.35$&   1.21& $ 20.93\pm  0.38$&   0.86& 1.14& $3.7769\pm0.0026$& $-0.086\pm0.003$& $ 0.153\pm0.015$& $4.464\pm0.042$& 0 0 0 0 0 0 2 1 0\\
19207 &  1250\p1     4 1& $23.57\pm 2.26$& $ 21.66\pm0.51$&   1.72& $ 22.11\pm  0.53$&   1.27& 0.70& $3.6580\pm0.0031$& $-0.566\pm0.026$& $-0.742\pm0.023$& $4.646\pm0.067$& 0 0 0 0 0 0 2 1 0\\
19261 &  1250\q1    74 1& $21.27\pm 1.03$& $ 22.62\pm0.35$&   2.22& $ 23.37\pm  0.60$&   1.12& 2.46& $3.8322\pm0.0032$& $ 0.009\pm0.002$& $ 0.778\pm0.013$& $4.316\pm0.026$& 0 0 0 1 1 1 2 1 0\\
19263 &  1250\r1   414 1& $19.70\pm 1.68$& $ 21.76\pm0.43$&   1.81& $ 21.87\pm  0.43$&   2.82& 0.83& $3.6879\pm0.0030$& $-0.357\pm0.019$& $-0.610\pm0.019$& $4.610\pm0.057$& 0 0 0 0 0 0 2 1 0\\
19316 &   674\q1    79 1& $24.90\pm 2.59$& $ 20.84\pm0.71$&   2.94& $ 21.21\pm  0.49$&   1.18& 0.60& $3.6158\pm0.0025$& $-0.883\pm0.022$& $-0.898\pm0.031$& $4.670\pm0.079$& 0 0 0 0 0 0 2 1 0\\
19365 &  1826\s1  1301 1& $10.68\pm 1.43$& $ 14.68\pm0.37$&   9.34& $ 14.82\pm  0.32$&   0.12& 2.56& $3.7651\pm0.0025$& $-0.108\pm0.004$& $ 0.300\pm0.022$& $4.495\pm0.029$& 1 0 0 1 0 0 2 4 0\\
19386 &    73\r1   698 1& $15.37\pm 0.97$& $ 11.82\pm0.40$&  29.64& $ 11.65\pm  0.32$&  10.87& 1.31&                  &                 &                 &                & 0 0 0 0 0 0 0 4 0\\
19441 &   666\q1    35 1& $29.78\pm 1.90$& $ 28.33\pm0.58$&   0.70& $ 27.83\pm  0.64$&   0.25& 0.69& $3.6558\pm0.0023$& $-0.583\pm0.019$& $-0.813\pm0.019$& $4.649\pm0.067$& 0 0 0 0 0 0 2 1 0\\
19449 &    76\r1   521 1& $12.14\pm 2.03$& $  8.75\pm0.70$&   5.22& $  8.64\pm  0.41$&   2.76& 0.97&                  &                 &                 &                & 0 0 0 0 0 0 0 7 0\\
19504 &  1255\r1   312 1& $23.22\pm 0.92$& $ 22.52\pm0.31$&   0.74& $ 22.06\pm  0.30$&   0.00& 1.36& $3.8235\pm0.0029$& $-0.002\pm0.002$& $ 0.550\pm0.012$& $4.336\pm0.036$& 0 0 0 0 0 1 2 1 0\\
19554 &    80\s1  1093 1& $25.89\pm 0.95$& $ 26.55\pm0.34$&   9.33& $ 26.83\pm  0.31$&   8.36& 1.51& $3.8433\pm0.0041$& $ 0.017\pm0.001$& $ 0.759\pm0.011$& $4.301\pm0.035$& 1 1 1 1 0 0 2 1 0\\
19591 &  1815\r1   309 1& $27.21\pm 2.11$& $ 25.23\pm0.54$&   0.91& $ 26.12\pm  0.38$&   0.00& 1.48& $3.6732\pm0.0033$& $-0.449\pm0.023$& $-0.478\pm0.021$& $4.626\pm0.038$& 0 0 1 2 1 3 2 1 0\\
\hline
\hline
\label{tab:data_1}
\end{tabular}
}
\end{center}
\end{table*}
\setlength{\textwidth}{177.6mm}
\setlength{\textheight}{56pc}
\end{landscape}

\begin{landscape}
\addtocounter{table}{-1}
\setlength{\textwidth}{56pc}
\setlength{\textheight}{177.6mm}
\begin{table*}[t]
\caption[]{Data for the 218 P98 Hyades candidates (continued).}
\begin{center}
{\scriptsize
\renewcommand{\arraystretch}{0.9}
\renewcommand{\tabcolsep}{5.7pt}
\begin{tabular}{lrrrrrrrrrrrrc}
\noalign{\vskip 0.07truecm}
\hline
\hline
\noalign{\vskip 0.07truecm}
\multicolumn{1}{c}{HIP}& \multicolumn{1}{c}{TYC}& \multicolumn{1}{c}{$\pi_{\rm Hip}$}& \multicolumn{2}{c}{$\pi_{\rm sec}$ \& $g$ [Hip]}& \multicolumn{2}{c}{$\pi_{\rm sec}$ \& $g$ [Tycho--2]}& \multicolumn{1}{c}{$M$}& \multicolumn{1}{c}{$\log T_{\rm eff}$}& \multicolumn{1}{c}{${\rm BC}_{V}$}& \multicolumn{1}{c}{$\log(L/L_\odot)$}& \multicolumn{1}{c}{$\log g$}& \multicolumn{1}{c}{a b c d e f g h i}\\[0.05truecm]
\noalign{\vskip 0.07truecm}
\hline
\noalign{\vskip 0.07truecm}
19781 &   679\r1   597 1& $21.91\pm 1.27$& $ 19.83\pm0.41$&   4.29& $ 20.20\pm  0.40$&   1.93& 1.01& $3.7513\pm0.0027$& $-0.141\pm0.005$& $-0.020\pm0.018$& $4.526\pm0.048$& 0 0 0 0 0 0 2 1 0\\
19786 &   675\r1   351 1& $22.19\pm 1.45$& $ 21.63\pm0.45$&   4.03& $ 21.41\pm  0.42$&   0.12& 1.07& $3.7643\pm0.0025$& $-0.109\pm0.004$& $ 0.052\pm0.018$& $4.497\pm0.045$& 0 0 0 0 0 0 2 1 0\\
19789 &  1263\r1   949 1& $18.12\pm 0.92$& $ 17.59\pm0.30$&   1.42& $ 17.21\pm  0.31$&   5.08& 1.36& $3.8244\pm0.0029$& $-0.001\pm0.002$& $ 0.588\pm0.015$& $4.334\pm0.037$& 0 0 0 0 0 0 2 1 1\\
19793 &  1815\r1   517 1& $21.69\pm 1.14$& $ 22.42\pm0.35$&   0.45& $ 22.40\pm  0.32$&   0.22& 1.08& $3.7602\pm0.0025$& $-0.119\pm0.004$& $ 0.025\pm0.014$& $4.506\pm0.044$& 0 0 0 0 0 0 2 1 0\\
19796 &   671\r1   211 1& $21.08\pm 0.97$& $ 22.04\pm0.38$&   4.02& $ 22.39\pm  0.46$&   2.17& 1.29& $3.7987\pm0.0027$& $-0.041\pm0.003$& $ 0.384\pm0.015$& $4.403\pm0.038$& 0 0 0 0 0 0 2 1 0\\
19808 &   675\r1   186 1& $22.67\pm 2.30$& $ 21.87\pm0.86$&   2.08& $ 22.84\pm  0.64$&   0.34& 0.69& $3.6532\pm0.0021$& $-0.601\pm0.018$& $-0.817\pm0.035$& $4.651\pm0.072$& 0 0 0 0 0 0 2 1 0\\
19834 &      \s1        & $31.94\pm 3.74$& $ 21.28\pm1.12$&   8.66&                  &       & 0.52& $3.5892\pm0.0040$& $-1.205\pm0.034$& $-0.899\pm0.048$& $4.737\pm0.098$& 0 0 0 0 0 0 2 4 0\\
19862 &      \s1        & $31.11\pm 2.76$& $ 21.81\pm0.77$&  12.83&                  &       & 0.59&                  &                 &                 &                & 0 0 0 0 0 0 0 4 0\\
19870 &  1263\s1  1000 1& $19.48\pm 0.99$& $ 20.64\pm0.34$&   1.61& $ 20.50\pm  0.33$&   1.85& 1.16& $3.7483\pm0.0024$& $-0.148\pm0.004$& $ 0.196\pm0.014$& $4.533\pm0.041$& 0 1 1 0 0 0 2 1 1\\
19877 &  1251\r1   128 1& $22.51\pm 0.82$& $ 21.54\pm0.32$&   1.56& $ 21.52\pm  0.31$&   0.03& 1.44& $3.8314\pm0.0048$& $ 0.008\pm0.003$& $ 0.705\pm0.013$& $4.318\pm0.038$& 1 0 0 1 0 0 2 1 1\\
19934 &  1276\q1    86 1& $19.48\pm 1.17$& $ 19.75\pm0.39$&   0.45& $ 19.55\pm  0.33$&   0.01& 0.95& $3.7240\pm0.0025$& $-0.218\pm0.009$& $-0.261\pm0.018$& $4.572\pm0.050$& 0 0 0 0 0 0 2 1 0\\
20019 &  1268\p1     6 1& $21.40\pm 1.24$& $ 21.16\pm0.38$&  11.02& $ 21.27\pm  0.36$&   2.59& 1.83& $3.7367\pm0.0025$& $-0.181\pm0.006$& $-0.008\pm0.016$& $4.553\pm0.030$& 0 0 1 0 0 3 2 1 1\\
20056 &  1268\r1   327 1& $21.84\pm 1.14$& $ 21.94\pm0.35$&   0.14& $ 21.73\pm  0.31$&   0.22& 2.05& $3.7543\pm0.0025$& $-0.134\pm0.004$& $ 0.257\pm0.014$& $4.520\pm0.027$& 0 0 1 0 0 0 2 1 0\\
20082 &  1264\r1   498 1& $20.01\pm 1.91$& $ 22.42\pm0.64$&  10.90& $ 22.39\pm  0.43$&   0.81& 0.87& $3.6923\pm0.0026$& $-0.333\pm0.015$& $-0.498\pm0.025$& $4.606\pm0.057$& 0 0 0 0 0 0 2 1 0\\
20086 &  1268\r1   707 1& $19.57\pm 1.86$& $ 23.56\pm0.51$&  28.15& $ 24.11\pm  0.47$&   5.75& 0.82& $3.6721\pm0.0022$& $-0.458\pm0.016$& $-0.667\pm0.020$& $4.628\pm0.057$& 0 1 0 0 6 0 2 4 0\\
20087 &  1276\s1  1622 1& $18.25\pm 0.82$& $ 18.31\pm0.69$&   0.19& $ 18.70\pm  0.29$&   0.00& 3.25& $3.8822\pm0.0036$& $ 0.028\pm0.000$& $ 1.103\pm0.033$& $4.327\pm0.038$& 0 0 1 1 3 0 1 1 1\\
20130 &  1272\r1   325 1& $23.53\pm 1.25$& $ 21.86\pm0.44$&   1.94& $ 21.46\pm  0.38$&   0.74& 0.97& $3.7392\pm0.0025$& $-0.174\pm0.005$& $-0.159\pm0.018$& $4.549\pm0.049$& 0 0 0 0 0 0 2 1 0\\
20146 &  1268\r1   352 1& $21.24\pm 1.32$& $ 21.62\pm0.42$&   0.23& $ 21.21\pm  0.38$&   0.61& 1.02& $3.7445\pm0.0039$& $-0.158\pm0.007$& $-0.096\pm0.017$& $4.541\pm0.049$& 0 0 0 0 0 0 2 1 0\\
20187 &   675\r1   372 1& $20.13\pm 2.02$& $ 31.69\pm0.69$&  36.59& $ 32.55\pm  0.49$&   2.84& 0.84&                  &                 &                 &                & 0 0 0 0 0 0 0 4 0\\
20205 &  1264\s1  1009 1& $21.17\pm 1.17$& $ 22.15\pm0.39$&   1.01& $ 22.26\pm  0.27$&   1.07& 2.30&                  &                 &                 &                & 0 0 0 0 0 0 0 1 1\\
20215 &  1264\r1   348 1& $23.27\pm 1.14$& $ 24.30\pm0.62$&   9.55& $ 22.51\pm  0.35$&   0.27& 2.22& $3.8000\pm0.0041$& $-0.038\pm0.004$& $ 0.402\pm0.022$& $4.399\pm0.034$& 0 0 1 1 1 1 2 1 1\\
20219 &   679\r1   750 1& $22.31\pm 0.92$& $ 22.28\pm0.35$&   0.24& $ 22.38\pm  0.37$&   0.09& 1.67& $3.8777\pm0.0035$& $ 0.034\pm0.001$& $ 0.955\pm0.014$& $4.169\pm0.033$& 1 0 0 1 0 0 1 1 0\\
20237 &  1272\r1   439 1& $22.27\pm 0.93$& $ 22.21\pm0.35$&   1.04& $ 22.08\pm  0.40$&   0.15& 1.19& $3.7858\pm0.0036$& $-0.068\pm0.004$& $ 0.249\pm0.014$& $4.440\pm0.042$& 0 0 0 0 0 0 2 1 0\\
20255 &  1268\s1  1268 1& $21.12\pm 0.77$& $ 23.55\pm0.32$&  68.15& $ 23.24\pm  0.34$&   9.98& 1.54& $3.8302\pm0.0032$& $ 0.006\pm0.002$& $ 0.708\pm0.012$& $4.321\pm0.033$& 0 1 1 0 0 0 2 4 1\\
20261 &  1264\s1  1010 1& $21.20\pm 0.99$& $ 21.02\pm0.35$&   0.20& $ 20.87\pm  0.32$&   0.39& 1.82& $3.8955\pm0.0032$& $ 0.036\pm0.000$& $ 1.132\pm0.015$& $4.099\pm0.031$& 0 0 0 0 0 0 1 1 0\\
20284 &   680\q1    27 1& $21.80\pm 0.85$& $ 20.59\pm0.35$&   2.73& $ 20.56\pm  0.36$&   1.90& 1.49& $3.8151\pm0.0032$& $-0.013\pm0.004$& $ 0.816\pm0.015$& $4.356\pm0.035$& 0 1 1 0 0 0 2 1 1\\
20319 &      \s1        & $11.64\pm 3.73$& $114.95\pm0.92$& 916.91&                  &       & 0.73&                  &                 &                 &                & 0 0 0 0 0 0 0 8 0\\
20349 &  1276\r1   251 1& $19.55\pm 0.89$& $ 20.21\pm0.36$&   2.49& $ 20.25\pm  0.32$&   2.13& 1.39& $3.8214\pm0.0044$& $-0.004\pm0.002$& $ 0.573\pm0.015$& $4.340\pm0.039$& 0 0 0 0 0 0 2 1 0\\
20350 &  1268\r1   295 1& $19.83\pm 0.89$& $ 21.38\pm0.34$&   3.47& $ 21.88\pm  0.35$&   0.30& 1.39& $3.8194\pm0.0041$& $-0.006\pm0.002$& $ 0.521\pm0.014$& $4.345\pm0.038$& 0 0 0 0 0 0 2 1 0\\
20357 &   680\r1   994 1& $19.46\pm 1.02$& $ 20.41\pm0.36$&   0.97& $ 20.32\pm  0.39$&   0.00& 1.44& $3.8279\pm0.0043$& $ 0.004\pm0.003$& $ 0.637\pm0.015$& $4.326\pm0.038$& 0 0 0 0 0 0 2 1 0\\
20400 &   680\r1   995 1& $21.87\pm 0.96$& $ 22.35\pm0.35$&   1.40& $ 22.30\pm  0.36$&   2.43& 1.64& $3.8566\pm0.0035$& $ 0.029\pm0.003$& $ 0.900\pm0.014$& $4.290\pm0.033$& 1 1 1 1 0 0 2 1 1\\
20419 &   676\r1   680 1& $19.17\pm 1.93$& $ 22.13\pm0.57$&   5.02& $ 22.67\pm  0.48$&   0.99& 0.85& $3.6574\pm0.0021$& $-0.570\pm0.017$& $-0.479\pm0.023$& $4.647\pm0.057$& 0 1 1 0 0 0 2 1 0\\
20440 &  1264\r1   755 1& $21.45\pm 2.76$& $ 21.72\pm0.65$&   0.97& $ 22.45\pm  0.50$&   2.31& 2.22& $3.7976\pm0.0041$& $-0.043\pm0.005$& $ 0.454\pm0.026$& $4.406\pm0.036$& 0 0 1 1 1 1 2 1 1\\
20441 &  1264\r1   758 1& $26.96\pm 1.40$& $ 25.47\pm2.27$&  16.17&                  &       & 1.04& $3.7553\pm0.0025$& $-0.131\pm0.004$& $-0.005\pm0.077$& $4.518\pm0.088$& 0 1 1 0 2 0 2 5 1\\
20455 &  1268\s1  1267 1& $21.29\pm 0.93$& $ 21.16\pm0.37$&   0.07& $ 21.14\pm  0.28$&   0.36& 2.30&                  &                 &                 &                & 1 1 1 1 0 0 0 1 1\\
20480 &  1276\r1   639 1& $20.63\pm 1.34$& $ 19.81\pm0.43$&   0.63& $ 19.80\pm  0.37$&   1.13& 0.97& $3.7363\pm0.0025$& $-0.182\pm0.006$& $-0.158\pm0.019$& $4.554\pm0.050$& 0 0 0 0 0 0 2 1 0\\
20482 &  1272\r1   912 1& $15.82\pm 1.44$& $ 19.23\pm0.49$&   5.96& $ 20.01\pm  0.41$&   1.64& 0.97& $3.7053\pm0.0027$& $-0.275\pm0.011$& $-0.316\pm0.023$& $4.593\pm0.051$& 0 1 1 0 3 0 2 4 0\\
20484 &  1264\s1  1008 1& $21.17\pm 0.80$& $ 20.86\pm0.36$&   0.78& $ 21.05\pm  0.34$&   0.01& 1.69& $3.8581\pm0.0034$& $ 0.031\pm0.003$& $ 0.991\pm0.015$& $4.290\pm0.033$& 0 1 1 0 0 0 2 1 1\\
20485 &  1264\r1   902 1& $21.08\pm 2.69$& $ 24.96\pm0.79$&   3.00& $ 24.43\pm  0.53$&   2.36& 0.74& $3.6472\pm0.0021$& $-0.642\pm0.018$& $-0.827\pm0.028$& $4.655\pm0.066$& 0 0 0 0 0 0 2 1 0\\
20491 &  1820\s1  1157 1& $20.04\pm 0.89$& $ 18.67\pm0.35$&   3.54& $ 18.46\pm  0.33$&   0.10& 1.29& $3.8134\pm0.0029$& $-0.016\pm0.001$& $ 0.490\pm0.016$& $4.361\pm0.039$& 0 0 0 0 0 0 2 1 0\\
20492 &   680\r1   194 1& $21.23\pm 1.80$& $ 21.03\pm0.52$&   1.34& $ 21.64\pm  0.43$&   2.27& 0.92& $3.7157\pm0.0047$& $-0.244\pm0.018$& $-0.294\pm0.023$& $4.582\pm0.056$& 0 0 0 0 0 0 2 1 0\\
20527 &   680\r1   889 1& $22.57\pm 2.78$& $ 22.61\pm0.75$&   3.56& $ 23.13\pm  0.61$&   0.21& 0.67& $3.6324\pm0.0024$& $-0.740\pm0.021$& $-0.870\pm0.030$& $4.666\pm0.072$& 0 0 0 0 0 0 2 1 0\\
20542 &  1269\s1  1244 1& $22.36\pm 0.88$& $ 22.14\pm0.38$&   4.20& $ 21.91\pm  0.32$&   1.91& 1.94& $3.9168\pm0.0031$& $ 0.026\pm0.005$& $ 1.275\pm0.015$& $4.069\pm0.030$& 1 1 0 1 0 0 1 1 1\\
20553 &   680\r1   146 1& $22.25\pm 1.52$& $ 20.49\pm0.46$&  69.87& $ 21.23\pm  0.38$&   4.71& 1.85& $3.7739\pm0.0039$& $-0.092\pm0.004$& $ 0.280\pm0.020$& $4.472\pm0.034$& 0 0 0 3 1 0 2 4 1\\
20557 &  1277\r1   747 1& $24.47\pm 1.06$& $ 23.65\pm0.40$&   0.76& $ 23.96\pm  0.35$&   0.39& 1.21& $3.7976\pm0.0038$& $-0.043\pm0.004$& $ 0.316\pm0.015$& $4.406\pm0.042$& 0 0 0 0 0 0 2 1 0\\
20563 &  1269\s1  1212 1& $19.35\pm 1.79$& $ 22.34\pm0.66$&   3.00& $ 22.20\pm  0.46$&   0.03& 0.83& $3.6799\pm0.0024$& $-0.401\pm0.015$& $-0.636\pm0.026$& $4.618\pm0.059$& 0 0 0 0 0 0 2 1 0\\
20567 &  1269\r1   806 1& $18.74\pm 1.17$& $ 19.86\pm0.43$&   2.22& $ 19.67\pm  0.47$&   0.17& 1.39& $3.8168\pm0.0053$& $-0.010\pm0.002$& $ 0.522\pm0.019$& $4.352\pm0.042$& 0 0 0 0 0 0 2 1 0\\
20577 &  1269\s1  1152 1& $20.73\pm 1.29$& $ 21.77\pm0.46$&   4.63& $ 22.30\pm  0.40$&  11.53& 1.16& $3.7753\pm0.0039$& $-0.089\pm0.004$& $ 0.142\pm0.018$& $4.468\pm0.044$& 0 1 2 0 0 0 2 1 1\\
20601 &    78\s1  1147 1& $14.97\pm 1.51$& $ 19.54\pm0.43$&  57.10& $ 19.50\pm  0.47$&  21.56& 1.61& $3.7363\pm0.0025$& $-0.182\pm0.006$& $-0.182\pm0.019$& $4.554\pm0.035$& 0 1 1 0 0 0 2 4 0\\
20605 &      \s1        & $24.41\pm 6.94$& $ 20.86\pm2.69$&   0.29&                  &       & 1.05& $3.5749\pm0.0056$& $-1.399\pm0.052$& $-0.845\pm0.114$& $4.778\pm0.123$& 0 0 0 3 1 1 2 1 0\\
20614 &  1273\s1  1106 1& $20.40\pm 0.74$& $ 21.65\pm0.33$&   6.79& $ 21.69\pm  0.34$&   0.05& 1.59& $3.8379\pm0.0045$& $ 0.013\pm0.001$& $ 0.834\pm0.013$& $4.308\pm0.035$& 0 0 0 0 0 0 2 1 1\\
20626 &  1820\q1    87 1& $15.92\pm 1.00$& $ 11.35\pm0.35$&  28.06& $ 11.26\pm  0.32$&  13.05& 1.26&                  &                 &                 &                & 0 0 0 0 0 0 0 7 0\\
20635 &  1277\s1  1626 1& $21.27\pm 0.80$& $ 21.14\pm0.32$&   0.04& $ 21.51\pm  0.31$&   0.06& 2.20& $3.9159\pm0.0030$& $ 0.035\pm0.005$& $ 1.548\pm0.013$& $3.848\pm0.027$& 1 0 0 1 0 0 1 1 0\\
20641 &  1277\s1  1627 1& $22.65\pm 0.84$& $ 22.56\pm0.33$&   0.09& $ 22.64\pm  0.31$&   0.68& 1.77& $3.8880\pm0.0032$& $ 0.035\pm0.000$& $ 1.067\pm0.013$& $4.123\pm0.030$& 1 0 0 1 0 0 1 1 0\\
20648 &  1269\s1  1246 1& $22.05\pm 0.77$& $ 21.80\pm0.36$&   0.46& $ 23.38\pm  0.50$&   0.06& 3.14& $3.9649\pm0.0061$& $-0.105\pm0.022$& $ 1.541\pm0.017$& $4.205\pm0.033$& 0 0 0 1 1 1 1 1 1\\
20661 &  1265\s1  1171 1& $21.47\pm 0.97$& $ 21.29\pm0.37$&   7.20& $ 21.08\pm  0.38$&   0.00& 2.47& $3.8000\pm0.0027$& $-0.038\pm0.003$& $ 0.681\pm0.015$& $4.399\pm0.026$& 0 0 1 1 1 0 2 1 1\\
20679 &  1269\r1   829 1& $20.79\pm 1.83$& $ 22.87\pm0.67$&   1.71& $ 20.85\pm  0.44$&   1.07& 1.64& $3.7003\pm0.0025$& $-0.294\pm0.013$& $-0.298\pm0.026$& $4.598\pm0.038$& 0 0 1 2 1 1 2 1 1\\
20686 &  1273\r1   138 1& $23.08\pm 1.22$& $ 22.13\pm0.51$&   1.07& $ 21.75\pm  0.38$&   0.03& 1.85& $3.7545\pm0.1736$& $-0.133\pm0.344$& $ 0.034\pm0.139$& $4.519\pm0.708$& 0 0 1 1 1 1 2 1 1\\
20711 &  1816\s1  1893 1& $21.07\pm 0.80$& $ 21.69\pm0.34$&   1.43& $ 22.13\pm  0.31$&   0.63& 2.17& $3.8768\pm0.0029$& $ 0.051\pm0.001$& $ 1.491\pm0.014$& $3.742\pm0.027$& 1 0 0 1 0 2 1 1 1\\
20712 &  1277\r1   808 1& $21.54\pm 0.97$& $ 20.94\pm0.39$&  10.46& $ 21.12\pm  0.34$&   5.18& 1.21& $3.7866\pm0.0033$& $-0.067\pm0.004$& $ 0.339\pm0.016$& $4.438\pm0.042$& 0 1 1 0 0 0 2 1 0\\
20713 &  1265\s1  1176 1& $20.86\pm 0.84$& $ 23.16\pm0.33$&  30.30& $ 22.48\pm  0.35$&   0.70& 2.12& $3.8798\pm0.0033$& $ 0.046\pm0.000$& $ 1.356\pm0.012$& $3.879\pm0.027$& 0 1 1 0 0 0 1 4 1\\
20719 &  1265\q1    59 1& $21.76\pm 1.46$& $ 20.41\pm0.43$&  35.22& $ 21.34\pm  0.44$&   3.64& 1.08& $3.7616\pm0.0025$& $-0.116\pm0.004$& $ 0.109\pm0.018$& $4.503\pm0.045$& 0 1 1 0 0 0 2 1 1\\
20741 &  1265\r1   241 1& $21.42\pm 1.54$& $ 22.18\pm0.46$&   0.41& $ 21.56\pm  0.40$&   0.37& 1.08& $3.7585\pm0.0025$& $-0.123\pm0.004$& $ 0.016\pm0.018$& $4.510\pm0.045$& 0 0 0 0 0 0 2 1 0\\
20745 &   676\r1   389 1& $28.27\pm 3.17$& $ 25.13\pm0.99$&   1.19& $ 25.22\pm  0.68$&   0.04& 1.19& $3.5906\pm0.0037$& $-1.189\pm0.032$& $-0.627\pm0.036$& $4.736\pm0.054$& 0 0 0 2 1 1 2 1 0\\
20751 &   672\r1   425 1& $23.03\pm 1.66$& $ 23.01\pm0.55$&  10.44& $ 22.33\pm  0.50$&   0.13& 0.85& $3.6829\pm0.0022$& $-0.384\pm0.014$& $-0.452\pm0.021$& $4.615\pm0.056$& 0 1 1 0 0 0 2 1 1\\
20762 &   680\r1   639 1& $21.83\pm 2.29$& $ 21.31\pm0.61$&   0.78& $ 21.61\pm  0.53$&   0.23& 0.73& $3.6638\pm0.0020$& $-0.521\pm0.016$& $-0.742\pm0.026$& $4.638\pm0.065$& 0 0 0 0 0 0 2 1 0\\
\hline
\hline
\end{tabular}
}
\end{center}
\end{table*}
\setlength{\textwidth}{177.6mm}
\setlength{\textheight}{56pc}
\end{landscape}

\begin{landscape}
\addtocounter{table}{-1}
\setlength{\textwidth}{56pc}
\setlength{\textheight}{177.6mm}
\begin{table*}[t]
\caption[]{Data for the 218 P98 Hyades candidates (continued).}
\begin{center}
{\scriptsize
\renewcommand{\arraystretch}{0.9}
\renewcommand{\tabcolsep}{5.7pt}
\begin{tabular}{lrrrrrrrrrrrrc}
\noalign{\vskip 0.07truecm}
\hline
\hline
\noalign{\vskip 0.07truecm}
\multicolumn{1}{c}{HIP}& \multicolumn{1}{c}{TYC}& \multicolumn{1}{c}{$\pi_{\rm Hip}$}& \multicolumn{2}{c}{$\pi_{\rm sec}$ \& $g$ [Hip]}& \multicolumn{2}{c}{$\pi_{\rm sec}$ \& $g$ [Tycho--2]}& \multicolumn{1}{c}{$M$}& \multicolumn{1}{c}{$\log T_{\rm eff}$}& \multicolumn{1}{c}{${\rm BC}_{V}$}& \multicolumn{1}{c}{$\log(L/L_\odot)$}& \multicolumn{1}{c}{$\log g$}& \multicolumn{1}{c}{a b c d e f g h i}\\[0.05truecm]
\noalign{\vskip 0.07truecm}
\hline
\noalign{\vskip 0.07truecm}
20815 &  1265\s1  1048 1& $21.83\pm 1.01$& $ 21.21\pm0.38$&   1.05& $ 21.27\pm  0.48$&   0.00& 1.21& $3.7923\pm0.0040$& $-0.055\pm0.005$& $ 0.303\pm0.016$& $4.421\pm0.042$& 0 0 0 0 0 0 2 1 0\\
20826 &   676\q1    62 1& $21.18\pm 1.04$& $ 22.34\pm0.42$&   1.50& $ 22.44\pm  0.45$&   0.10& 1.21& $3.7858\pm0.0026$& $-0.068\pm0.003$& $ 0.232\pm0.016$& $4.440\pm0.041$& 0 0 0 0 0 0 2 1 0\\
20827 &   680\r1   104 1& $17.29\pm 2.23$& $ 20.52\pm0.55$&   2.72& $ 20.80\pm  0.48$&   0.10& 0.93& $3.7014\pm0.0026$& $-0.289\pm0.012$& $-0.402\pm0.024$& $4.596\pm0.053$& 0 0 0 0 0 0 2 1 0\\
20842 &  1277\s1  1628 1& $20.85\pm 0.86$& $ 20.24\pm0.35$&   0.55& $ 20.39\pm  0.32$&   1.23& 1.67& $3.8821\pm0.0051$& $ 0.035\pm0.000$& $ 0.982\pm0.015$& $4.159\pm0.036$& 1 0 0 1 0 0 1 1 0\\
20850 &   680\r1   423 1& $21.29\pm 1.91$& $ 21.61\pm0.48$&   0.61& $ 21.25\pm  0.45$&   1.22& 0.94& $3.7189\pm0.0026$& $-0.234\pm0.010$& $-0.285\pm0.020$& $4.578\pm0.051$& 0 0 0 0 0 1 2 1 0\\
20873 &   681\s1  1152 1& $18.42\pm 1.93$& $ 22.06\pm0.62$&   3.87& $ 21.46\pm  0.37$&   1.92& 1.69& $3.8536\pm0.0045$& $ 0.027\pm0.003$& $ 0.841\pm0.024$& $4.292\pm0.040$& 0 1 0 0 5 0 2 1 0\\
20885 &  1265\s1  1170 1& $20.66\pm 0.85$& $ 21.29\pm0.37$&  48.26& $ 21.93\pm  0.35$&   1.24& 2.32&                  &                 &                 &                & 1 1 1 1 0 0 0 1 1\\
20889 &  1273\s1  1104 1& $21.04\pm 0.82$& $ 21.93\pm0.36$&   1.38& $ 21.83\pm  0.28$&   0.19& 2.30&                  &                 &                 &                & 1 0 0 1 0 0 0 1 1\\
20890 &  1273\r1   711 1& $20.09\pm 1.11$& $ 20.92\pm0.45$&   4.24& $ 20.54\pm  0.40$&   5.97& 1.02& $3.7401\pm0.0035$& $-0.171\pm0.007$& $-0.122\pm0.019$& $4.548\pm0.049$& 0 1 1 0 0 0 2 1 0\\
20894 &  1265\s1  1172 1& $21.89\pm 0.83$& $ 22.24\pm0.36$&   0.26& $ 22.35\pm  0.36$&   0.22& 4.32&                  &                 &                 &                & 1 0 1 1 0 2 0 1 1\\
20899 &  1269\q1    22 1& $21.09\pm 1.08$& $ 21.62\pm0.40$&   0.28& $ 22.47\pm  0.40$&   1.68& 1.14& $3.7726\pm0.0025$& $-0.094\pm0.003$& $ 0.134\pm0.016$& $4.475\pm0.043$& 0 0 0 0 0 0 2 1 0\\
20901 &   677\s1  1116 1& $20.33\pm 0.84$& $ 21.37\pm0.35$&   2.29& $ 21.49\pm  0.43$&   0.69& 1.94& $3.8977\pm0.0032$& $ 0.038\pm0.000$& $ 1.213\pm0.014$& $4.055\pm0.029$& 0 0 0 0 0 0 1 1 1\\
20916 &  1265\r1   791 1& $20.58\pm 1.74$& $ 18.75\pm0.57$&   2.93& $ 21.44\pm  0.35$&   1.14& 2.47& $3.7925\pm0.0029$& $-0.054\pm0.003$& $ 0.738\pm0.026$& $4.421\pm0.034$& 0 1 1 1 1 1 2 1 1\\
20935 &  1269\r1   294 1& $23.25\pm 1.04$& $ 21.85\pm0.39$&   2.18& $ 21.21\pm  0.41$&   0.06& 1.26& $3.7953\pm0.0027$& $-0.048\pm0.003$& $ 0.431\pm0.015$& $4.412\pm0.039$& 0 1 1 0 3 0 2 1 0\\
20948 &  1269\r1   557 1& $21.59\pm 1.09$& $ 21.83\pm0.41$&   0.13& $ 21.72\pm  0.37$&   0.01& 1.31& $3.8165\pm0.0029$& $-0.011\pm0.001$& $ 0.465\pm0.016$& $4.352\pm0.039$& 1 0 0 1 0 0 2 1 0\\
20949 &  1837\r1   270 1& $17.08\pm 1.18$& $ 17.28\pm0.39$&   2.95& $ 17.52\pm  0.37$&   0.40& 0.98& $3.7345\pm0.0025$& $-0.188\pm0.006$& $-0.178\pm0.020$& $4.557\pm0.050$& 0 0 0 0 0 0 2 1 0\\
20951 &  1269\r1   697 1& $24.19\pm 1.76$& $ 22.30\pm0.56$&   1.21& $ 21.93\pm  0.39$&   0.23& 0.91& $3.7205\pm0.0026$& $-0.229\pm0.009$& $-0.287\pm0.022$& $4.576\pm0.054$& 1 0 0 1 0 0 2 1 0\\
20978 &  1265\r1   883 1& $24.71\pm 1.27$& $ 22.05\pm0.42$&   4.76& $ 21.73\pm  0.40$&   2.22& 0.88& $3.7137\pm0.0034$& $-0.250\pm0.013$& $-0.321\pm0.017$& $4.584\pm0.054$& 0 0 0 0 0 0 2 1 0\\
20995 &  1265\s1  1175 1& $22.93\pm 1.25$& $ 22.31\pm0.42$&   1.72& $ 23.04\pm  0.52$&   0.82& 2.65& $3.8539\pm0.0039$& $ 0.027\pm0.003$& $ 0.959\pm0.016$& $4.291\pm0.028$& 0 0 1 1 1 1 2 1 1\\
21008 &  1273\r1   583 1& $19.94\pm 0.93$& $ 19.10\pm0.38$&  14.49& $ 19.64\pm  0.36$&   0.56& 1.31& $3.8112\pm0.0042$& $-0.019\pm0.002$& $ 0.508\pm0.017$& $4.367\pm0.041$& 0 1 1 0 0 0 2 1 1\\
21029 &  1265\s1  1169 1& $22.54\pm 0.77$& $ 21.86\pm0.33$&   2.42& $ 21.69\pm  0.35$&   1.22& 1.94& $3.9110\pm0.0031$& $ 0.035\pm0.002$& $ 1.291\pm0.013$& $4.030\pm0.029$& 1 0 0 1 0 0 1 1 0\\
21036 &   681\s1  1153 1& $21.84\pm 0.89$& $ 22.42\pm0.36$&   0.53& $ 22.64\pm  0.38$&   0.04& 1.74& $3.8843\pm0.0033$& $ 0.035\pm0.000$& $ 1.021\pm0.014$& $4.146\pm0.032$& 1 0 0 1 0 0 1 1 1\\
21039 &  1265\s1  1174 1& $22.55\pm 1.09$& $ 21.72\pm0.39$&   1.17& $ 21.84\pm  0.34$&   0.71& 1.69& $3.8858\pm0.0033$& $ 0.035\pm0.000$& $ 1.020\pm0.016$& $4.140\pm0.033$& 1 1 1 1 0 0 1 1 1\\
21053 &  1265\r1   763 1& $24.28\pm 0.79$& $ 20.93\pm0.35$&  24.82& $ 21.67\pm  0.38$&   0.00& 1.36& $3.8232\pm0.0029$& $-0.002\pm0.002$& $ 0.658\pm0.015$& $4.337\pm0.037$& 1 0 0 1 0 0 2 4 1\\
21066 &   673\r1   700 1& $22.96\pm 0.99$& $ 21.89\pm0.39$&   1.66& $ 21.40\pm  0.53$&   1.31& 1.26& $3.8106\pm0.0037$& $-0.020\pm0.001$& $ 0.414\pm0.015$& $4.369\pm0.041$& 0 0 0 0 0 0 2 1 0\\
21092 &      \s1        & $19.64\pm 9.61$& $ 33.98\pm2.97$&   3.45&                  &       & 1.00&                  &                 &                 &                & 0 0 0 3 1 0 0 7 0\\
21099 &  1273\r1   428 1& $21.81\pm 1.25$& $ 21.74\pm0.47$&   1.05& $ 22.16\pm  0.38$&   0.20& 0.99& $3.7416\pm0.0035$& $-0.167\pm0.007$& $-0.145\pm0.019$& $4.545\pm0.050$& 0 0 0 0 0 0 2 1 0\\
21112 &   681\r1   829 1& $19.46\pm 1.02$& $ 19.62\pm0.43$&   2.29& $ 18.82\pm  0.45$&   0.32& 1.19& $3.7914\pm0.0050$& $-0.057\pm0.006$& $ 0.224\pm0.019$& $4.424\pm0.046$& 0 0 0 0 0 0 2 1 0\\
21123 &  1269\s1  1027 1& $23.41\pm 1.65$& $ 22.15\pm0.49$&   1.38& $ 22.80\pm  0.47$&   1.13& 0.83& $3.6910\pm0.0059$& $-0.339\pm0.035$& $-0.469\pm0.024$& $4.607\pm0.062$& 0 1 1 0 3 0 2 1 0\\
21137 &  1265\s1  1173 1& $22.25\pm 1.14$& $ 22.86\pm0.35$&  12.43& $ 22.78\pm  0.36$&   4.27& 1.54& $3.8498\pm0.0041$& $ 0.022\pm0.001$& $ 0.767\pm0.013$& $4.294\pm0.035$& 0 1 1 0 0 0 2 1 1\\
21138 &  1265\r1   924 1& $15.11\pm 4.75$& $ 21.12\pm1.17$&   2.27&                  &       & 0.76& $3.6347\pm0.0034$& $-0.726\pm0.029$& $-0.869\pm0.050$& $4.664\pm0.077$& 0 0 0 0 0 0 2 4 1\\
21152 &    90\q1    33 1& $23.13\pm 0.92$& $ 23.87\pm0.41$&   2.77& $ 23.79\pm  0.48$&   2.69& 1.41& $3.8256\pm0.0042$& $ 0.001\pm0.003$& $ 0.594\pm0.015$& $4.331\pm0.038$& 0 0 0 0 0 0 2 1 0\\
21179 &   677\r1   290 1& $17.55\pm 2.97$& $ 21.89\pm0.99$&   2.53& $ 21.85\pm  0.68$&   1.57& 0.72& $3.6553\pm0.0021$& $-0.586\pm0.017$& $-0.948\pm0.040$& $4.649\pm0.073$& 0 0 1 0 0 3 2 1 0\\
21256 &  1278\s1  1315 1& $24.98\pm 1.95$& $ 23.21\pm0.62$&   0.87& $ 23.43\pm  0.44$&   0.43& 0.67& $3.6459\pm0.0022$& $-0.651\pm0.019$& $-0.849\pm0.024$& $4.656\pm0.070$& 0 0 0 0 0 0 2 1 0\\
21261 &  1274\s1  1346 1& $21.06\pm 2.21$& $ 21.61\pm0.68$&   0.30& $ 21.23\pm  0.47$&   0.11& 0.70& $3.6547\pm0.0021$& $-0.590\pm0.018$& $-0.831\pm0.028$& $4.650\pm0.069$& 0 0 0 0 0 0 2 1 0\\
21267 &   681\r1   651 1& $22.80\pm 0.98$& $ 21.87\pm0.43$&   1.02& $ 21.92\pm  0.40$&   0.02& 1.36& $3.8229\pm0.0035$& $-0.002\pm0.002$& $ 0.572\pm0.017$& $4.337\pm0.039$& 0 0 0 0 0 0 2 1 0\\
21273 &   681\s1  1151 1& $21.39\pm 1.24$& $ 22.41\pm0.51$&   2.66& $ 22.27\pm  0.35$&   2.85& 2.02& $3.8824\pm0.0030$& $ 0.044\pm0.000$& $ 1.317\pm0.020$& $3.907\pm0.032$& 0 1 1 0 3 0 1 1 1\\
21280 &  1266\r1   398 1& $24.02\pm 1.68$& $ 22.93\pm0.66$&  24.24& $ 22.04\pm  0.38$&   4.96& 1.69& $3.7173\pm0.0101$& $-0.239\pm0.038$& $-0.119\pm0.029$& $4.580\pm0.056$& 0 0 1 2 1 0 2 1 1\\
21317 &  1266\r1   278 1& $23.19\pm 1.30$& $ 22.01\pm0.46$&   3.42& $ 21.18\pm  0.42$&   1.81& 1.08& $3.7667\pm0.0034$& $-0.105\pm0.005$& $ 0.095\pm0.018$& $4.491\pm0.046$& 0 0 0 0 0 0 2 1 0\\
21459 &  1830\s1  2128 1& $22.60\pm 0.76$& $ 23.61\pm0.36$&   2.97& $ 23.45\pm  0.34$&   1.30& 1.51& $3.8373\pm0.3266$& $ 0.013\pm0.936$& $ 0.743\pm0.374$& $4.309\pm1.359$& 0 0 0 0 0 0 2 1 1\\
21474 &  1266\s1  1214 1& $22.99\pm 0.95$& $ 20.59\pm0.42$&   8.36& $ 21.17\pm  0.43$&   8.13& 1.34& $3.8191\pm0.0052$& $-0.007\pm0.004$& $ 0.618\pm0.018$& $4.346\pm0.043$& 0 1 1 0 0 0 2 1 0\\
21482 &  1838\r1   564 1& $56.02\pm 1.21$& $ 51.94\pm0.40$&  14.79& $ 51.20\pm  0.33$&   0.35& 0.78& $3.6709\pm0.0023$& $-0.466\pm0.016$& $-0.586\pm0.009$& $4.629\pm0.057$& 0 1 1 0 0 5 2 3 0\\
21543 &  1266\r1   112 1& $23.54\pm 1.29$& $ 19.72\pm1.26$&  13.22& $ 22.38\pm  0.36$&   0.00& 1.14& $3.7758\pm0.0031$& $-0.088\pm0.004$& $ 0.332\pm0.056$& $4.467\pm0.069$& 0 1 1 0 2 0 2 4 1\\
21588 &  1266\s1  1417 1& $21.96\pm 1.04$& $ 25.54\pm0.46$&  34.53& $ 19.96\pm  0.36$&   0.25& 1.62& $3.8575\pm0.0037$& $ 0.030\pm0.003$& $ 0.760\pm0.016$& $4.290\pm0.034$& 0 1 0 1 2 0 2 4 1\\
21589 &   690\s1  1547 1& $21.79\pm 0.79$& $ 22.30\pm0.40$&   1.20& $ 22.44\pm  0.44$&   0.06& 2.15& $3.9225\pm0.0034$& $ 0.018\pm0.009$& $ 1.484\pm0.016$& $3.928\pm0.029$& 1 0 1 1 0 0 1 1 1\\
21637 &  1830\s1  1020 1& $22.60\pm 0.91$& $ 23.30\pm0.38$&   1.28& $ 23.43\pm  0.35$&   0.06& 1.16& $3.7814\pm0.0026$& $-0.078\pm0.003$& $ 0.191\pm0.014$& $4.452\pm0.041$& 0 0 0 0 0 0 2 1 0\\
21654 &   694\r1   225 1& $20.81\pm 1.30$& $ 22.80\pm0.44$&   2.59& $ 22.40\pm  0.46$&   0.55& 1.11& $3.7606\pm0.0025$& $-0.118\pm0.004$& $ 0.046\pm0.017$& $4.505\pm0.044$& 0 1 1 0 0 0 2 1 0\\
21670 &   682\s1  1726 1& $19.44\pm 0.86$& $ 20.46\pm0.42$&   8.75& $ 20.71\pm  0.49$&   2.71& 1.84& $3.8852\pm0.0032$& $ 0.037\pm0.000$& $ 1.107\pm0.018$& $4.088\pm0.032$& 1 0 0 1 0 0 1 1 1\\
21683 &  1266\s1  1418 1& $20.51\pm 0.82$& $ 18.33\pm0.42$&  11.92& $ 18.70\pm  0.38$&   8.74& 2.05& $3.9133\pm0.0029$& $ 0.039\pm0.004$& $ 1.486\pm0.020$& $3.868\pm0.031$& 1 0 0 1 0 0 1 1 1\\
21723 &   690\r1   945 1& $23.95\pm 1.63$& $ 22.95\pm0.68$&   0.44& $ 22.98\pm  0.56$&   0.04& 0.76& $3.6761\pm0.0022$& $-0.429\pm0.015$& $-0.668\pm0.026$& $4.623\pm0.064$& 0 0 0 0 0 0 2 1 0\\
21741 &  1830\s1  1358 1& $15.96\pm 1.36$& $ 16.58\pm0.44$&   0.73& $ 17.13\pm  0.39$&   2.45& 0.97& $3.7245\pm0.0025$& $-0.217\pm0.009$& $-0.214\pm0.023$& $4.572\pm0.051$& 0 0 0 0 0 0 2 1 0\\
21762 &  1267\r1   210 1& $23.65\pm 2.53$& $ 21.15\pm1.03$&   1.97& $ 21.49\pm  0.49$&   0.04& 1.96& $3.6722\pm0.0022$& $-0.457\pm0.015$& $-0.358\pm0.043$& $4.627\pm0.049$& 0 0 1 2 1 1 2 1 0\\
21788 &  1830\r1   545 1& $19.48\pm 1.26$& $ 16.55\pm0.46$&  63.40& $ 15.59\pm  0.37$&  93.20& 0.99& $3.7525\pm0.0025$& $-0.138\pm0.004$& $-0.020\pm0.024$& $4.524\pm0.051$& 0 0 0 0 0 0 2 4 0\\
22044 &   687\s1  1627 1& $20.73\pm 0.88$& $ 22.73\pm0.46$&   6.72& $ 23.03\pm  0.45$&   0.38& 1.79& $3.8888\pm0.0033$& $ 0.033\pm0.000$& $ 1.014\pm0.018$& $4.184\pm0.033$& 1 0 0 1 0 0 1 1 0\\
22157 &   691\s1  1509 1& $12.24\pm 0.86$& $ 15.35\pm0.44$&  23.34& $ 15.59\pm  0.46$&   2.83& 2.22& $3.9013\pm0.0031$& $ 0.042\pm0.001$& $ 1.367\pm0.025$& $3.974\pm0.034$& 1 0 1 1 0 0 1 4 1\\
22177 &    88\s1  1355 1& $22.45\pm 2.32$& $ 22.01\pm0.91$&   0.07& $ 21.85\pm  0.68$&   0.01& 0.67& $3.6355\pm0.0024$& $-0.720\pm0.021$& $-0.867\pm0.037$& $4.664\pm0.075$& 0 0 0 0 0 0 2 1 0\\
22203 &  1267\s1  1102 1& $19.42\pm 1.09$& $ 21.28\pm0.51$&   4.05& $ 20.74\pm  0.49$&   1.08& 1.08& $3.7582\pm0.0025$& $-0.124\pm0.004$& $-0.028\pm0.021$& $4.511\pm0.046$& 0 1 1 0 0 0 2 1 0\\
22221 &   683\r1   688 1& $26.26\pm 1.04$& $ 22.75\pm0.55$&  51.38& $ 25.22\pm  0.66$&   0.01& 1.16& $3.7858\pm0.0025$& $-0.068\pm0.003$& $ 0.316\pm0.021$& $4.440\pm0.044$& 0 1 1 0 2 5 2 4 1\\
22224 &  1271\s1  1301 1& $24.11\pm 1.72$& $ 22.88\pm0.71$&   1.48& $ 22.66\pm  0.49$&   0.10& 0.82& $3.6946\pm0.0026$& $-0.322\pm0.015$& $-0.532\pm0.028$& $4.603\pm0.061$& 0 1 1 0 0 0 2 1 0\\
22253 &  1831\s1  1894 1& $15.74\pm 1.98$& $ 18.30\pm0.56$&   1.97& $ 18.96\pm  0.54$&   0.42& 0.79& $3.6695\pm0.0022$& $-0.476\pm0.016$& $-0.712\pm0.027$& $4.631\pm0.062$& 0 0 0 0 0 0 2 1 0\\
22265 &  1271\s1  1229 1& $19.81\pm 1.43$& $ 20.01\pm0.63$&  10.16& $ 20.42\pm  0.46$&   1.77& 1.04& $3.7447\pm0.0025$& $-0.158\pm0.004$& $-0.057\pm0.027$& $4.540\pm0.051$& 0 1 1 0 0 0 2 1 0\\
22271 &  1835\r1   809 1& $22.07\pm 2.03$& $ 27.33\pm0.58$&  10.63& $ 27.78\pm  0.57$&  11.22& 0.71& $3.6590\pm0.0021$& $-0.558\pm0.017$& $-0.996\pm0.020$& $4.645\pm0.065$& 0 0 0 0 0 0 2 4 0\\
22350 &  1292\r1   770 1& $19.30\pm 1.67$& $ 20.60\pm1.55$&   1.02& $ 21.99\pm  0.40$&   2.35& 0.96& $3.7181\pm0.0026$& $-0.236\pm0.010$& $-0.259\pm0.066$& $4.579\pm0.080$& 0 1 1 0 2 0 2 1 1\\
22380 &  1284\s1  1397 1& $21.38\pm 1.46$& $ 20.91\pm0.59$&   7.61& $ 22.05\pm  0.48$&   2.28& 0.94& $3.7201\pm0.0026$& $-0.230\pm0.009$& $-0.242\pm0.025$& $4.577\pm0.053$& 0 0 0 0 0 0 2 1 1\\
22394 &  1835\r1   658 1& $18.96\pm 1.62$& $ 20.37\pm0.50$&   1.84& $ 20.82\pm  0.45$&   0.80& 1.52& $3.6796\pm0.0022$& $-0.403\pm0.014$& $-0.362\pm0.022$& $4.618\pm0.037$& 0 1 1 0 0 0 2 1 0\\
22422 &  1280\r1   485 1& $19.68\pm 0.96$& $ 20.80\pm0.45$&   6.63& $ 21.20\pm  0.45$&   3.56& 1.19& $3.7808\pm0.0026$& $-0.079\pm0.003$& $ 0.206\pm0.019$& $4.454\pm0.042$& 0 0 0 0 0 0 2 1 0\\
\hline
\hline
\end{tabular}
}
\end{center}
\end{table*}
\setlength{\textwidth}{177.6mm}
\setlength{\textheight}{56pc}
\end{landscape}

\begin{landscape}
\addtocounter{table}{-1}
\setlength{\textwidth}{56pc}
\setlength{\textheight}{177.6mm}
\begin{table*}[t]
\caption[]{Data for the 218 P98 Hyades candidates (continued).}
\begin{center}
{\scriptsize
\renewcommand{\arraystretch}{0.9}
\renewcommand{\tabcolsep}{5.7pt}
\begin{tabular}{lrrrrrrrrrrrrc}
\noalign{\vskip 0.07truecm}
\hline
\hline
\noalign{\vskip 0.07truecm}
\multicolumn{1}{c}{HIP}& \multicolumn{1}{c}{TYC}& \multicolumn{1}{c}{$\pi_{\rm Hip}$}& \multicolumn{2}{c}{$\pi_{\rm sec}$ \& $g$ [Hip]}& \multicolumn{2}{c}{$\pi_{\rm sec}$ \& $g$ [Tycho--2]}& \multicolumn{1}{c}{$M$}& \multicolumn{1}{c}{$\log T_{\rm eff}$}& \multicolumn{1}{c}{${\rm BC}_{V}$}& \multicolumn{1}{c}{$\log(L/L_\odot)$}& \multicolumn{1}{c}{$\log g$}& \multicolumn{1}{c}{a b c d e f g h i}\\[0.05truecm]
\noalign{\vskip 0.07truecm}
\hline
\noalign{\vskip 0.07truecm}
22496 &  1284\r1   332 1& $22.96\pm 1.17$& $ 24.38\pm0.46$&  20.57& $ 24.13\pm  0.40$&   3.11& 1.24& $3.7849\pm0.0026$& $-0.070\pm0.003$& $ 0.312\pm0.016$& $4.443\pm0.040$& 0 1 1 0 2 0 2 1 1\\
22505 &  1280\r1   662 1& $23.64\pm 0.99$& $ 21.86\pm0.47$&  13.07& $ 21.54\pm  0.53$&   5.78& 1.10& $3.7403\pm0.0034$& $-0.171\pm0.006$& $ 0.195\pm0.019$& $4.548\pm0.046$& 0 1 1 0 6 0 2 1 1\\
22524 &  1280\s1  1207 1& $19.30\pm 0.95$& $ 20.32\pm0.44$&   5.18& $ 20.65\pm  0.50$&   1.08& 1.29& $3.7925\pm0.0029$& $-0.054\pm0.003$& $ 0.388\pm0.019$& $4.421\pm0.040$& 0 1 1 0 0 0 2 1 0\\
22550 &   688\q1    43 1& $20.15\pm 1.14$& $ 21.32\pm0.53$&   3.99& $ 21.21\pm  0.58$&   0.03& 2.42& $3.7906\pm0.0034$& $-0.058\pm0.004$& $ 0.548\pm0.022$& $4.426\pm0.031$& 0 0 1 1 1 1 2 1 1\\
22565 &  1288\s1  1706 1& $17.27\pm 0.82$& $ 19.29\pm0.41$&   7.81& $ 19.48\pm  0.35$&   2.93& 2.05& $3.8970\pm0.0031$& $ 0.040\pm0.000$& $ 1.277\pm0.018$& $4.012\pm0.031$& 1 0 0 1 0 2 1 1 1\\
22566 &  1280\s1  1110 1& $17.14\pm 1.00$& $ 16.44\pm0.45$&   6.09& $ 16.38\pm  0.52$&   3.66& 1.21& $3.7951\pm0.0027$& $-0.049\pm0.003$& $ 0.326\pm0.024$& $4.413\pm0.044$& 0 0 0 0 0 0 2 1 0\\
22607 &   696\s1  1789 1& $23.91\pm 1.04$& $ 25.81\pm0.47$&  28.09& $ 22.11\pm  0.54$&   5.74& 3.55& $3.8020\pm0.0027$& $-0.034\pm0.002$& $ 0.569\pm0.016$& $4.393\pm0.023$& 0 0 1 1 1 1 2 1 1\\
22654 &  1288\q1    32 1& $18.93\pm 2.02$& $ 19.18\pm0.71$&   2.48& $ 19.91\pm  0.61$&   5.16& 0.79& $3.6766\pm0.0033$& $-0.425\pm0.022$& $-0.613\pm0.033$& $4.622\pm0.066$& 0 0 0 0 0 0 2 1 0\\
22850 &  1288\s1  1591 1& $14.67\pm 0.95$& $ 15.90\pm0.41$&   3.36& $ 16.15\pm  0.44$&   0.67& 1.72& $3.8748\pm0.0043$& $ 0.032\pm0.002$& $ 0.936\pm0.022$& $4.188\pm0.038$& 0 0 0 0 0 0 1 1 0\\
23044 &    85\s1  1179 1& $12.62\pm 1.89$& $ 14.91\pm0.79$&  55.75& $ 14.56\pm  0.50$& 120.35& 2.15& $3.7707\pm0.0025$& $-0.097\pm0.003$& $ 0.194\pm0.046$& $4.480\pm0.051$& 0 0 0 1 1 1 2 4 0\\
23069 &   696\r1   390 1& $19.66\pm 1.62$& $ 18.07\pm0.61$&   1.14& $ 18.29\pm  0.50$&   0.00& 0.98& $3.7410\pm0.0025$& $-0.169\pm0.005$& $-0.104\pm0.029$& $4.546\pm0.054$& 0 0 0 0 0 0 2 1 0\\
23205 &  2388\q1    60 1& $10.73\pm 1.66$& $ 15.13\pm0.61$&  54.39& $ 15.11\pm  0.53$&  59.54& 2.28& $3.7269\pm0.1269$& $-0.210\pm0.440$& $ 0.387\pm0.180$& $4.568\pm0.539$& 0 0 0 1 1 1 2 8 1\\
23214 &  1281\r1   792 1& $23.09\pm 0.83$& $ 23.31\pm0.41$&   2.17& $ 23.38\pm  0.44$&   0.23& 1.31& $3.8168\pm0.0044$& $-0.010\pm0.001$& $ 0.467\pm0.015$& $4.352\pm0.041$& 1 0 0 1 0 0 2 1 0\\
23312 &   106\r1   892 1& $16.77\pm 1.79$& $ 19.38\pm0.65$&   4.12& $ 20.26\pm  0.65$&   1.85& 0.91& $3.6964\pm0.0121$& $-0.313\pm0.065$& $-0.435\pm0.039$& $4.602\pm0.078$& 0 0 0 0 0 0 2 1 0\\
23497 &  1293\s1  2747 1& $20.01\pm 0.91$& $ 19.06\pm0.39$&   3.47& $ 18.93\pm  0.34$&   1.31& 2.10& $3.9115\pm0.0029$& $ 0.040\pm0.003$& $ 1.472\pm0.018$& $3.886\pm0.030$& 0 0 0 0 0 0 1 1 1\\
23498 &   697\s1  1892 1& $18.44\pm 1.66$& $ 18.47\pm0.58$&   0.03& $ 18.41\pm  0.60$&   0.67& 0.98& $3.7347\pm0.0025$& $-0.187\pm0.006$& $-0.160\pm0.027$& $4.556\pm0.053$& 0 1 0 0 2 0 2 1 0\\
23662 &   110\r1   641 1& $16.69\pm 1.12$& $ 14.19\pm0.57$&  35.88& $ 14.93\pm  0.75$&  49.54& 1.26& $3.8000\pm0.0040$& $-0.038\pm0.003$& $ 0.534\pm0.035$& $4.399\pm0.052$& 0 0 0 0 3 0 2 7 0\\
23701 &   110\s1  1206 1& $13.78\pm 2.08$& $ 18.61\pm0.78$&   6.44& $ 16.39\pm  0.76$&   0.38& 0.95& $3.6982\pm0.0025$& $-0.305\pm0.013$& $-0.463\pm0.037$& $4.600\pm0.059$& 0 1 1 0 2 5 2 4 1\\
23750 &  1286\s1  2135 1& $18.78\pm 1.40$& $ 18.85\pm0.52$&   0.18& $ 18.79\pm  0.50$&   0.08& 1.01& $3.7425\pm0.0037$& $-0.164\pm0.007$& $-0.115\pm0.024$& $4.544\pm0.051$& 0 0 0 0 0 0 2 1 0\\
23772 &  1853\r1   313 1& $12.00\pm 1.87$& $ 10.01\pm0.65$&  52.29& $  9.25\pm  0.45$& 229.31& 1.84&                  &                 &                 &                & 0 0 2 3 1 0 0 4 1\\
23983 &   702\s1  2789 1& $18.54\pm 0.83$& $ 19.01\pm0.51$&   0.44& $ 19.20\pm  0.52$&   0.34& 1.87& $3.8872\pm0.0031$& $ 0.038\pm0.000$& $ 1.151\pm0.023$& $4.059\pm0.035$& 1 1 1 1 0 0 1 1 1\\
24019 &  1853\s1  1958 1& $18.28\pm 1.30$& $ 17.94\pm0.45$&   5.61& $ 17.53\pm  0.68$&   1.22& 1.69& $3.8578\pm0.0034$& $ 0.030\pm0.003$& $ 1.007\pm0.022$& $4.290\pm0.036$& 1 0 0 1 1 3 2 1 1\\
24021 &   111\s1  1835 1& $21.39\pm 1.21$& $ 17.99\pm0.63$& 169.14& $ 15.30\pm  0.67$&  67.37& 1.21& $3.8128\pm0.0041$& $-0.017\pm0.001$& $ 0.431\pm0.030$& $4.363\pm0.050$& 0 0 0 0 0 0 2 7 0\\
24116 &  1290\r1   502 1& $11.56\pm 1.19$& $ 12.37\pm0.47$&   1.22& $ 12.49\pm  0.46$&   2.22& 1.41& $3.8182\pm0.0044$& $-0.008\pm0.002$& $ 0.581\pm0.033$& $4.348\pm0.048$& 0 0 0 0 0 0 2 1 0\\
24923 &   707\s1  1553 1& $18.26\pm 1.58$& $ 17.78\pm0.71$&   0.23& $ 18.39\pm  0.66$&   0.23& 0.98& $3.7347\pm0.0072$& $-0.187\pm0.018$& $-0.139\pm0.035$& $4.556\pm0.064$& 0 0 0 0 0 0 2 1 0\\
25639 &  2407\r1   211 1& $11.58\pm 1.13$& $ 14.23\pm0.35$& 223.09& $ 16.05\pm  0.44$& 185.78& 1.26& $3.7869\pm0.0052$& $-0.066\pm0.006$& $ 0.218\pm0.021$& $4.437\pm0.046$& 0 0 0 0 0 0 2 5 0\\
25694 &  1309\s1  2465 1& $11.17\pm 1.28$& $  9.20\pm0.62$& 133.50& $  7.63\pm  0.51$& 187.68& 1.36&                  &                 &                 &                & 0 0 0 0 0 0 0 7 0\\
25871 &  1856\r1   805 1& $11.55\pm 0.91$& $ 11.90\pm0.45$&  15.25& $ 12.26\pm  0.54$&   0.80& 1.51&                  &                 &                 &                & 0 0 0 0 0 1 0 2 0\\
26159 &  1309\s1  1314 1& $11.13\pm 1.39$& $  6.53\pm0.62$&  13.74& $  6.85\pm  0.57$&   1.49& 1.24&                  &                 &                 &                & 0 0 0 0 6 0 0 7 0\\
26382 &  1302\r1   362 1& $18.56\pm 0.86$& $ 19.71\pm0.65$&   4.61& $ 19.94\pm  0.63$&   0.04& 1.82& $3.8925\pm0.0036$& $ 0.034\pm0.000$& $ 1.081\pm0.029$& $4.139\pm0.040$& 0 0 0 0 0 0 1 1 0\\
26844 &  1298\s1  1046 1& $46.51\pm 2.35$& $ 39.13\pm1.46$&  23.57& $ 37.66\pm  1.09$&  10.72& 0.52&                  &                 &                 &                & 0 0 0 0 0 0 0 6 0\\
27431 &   715\s1  1157 1& $13.11\pm 0.87$& $ 11.04\pm1.15$& 119.46& $  9.01\pm  1.52$&  17.27& 1.54&                  &                 &                 &                & 0 0 0 0 0 0 0 8 0\\
27791 &  2414\s1  1204 1& $11.50\pm 6.04$& $ 12.60\pm1.80$&   2.39& $ 14.57\pm  0.41$& 617.15& 2.43&                  &                 &                 &                & 0 0 0 1 7 1 0 7 0\\
28356 &   121\r1   432 1& $14.87\pm 0.98$& $ 13.44\pm1.26$&   7.61& $ 12.40\pm  1.29$&   6.73& 1.31& $3.8137\pm0.0043$& $-0.015\pm0.002$& $ 0.536\pm0.081$& $4.360\pm0.090$& 0 0 0 0 0 0 2 1 0\\
28469 &  1876\r1   439 1& $10.52\pm 0.99$& $  8.47\pm0.45$&  65.79& $  8.93\pm  0.55$&  56.85& 1.57&                  &                 &                 &                & 0 0 0 0 0 0 0 4 0\\
28774 &  2419\s1  1011 1& $12.81\pm12.80$& $  2.48\pm7.03$&   0.93& $  1.17\pm  0.85$&   0.25& 1.04&                  &                 &                 &                & 1 0 0 1 1 1 0 8 0\\
\hline
\hline
\end{tabular}
}
\end{center}
\null\vskip-0.40truecm
{\it\noindent Notes to Table~A1:\/}
12031: white dwarf (footnote~20); 14838: giant $\delta$ Ari (\S\S
\ref{subsec:HRD_4} and \ref{subsec:trans_other_giants}); `$\Delta\mu$
binary' (Wielen et al.\ 2000); 15720: M dwarf (Figure~\ref{fig:B+L});
17962: white dwarf (\S \ref{subsec:HRD_5}); 18018: Tycho--1 number
suppressed in Tycho--2 catalogue; 19789: `$\Delta\mu$ binary' (Wielen
et al.\ 2000); 19870: `$\Delta\mu$ binary' (Wielen et al.\ 2000);
19877: X-ray star (Figure~\ref{fig:BVgap}); 20019: spectroscopic
binary (e.g., McClure 1982; Torres et al.\ 1997a; Lastennet et al.\
1999; Lebreton 2000); 20087: spectroscopic binary
(Table~\ref{tab:orbpar}; cf., e.g., Torres et al.\ 1997a; Lastennet et
al.\ 1999); $M_1 = 1.72 \pm 0.27~M_\odot$ and $M_2 = 1.31 \pm
0.21~M_\odot$ (S\"oderhjelm 1999); 20205: giant (\S\S
\ref{subsec:HRD_4} and \ref{subsec:trans_giants}); 20215: $\pi_{\rm
trigonometric} = 23.3 \pm 1.2$~mas (S\"oderhjelm 1999); `$\Delta\mu$
binary' (Wielen et al.\ 2000); 20255: `$\Delta\mu$ binary' (Wielen et
al.\ 2000); 20284: white dwarf (\S \ref{subsec:HRD_5}); 20400:
Table~\ref{tab:turn-off}; 20440: entry with resolved Tycho--2
photometry; 23205: entry with resolved Tycho--2 photometry; 20441:
`$\Delta\mu$ binary' (Wielen et al.\ 2000); no Tycho--2 proper motion;
20455: giant (\S\S \ref{subsec:HRD_4} and
\ref{subsec:trans_other_giants}); 20484: Table~\ref{tab:turn-off};
20542: companion of $\delta^1$ Tau (\ref {subsec:trans_other_giants});
20553: `$\Delta\mu$ binary' (Wielen et al.\ 2000); 20577: footnote~12;
20614: Table~\ref{tab:turn-off} and \S 9.2 in P98; `$\Delta\mu$
binary' (Wielen et al.\ 2000); 20648: blue straggler (\S
\ref{subsec:HRD_3}); `$\Delta\mu$ binary' (Wielen et al.\ 2000);
20661: spectroscopic binary (Table~\ref{tab:orbpar}; cf., e.g.,
Peterson \& Solensky 1987; Torres et al.\ 1997b); `$\Delta\mu$ binary'
(Wielen et al.\ 2000); 20679: `$\Delta\mu$ binary' (Wielen et al.\
2000); 20686: $\pi_{\rm trigonometric} = 24.1 \pm 1.1$~mas
(S\"oderhjelm 1999); 20711: Table~\ref{tab:turn-off}; 20713:
Table~\ref{tab:turn-off}; `$\Delta\mu$ binary' (Wielen et al.\ 2000;
cf.\ Simon \& Ayres 2000); 20719: `$\Delta\mu$ binary' (Wielen et al.\
2000); 20751: `$\Delta\mu$ binary' (Wielen et al.\ 2000); 20885: giant
(\S\S \ref{subsec:HRD_4} and \ref{subsec:trans_other_giants});
`$\Delta\mu$ binary' (Wielen et al.\ 2000); 20889: giant (\S\S
\ref{subsec:HRD_4} and \ref{subsec:trans_giants}); 20894: `resolved
spectroscopic binary' $\theta^2$ Tau (Table~\ref{tab:orbpar} and \S\S
\ref{subsec:HRD_3} and \ref{subsec:trans_theta2}); 20901:
Table~\ref{tab:turn-off} and \S 9.2 in P98; 20916: $\pi_{\rm
trigonometric} = 24.5 \pm 1.6$~mas (S\"oderhjelm 1999); `$\Delta\mu$
binary' (Wielen et al.\ 2000); 20995: `$\Delta\mu$ binary' (Wielen et
al.\ 2000); entry with resolved Tycho--2 photometry; 21008:
`$\Delta\mu$ binary' (Wielen et al.\ 2000); 21036: chromospherically
active star (\S \ref{subsec:HRD_2}); 21039: Table~\ref{tab:turn-off};
21053: `$\Delta\mu$ binary' (Wielen et al.\ 2000); 21138: Tycho--1
number suppressed in Tycho--2 catalogue; 21137: `$\Delta\mu$ binary'
(Wielen et al.\ 2000); 21273: Table~\ref{tab:turn-off}; 21280:
$\pi_{\rm trigonometric} = 26.0 \pm 1.5$~mas (S\"oderhjelm 1999);
`$\Delta\mu$ binary' (Wielen et al.\ 2000); 21459: `$\Delta\mu$
binary' (Wielen et al.\ 2000); 21543: `$\Delta\mu$ binary' (Wielen et
al.\ 2000); 21588: `$\Delta\mu$ binary' (Wielen et al.\ 2000); 21589:
Table~\ref{tab:turn-off}; 21670: Table~\ref{tab:turn-off} and \S 9.2
in P98; `$\Delta\mu$ binary' (Wielen et al.\ 2000); 21683:
Table~\ref{tab:turn-off}; 22157: Table~\ref{tab:turn-off}; 22221:
`$\Delta\mu$ binary' (Wielen et al.\ 2000); 22350: `$\Delta\mu$
binary' (Wielen et al.\ 2000); 22380: `$\Delta\mu$ binary' (Wielen et
al.\ 2000); 22496: `$\Delta\mu$ binary' (Wielen et al.\ 2000); 22505:
$\pi_{\rm trigonometric} = 20.0 \pm 1.0$~mas (S\"oderhjelm 1999);
22550: $\pi_{\rm trigonometric} = 20.21 \pm 1.08$~mas (S\"oderhjelm
1999); 22565: Table~\ref{tab:turn-off}; `$\Delta\mu$ binary' (Wielen
et al.\ 2000); 22607: $\pi_{\rm trigonometric} = 22.9 \pm 1.0$~mas
(S\"oderhjelm 1999); `$\Delta\mu$ binary' (Wielen et al.\ 2000);
23497: \S \ref{subsec:HRD_3} and Figure~\ref{fig:mass}; 23701:
`$\Delta\mu$ binary' (Wielen et al.\ 2000); 23772: `$\Delta\mu$
binary' (Wielen et al.\ 2000); 23983: Table~\ref{tab:turn-off}; 24019:
Table~\ref{tab:turn-off}; 24020: Table~\ref{tab:turn-off}.
\end{table*}
\setlength{\textwidth}{177.6mm}
\setlength{\textheight}{56pc}
\end{landscape}

\begin{landscape}
\setlength{\textwidth}{56pc}
\setlength{\textheight}{177.6mm}
\begin{table*}[ht]
\caption[]{The 15 new Hyades candidate members (\S
\ref{subsec:addi_mems}) in (extended) P98 table 2 format, but
excluding columns (b)--(m). Only HIP 19757 is a likely new member (\S
\ref{subsec:Hip_additional_members}). Columns (cf.\ P98):
{(a)}~Hipparcos and Tycho identifier;$\;$ {(n, o)}~Hipparcos parallax
and error (mas);$\;$ {(p, q)}~radial velocity and error
(km~s$^{-1}$);$\;$ {(r)}~source of radial velocity (`S' means data
from SIMBAD [{\tt http://cdsweb.u-strasbg.fr/Simbad.html}]);$\;$
{(s)}~`RV' $=$ radial velocity variable (from Duflot et al.\
1995);$\;$ {(t)}~`H$\,$I$\,$M' $=$ star was pre\-viously known, or
classified by Hipparcos, to have resolved components (from
field~H56);$\;$ {(u)}~`C$\,$G$\,$O$\,$V$\,$X' $=$ relevant part of the
Hipparcos Double and Multiple Systems Annex (DMSA; field~H59)
supplemented by `S' $=$ suspected Hipparcos binary (field~H61);$\;$
{(v)}~three-dimensional distance from the cluster center of mass
defined by 180 stars (table~3 in P98; in pc), based on Hipparcos
trigonometric parallax;$\;$ {(w)}~kinematic P98 membership statistic
$c$ (their eq.~16);$\;$ {(x)}~final membership assigned by P98 (`0'
$=$ non-member, `1' $=$ member);$\;$ {(y)}~Hoogerwerf \& Aguilar
(1999) membership `probability' $p_{\rm HA}$ (denoted $S$ in their
paper), based on proper motion and parallax data (their \S 2.4);$\;$
{(z)}~de Bruijne (1999a) membership probability $p_{\rm dB}$, based on
proper motion data only (his \S 2.4.4). The last four columns provide
the Hipparcos and Tycho--2 secular parallaxes $\pi_{\rm sec,
Hip/Tycho-2}$ and corresponding goodness-of-fit parameters $g_{\rm
Hip/Tycho-2}$ (\S \ref{sec:sec_pars}).}
\begin{center}
\renewcommand{\arraystretch}{0.9}
\renewcommand{\tabcolsep}{5.0pt}
\begin{tabular}{lrrcrrrrrrrrrr}
\noalign{\vskip -0.20truecm}
\hline
\hline
\noalign{\vskip 0.07truecm}
\multicolumn{1}{c}{HIP \& TYC}& \multicolumn{1}{c}{Parallax}& \multicolumn{1}{c}{Radial velocity}& \multicolumn{1}{c}{Multi-}& \multicolumn{5}{c}{Membership}& \multicolumn{4}{c}{Secular parallaxes}\\
& \multicolumn{1}{c}{$\pi_{\rm Hip}\pm\sigma_{\pi, {\rm Hip}}$}& \multicolumn{1}{c}{$v_{\rm rad}\pm\sigma_{v_{\rm rad}}$}& \multicolumn{1}{c}{plicity}& & \multicolumn{1}{c}{$c$}& & \multicolumn{1}{c}{$p_{\rm HA}$}& \multicolumn{1}{c}{$p_{\rm dB}$}& \multicolumn{2}{c}{Hipparcos}& \multicolumn{2}{c}{Tycho--2}\\
\multicolumn{1}{c}{(a)}& \multicolumn{1}{c}{(n, o)}& \multicolumn{1}{c}{(p, q, r)}& \multicolumn{1}{c}{(s, t, u)}& \multicolumn{1}{c}{(v)}& \multicolumn{1}{c}{(w)}& \multicolumn{1}{c}{(x)}& \multicolumn{1}{c}{(y)}& \multicolumn{1}{c}{(z)}& \multicolumn{1}{c}{$\pi_{\rm sec, Hip}$}& \multicolumn{1}{c}{$g_{\rm Hip}$}& \multicolumn{1}{c}{$\pi_{\rm sec, Tycho-2}$}& \multicolumn{1}{c}{$g_{\rm Tycho-2}$}\\
\noalign{\vskip 0.07truecm}
\hline
\noalign{\vskip 0.07truecm}
14232 = 1794~01818~1          & $ 20.01\pm0.88$& $+10.80\pm2.00$(S)& \phantom{RV~I~C}          & 19.20&      &  &     & 0.76& $  9.42\pm0.23$& $181.88$&$  8.71\pm 0.28$ &$  0.04$\\
18965 = 1826~00378~1          & $ 28.44\pm1.12$& $-10.30\pm2.00$(S)& \phantom{RV~I~C}          & 15.17&      &  &     & 0.95& $ 41.59\pm0.28$& $170.83$&$ 43.48\pm 0.32$ &$  7.09$\\
19757 = 0671~00217~1          & $ 16.56\pm4.48$&                   & \phantom{RV~I~C}          & 15.40&      &  & 0.10& 0.65& $ 20.19\pm1.04$& $  2.67$&$ 20.52\pm 0.71$ &$  5.26$\\
19981 = 1815~01907~1          & $ 30.56\pm1.52$& $+28.82\pm0.20$(1)& RV~I~\phantom{C}          & 14.80& 21.14& 0&     & 0.51& $ 22.26\pm0.42$& $ 36.83$&$ 22.04\pm 0.39$ &$  0.04$\\
20616 = 0676~00438~1          & $ 21.00\pm1.37$&                   & \phantom{RV~I~C}          &  3.91&      &  &     & 0.06& $  9.49\pm0.42$& $137.63$&$  9.56\pm 0.42$ &$ 76.52$\\
20693 = 0081~01629~1          & $ 22.03\pm0.90$& $+29.67\pm0.30$(1)& \phantom{RV~I~C}          &  9.31& 17.54& 0& 0.19& 0.55& $ 19.67\pm0.37$& $ 13.47$&$ 18.98\pm 0.47$ &$  5.29$\\
20777 = 1820~01418~1$^{\rm a}$&$ 25.72\pm6.36$&                    & \phantom{RV~I}~V          & 10.45&      &  &     & 0.01& $  5.56\pm1.42$& $ 23.77$&$              $ &$      $\\
20895 = 1833~00567~1          & $ 25.00\pm3.24$& $-16.00\pm9.99$(S)& \phantom{RV}~$\!$H~$\!\!\!\:$C & 9.98&  &  &     & 0.60& $ 14.38\pm0.79$& $ 11.21$&$ 15.60\pm 0.61$ &$ 11.19$\\
20904 = 2372~02101~1$^{\rm b}$& $ 18.42\pm1.61$& $-36.60\pm2.00$(S)& RV~I~C                    & 14.21&      &  &     & 0.79& $  5.77\pm0.63$& $ 64.88$&$  5.25\pm 0.37$ &$ 18.02$\\
21475 = 0690~00797~1          & $ 18.93\pm1.75$&                   & \phantom{RV}~I~\phantom{C}&  7.50& 20.92& 0&     & 0.37& $ 34.96\pm0.58$& $105.05$&$ 36.10\pm 0.51$ &$  6.29$\\
21760 = 1842~01264~1          & $ 13.17\pm0.94$& $+24.80\pm2.00$(S)& \phantom{RV~I~C}          & 32.39&      &  & 0.22&     & $ 11.55\pm0.35$& $  3.77$&$ 12.28\pm 0.36$ &$  0.04$\\
21961 = 1830~02127~1          & $ 20.03\pm0.71$& $ +7.70\pm2.00$(S)& \phantom{RV~I~C}          &  7.70&      &  &     & 0.28& $  5.50\pm0.40$& $556.20$&$  4.67\pm 0.32$ &$131.19$\\
22449 = 0096~01462~1          & $124.60\pm0.95$& $+24.10\pm0.90$(S)& RV~I~\phantom{C}          & 38.99&      &  & 0.67&     & $122.54\pm0.48$& $ 26.42$&$113.44\pm 0.38$ &$  0.04$\\
24480 = 1291~00385~1          & $ 16.54\pm1.40$& $+21.70\pm2.00$(S)& \phantom{RV~I}~G          & 17.51&      &  & 0.15& 0.21& $ 19.30\pm0.66$& $ 10.38$&$ 20.61\pm 0.48$ &$ 27.80$\\
25730 = 1860~00628~1$^{\rm c}$& $ 11.24\pm0.84$& $+13.20\pm2.00$(S)& \phantom{RV~I~C}          & 47.24&      &  & 0.25&     & $ 12.71\pm0.39$& $  8.60$&$ 12.98\pm 0.41$ &$ 17.69$\\
\hline
\hline\\[-0.60truecm]
\end{tabular}
\label{tab:additional_members}
\end{center}
$^{\rm a}$: T Tauri-type star with two components (SIMBAD); Tycho--1 number suppressed in Tycho--2 catalogue.\\
$^{\rm b}$: A component in double system (SIMBAD).\\
$^{\rm c}$: T Tauri-type star (SIMBAD).\\
\ \\
\ \\
\end{table*}
\setlength{\textwidth}{177.6mm}
\setlength{\textheight}{56pc}
\end{landscape}

\end{document}